\documentclass[%
 aip,
 amsmath,amssymb,
reprint,%
floatfix,
]{revtex4-1}

\usepackage{graphicx}
\usepackage{dcolumn}
\usepackage{comment}
\usepackage{bm}
\usepackage{multirow}
\usepackage{csquotes}
\usepackage{algpseudocode}
\usepackage[mathscr]{euscript}
\usepackage{amsmath}
\usepackage{bbm}
\usepackage{xcolor}

\newcommand\bra[1] {\langle {#1} |}
\newcommand\ket[1] {| {#1} \rangle}
\newcommand\braket[2] {\langle {#1} | {#2} \rangle}

\newcommand{\themi}{M_I}

\newcommand{\thezi}{Z_I}
\newcommand{\thepi}{\mathbf{P}_I}

\newcommand{\thepii}{\boldsymbol{\Pi}_I}

\newcommand{\thechii}{\boldsymbol{\chi}_I}
\newcommand{\theri}{\mathbf{R}_I}

\newcommand{\theai}{\mathbf{A}(\mathbf{R}_I)}

\newcommand{\theb}{\mathbf{B}}

\newcommand{\thennuc}{N_\mathrm{nuc}}

\newcommand{\theoph}{H}
\newcommand{\theophel}{H_\mathrm{el}}
\newcommand{\theoptnuc}{T_\mathrm{nuc}}

\newcommand{\theG}{\mathbf{G}}
\newcommand{\thenel}{N_\mathrm{el}}
\newcommand{\thee}{e}
\newcommand{\theari}{\mathbf{A}(\mathbf{r}_i)}
\newcommand{\thepiel}{\mathbf{p}_i}

\newcommand{\theFZone}{\begin{pmatrix}
B_{z}\mathbf{S} & (B_x - \mathrm iB_y)\mathbf{S} \\
(B_x + \mathrm iB_y)\mathbf{S} & -B_{z}\mathbf{S}
\end{pmatrix}}

\def\tPhi{\tilde \Phi}
\def\Ia{{(I\alpha)}}
\def\Jb{{(J\beta)}}
\def\Iaa{{I\alpha}}
\def\Jbb{{J\beta}}

\def\trMW0{\tr \!\Mcal^v_0 W_N}

\def\trNW0{\tr \!\Ncal^\rho_0 W_N}

\def\a{\tilde a}

\def\x0{{\mathbf x}_0}

\begin{document}

\title{Analytic Calculation of the Berry Curvature and Diagonal Born--Oppenheimer Correction for Molecular Systems in Uniform Magnetic Fields
}

\author{Tanner Culpitt}
\email{t.p.culpitt@kjemi.uio.no}
\affiliation
{Hylleraas Centre for Quantum Molecular Sciences,  Department of Chemistry, 
University of Oslo, P.O. Box 1033 Blindern, N-0315 Oslo, Norway}
\author{Laurens D. M. Peters}
\affiliation
{Hylleraas Centre for Quantum Molecular Sciences,  Department of Chemistry, 
University of Oslo, P.O. Box 1033 Blindern, N-0315 Oslo, Norway}
\author{Erik I. Tellgren}
\affiliation
{Hylleraas Centre for Quantum Molecular Sciences,  Department of Chemistry, 
University of Oslo, P.O. Box 1033 Blindern, N-0315 Oslo, Norway}
\author{Trygve Helgaker}
\affiliation
{Hylleraas Centre for Quantum Molecular Sciences,  Department of Chemistry, 
University of Oslo, P.O. Box 1033 Blindern, N-0315 Oslo, Norway}


\begin{abstract}
 The diagonal nonadiabatic term arising from the Born--Oppenheimer wave-function ansatz contains contributions from a vector and scalar potential. The former is provably zero when the wave function can be taken to be real valued, and the latter, known as the diagonal Born--Oppenheimer correction (DBOC), is typically small in magnitude. Therefore, unless high accuracy is sought, the diagonal nonadiabatic term is usually neglected when calculating molecular properties. In the presence of a magnetic field, the wave function is generally complex, and the geometric vector potential gives rise to a screening force that is qualitatively important for molecular dynamics. This screening force is written in terms of the Berry curvature and is added to the bare Lorentz force acting on the nuclei in the presence of the field. In this work, we derive analytic expressions for the Berry curvature and DBOC using both first and second quantization formalisms for the case of generalized and restricted Hartree--Fock theories in a uniform magnetic field. The Berry curvature and DBOC are calculated as a function of the magnetic field strength and the bond distance for the ground-state singlets of H$_2$, LiH, BH, and CH$^+$. We also examine the stability and time-reversal symmetry of the underlying self-consistent field solutions. The character of the DBOC and Berry curvature is found to depend upon the magnetic field and varies between molecules. We also identify instances of broken time-reversal symmetry for the dissociation curves of BH and CH$^+$.
\end{abstract}

\maketitle

\section{Introduction}

A crucial component for performing Born--Oppenheimer molecular dynamics (BOMD) in the presence of a magnetic field is the Berry curvature,\cite{Berry1984,Resta2000,Zwanziger1990,Moore1991,Mead1992} which is used to calculate a screening force due to the electrons in the molecular system.\cite{Schmelcher1988,Yin1992,Yin1994,Peternelj1993,Ceresoli2007,Culpitt2021,Peters2021} This screening force serves to counteract the bare Lorentz force acting on the nuclei by accounting for electronic shielding. For example, in neutral systems, exact cancellation of the bare Lorentz force and screening force is achieved for center-of-mass motion.\cite{Yin1992,Yin1994,Ceresoli2007,Culpitt2021,Peters2021} It is therefore qualitatively important to include the Berry curvature for dynamics simulations in a magnetic field. 

Recently, the Berry curvature was calculated for molecular systems using finite difference,\cite{Culpitt2021} accounting for the arbitrary global phase of the perturbed electronic wave functions. This scheme was applied to the dynamics of the H$_2$ molecule in uniform magnetic fields.\cite{Peters2021} While the finite difference scheme was shown to be successful in calculating the Berry curvature, it is advantageous to calculate the Berry curvature analytically. Analytic calculation circumvents possible issues related to stability, step size, and also obviates the need to account for the global phase of the wave function.

Another related quantity, the diagonal Born--Oppenheimer correction (DBOC), is a scalar potential that modifies the Born--Oppenheimer (BO) potential used in BOMD. The DBOC has been calculated both numerically\cite{Sellers1984,Cencek1997,Valeev2003,Schneider2019} and analytically\cite{Handy1986,Dinelli1995,Ioannou1996} for molecules in the absence of a field. It is typically small and has been shown to have a mostly negligible impact on many quantities of interest.\cite{Sellers1984,Handy1986,Dinelli1995,Ioannou1996,Cencek1997,Valeev2003,Schneider2019} As such, it is usually neglected when performing calculations in quantum chemistry, unless a high level of quantitative accuracy is sought. However, to the best of our knowledge,  little is known about the behavior of the DBOC in a magnetic field. For this reason, it is desirable to calculate the DBOC and study its dependence on the magnetic field.  

Here, we present derivations of  the analytic Berry curvature and DBOC in a uniform magnetic field, using both the first- and second-quantization formalisms.The resulting expressions were implemented in the software package {\sc london}.\cite{LondonProgram} The {\sc London} program has the capability to perform \emph{ab initio} molecular electronic-structure calculations in the presence of a magnetic field, using London atomic orbitals (also known as gauge-including atomic orbitals (GIAOs))\cite{London1937,Hameka1958,Ditchfield1976,Helgaker1991,Tellgren2008,Tellgren2012,Irons2017,Pausch2020} for gauge-origin invariant calculation of energies and molecular properties at various levels of theory including Hartree--Fock theory,\cite{Tellgren2008,Tellgren2009,Tellgren2012} (current-)density-functional theory,\cite{TELLGREN_JCP140_034101,FURNESS_JCTC11_4169}  full-configuration-interaction theory,\cite{Lange2012,Austad2020} coupled-cluster theory,\cite{Stopkowicz2015} and linear-response theory.\cite{Sen2019}
 
This work is organized as follows. Section II contains a presentation of the effective nuclear Hamiltonian in a magnetic field and the attendant equations of motion, as well as an overview of the generally complex coupled perturbed Hartree--Fock (CPHF) equations and the firt- and second-quantization derivations of the analytic Berry curvature and DBOC. Section III presents Berry curvature and DBOC results for H$_2$, LiH, BH, and CH$^+$ at the Restricted Hartree--Fock (RHF) level of theory. The work is summarized and future directions are given in Section~IV.

\section{Theory}

We consider a joint system of nuclei and electrons. Throughout this work, $I$ and $J$  serve as indices for the $\thennuc$ nuclei, and their lowercase counterparts $i$ and $j$ will serve as indices for the $\thenel$ electrons. We use the notation $\themi$, $\thezi$, and $\theri$ for the  mass, atom number, and position of nucleus $I$, respectively.  We use $\mathbf{r}_i$ and $\mathbf{p}_i$ for the position operator and momentum operator of electron $i$, respectively. 
The vectors of collective nuclear and electronic coordinates are denoted by $\mathbf{R}$ and $\mathbf r$, respectively.
The vector potential of a uniform magnetic field $\mathbf B$ at position $\mathbf u$ is given by $\mathbf A(\mathbf u) = \frac{1}{2} \theb \times (\mathbf u-\theG)$, where $\theG$ is the gauge origin. 

\subsection{Screened Lorentz Force and Berry Curvature}

All relevant equations in this subsection have been previously derived.\cite{Culpitt2021,Peters2021} Our purpose is to present the relevant equations and quantities that will be of interest for subsequent sections. 

The nonrelativistic Hamiltonian of a molecular system in a uniform magnetic can be written according to 
\begin{equation}
\theoph_\text{mol} = \theoptnuc + \theophel + V_\text{nuc}\ .
\label{ham_000}
\end{equation}
Here the nuclear kinetic energy operator is given by
\begin{equation}
\theoptnuc = \sum \limits_{I=1}^{\thennuc} \dfrac{\Pi_I^2}{2\themi} , \quad\thepii = \thepi - \thezi e \theai \ ,
\label{ham_001}
\end{equation}
where $\thepi = - \mathrm i\hbar \partial/\partial \theri$ is the canonical
momentum and $\thepii$ the physical momentum of nucleus $I$, while the
nuclear repulsion operator is given by
\begin{equation}
V_\text{nuc} = \sum_{I>J=1}^{N_\text{nuc}}\frac{Z_I Z_J e^2}{4 \pi \varepsilon_0\vert \mathbf R_I - \mathbf R_J\vert}
\end{equation}
where $e$ is the elementary charge and $\varepsilon_0$ the vacuum permittivity. The  electronic Hamiltonian is given by
\begin{align}
\theophel &= \frac{1}{2m_\text{e}} \sum_{i=1}^{\thenel} (\thepiel + \thee \theari)^2 
 \nonumber \\ &
 + \sum_{i>j=1}^{N_\text{el}}\frac{e^2}{4 \pi \varepsilon_0\vert \mathbf r_i - \mathbf r_j\vert}  - \sum_{i=1}^{N_\text{el}}\sum_{I=1}^{N_\text{nuc}}\frac{Z_I e^2}{4 \pi \varepsilon_0\vert \mathbf r_i - \mathbf R_I\vert}.
\label{v_nn} 
\end{align}
The  molecular wave function satisfies the time-dependent Schr\"odinger equation
\begin{equation}
H_\text{mol} \ket{\Psi} = \mathrm{i} \hbar \frac{\partial }{\partial t} \ket{\Psi} .
\label{molsch}
\end{equation}

In the Born--Oppenheimer (BO) approximation, the total ground-state wave function can be written as a product of nuclear and electronic wave functions
\begin{align}
\Psi(\mathbf{r},\mathbf{R},t) = \psi(\mathbf{r};\mathbf{R})\Theta(\mathbf{R},t)
\label{wf_def} \ ,
\end{align}
where $\Theta(\mathbf{R},t)$ is the nuclear wave function and $\psi(\mathbf{r};\mathbf{R})$ is the electronic wave function. From here onward, we suppress the arguments of the wave functions. The BO nuclear Schr\"odinger equation in a magnetic field takes the form
\begin{align}
    H \ket{\Theta} = (T + U) \ket{\Theta} = 
    \mathrm{i} \hbar \frac{\partial}{\partial t} \ket{\Theta}
    \label{HTUop} \ ,
\end{align}
where the kinetic- and potential-energy operators, respectively, are
\begin{align}
T &=\sum_{I=1}^{N_\text{nuc}}\frac{1}{2M_I}\overline \Pi_I^2 \,,
\label{Top}
\\
U &=  U_\text{BO} + U_\text{DBOC} \,.
\label{Uop}
\end{align}
In Eq.\,\eqref{Top}, $\overline {\boldsymbol \Pi}_I$ is the effective nuclear physical momentum
\begin{align}
    \overline {\boldsymbol \Pi}_I &= \boldsymbol \Pi_I + \boldsymbol \chi_I
    \nonumber \\ &= \mathbf P_I - Z_I e \mathbf A(\mathbf R_I) + \boldsymbol \chi_I
    \label{eff_mom}
\end{align}
where  $\boldsymbol \chi_I$ is the geometric vector potential
\begin{align}
\thechii &= \bra{\psi} \mathbf{P}_I \ket{\psi} \ .
\label{gvp_def}
\end{align}
In Eq.\,\eqref{Uop}, $U_\text{BO}$ is the BO scalar potential obtained from solving the electronic Schr\"odinger equation at a given nuclear configuration and $U_\text{DBOC}$ is the diagonal BO correction (DBOC)
\begin{align}
U_\text{DBOC} = \sum_{I=1}^{N_\text{nuc}}\frac{1}{2M_I} \overline \Delta_I \ ,
\label{U_dboc}
\end{align}
where
\begin{align}
    \overline \Delta _I &= \braket{\mathbf{P}_I \psi}{\mathbf{P}_I \psi} - \chi_I^2 \ .
\label{dboc_bar}
\end{align}

Note that $\overline \Delta_I$ is invariant to geometric gauge transformations of the electronic wave function $\psi$ of the sort
 \begin{equation}
 \psi'  = \mathrm{e}^{-\mathrm{i}  F(\mathbf{R})/\hbar} \psi \,,
\end{equation}
where the gauge function $F$ is a real-valued differentiable function of the nuclear coordinates $\mathbf{R}$.

Starting from Eq.\,\eqref{HTUop} and using Ehrenfest's theorem as well as locality assumptions regarding the nuclear wave function, the nuclear equations of motion become\cite{Culpitt2021,Peters2021}
\begin{align}
M_{I} \, \ddot{{\mathbf R}}_{I}
= \mathbf F^{\text{BO}}_{I}({\mathbf{R}})
+\mathbf F^{\text{L}}_{I}(\dot{\mathbf{R}})
+\mathbf F^{\text{B}}_{I}({\mathbf{R}},\dot{\mathbf{R}})
\label{eom_t015}
\end{align}
where we have introduced the BO force
\begin{align}
\mathbf{F}^{\text{BO}}_{I}({\mathbf{R}})
= - \boldsymbol \nabla_I  U_\text{BO}({\mathbf R}) \, ,
\label{BOforce}
\end{align}
the (bare) Lorentz force
\begin{equation}
\mathbf{F}^{\text{L}}_{I}(\dot{\mathbf{R}}) =- e Z_{I}  \, \mathbf B \times
\dot{{\mathbf{R}}}_{I} \, ,
\end{equation}
and the Berry (screening) force
\begin{equation}
\mathbf{F}^{\text{B}}_{I}(\mathbf{R},\dot{\mathbf{R}}) =\sum_{J} \boldsymbol{\Omega}_{IJ}( {{\mathbf{R}}}) \, \dot{{\mathbf{R}}}_{J} \, .
\label{BerryForce}
\end{equation}
The $\boldsymbol \nabla_I$ in~Eq.\,\eqref{BOforce} differentiates with respect 
to ${\mathbf R}_{I}$ and $\mathbf{\Omega}_{IJ}$ in Eq.\,\eqref{BerryForce} is the Berry curvature
\begin{align}
    \Omega_{I\alpha J\beta} &= \nabla_{J\beta}\chi_{I\alpha} - \nabla_{I\alpha} \chi_{J\beta} \nonumber \\&= \mathrm i \hbar 
    \big[
    \braket{\nabla_{I\alpha} \psi}{\nabla_{J\beta} \psi} -  \braket{\nabla_{J\beta} \psi}{\nabla_{I\alpha} \psi}
    \big] \nonumber \\
&= -2\hbar\text{Im}\braket{\nabla_{I\alpha} \psi}{\nabla_{J\beta} \psi}
    \label{eom_t006} \ .
\end{align}
where $I\alpha$ is a composite nuclear-Cartesian index.
The screened Lorentz force on nucleus $I$ is the sum of the bare Lorentz force and the Berry force on this nucleus. Henceforth, we omit the arguments ${\mathbf R}$ and $\dot{\mathbf R}$ to the forces. Finally, note that the Berry curvature is gauge invariant.\cite{Culpitt2021} For further information on the Berry curvature and its physical meaning in the context of molecular dynamics, see Refs.~\onlinecite{Ceresoli2007},~\onlinecite{Culpitt2021}, and~\onlinecite{Peters2021}.
\subsection{ Diagonal Nonadiabatic Matrix Elements in First Quantization}

The coupled perturbed Hartree--Fock (CPHF) equations are used to calculate the derivatives of orbital coefficients, which will be necessary when calculating the analytic Berry curvature and DBOC. The CPHF equations are well known,\cite{Gerratt1968,Pople1979,Takada1983,Osamura1986} and there is relatively little modification to the structure of the equations in the presence of a uniform magnetic field. Nevertheless, here we give an overview of the derivation for the generally complex CPHF equations in the spinor basis, for the Generalized Hartree--Fock (GHF) level of theory, later specialized to the Restricted Hartree--Fock (RHF) level of theory. 

Before deriving working equations, we note that there are different conventions concerning nomenclature for coupling elements and the Born--Oppenheimer approximation. Some authors use ``Born--Oppenheimer" to refer to the case where all nonadiabatic coupling elements are neglected, see for example Refs.~\onlinecite{Stanke2017_bookverbatim} and~\onlinecite{Malhado2014}. On the other hand, some authors use ``Born--Oppenheimer" to refer to the case where the diagonal nonadiabatic coupling elements are included in the nuclear equation, see Refs.~\onlinecite{Worth_2004} and~\onlinecite{Mead1979}. Additionally, ``nonadiabatic" coupling elements are understood by certain authors to mean coupling between different electronic states exclusively, but couplings between the same electronic state (diagonal) may be referred to as ``nonadiabatic" as well, see Refs.~\onlinecite{Schmelcher1988},~\onlinecite{Stanke2017_bookverbatim},~\onlinecite{Malhado2014}, and~\onlinecite{Worth_2004}. Here, we use the convention that ``Born--Oppenheimer" refers to the nuclear equation neglecting all couplings, and we refer to the diagonal couplings as ``nonadiabatic".

\subsubsection{Coupled Perturbed Hartree--Fock Theory}

In what follows, $p,q,r,s$ refer to general spinor indices, $i,j,k,l$ refer to occupied spinors, and $a,b,c,d$ refer to virtual spinors. Throughout this section, superscript Greek letters $\tau$,$\kappa$,$\eta$, and $\xi$ all refer to a spin index, which is to say $\tau,\kappa,\eta,\xi \ \in \ \{\uparrow,\downarrow\}$.

We adopt the chemist's notation for two particle integrals
\begin{align}
(\mu\nu|\lambda\sigma)=\int \!\mathrm  d\mathbf{r}_1 \mathrm d\mathbf{r}_2\phi_{\mu}^{*}(\mathbf{r}_1)\phi_{\nu}^{}(\mathbf{r}_1)r_{12}^{-1}\phi_{\lambda}^{*}(\mathbf{r}_2)\phi_{\sigma}^{}(\mathbf{r}_2)
\label{cphf_tpi},
\end{align}
and adopt the Coulomb--exchange shorthand notation
\begin{align}
(\mu\nu\Vert \lambda\sigma)= (\mu\nu|\lambda\sigma) - (\mu\sigma|\lambda\nu)
\label{cphf_tpi1} \ .
\end{align}
A spinor is written as a linear combination of basis functions $\phi(\mathbf{r})$ according to
\begin{align}
\Phi_{i}(\mathbf{r})
= \sum \limits_{\mu\tau} c^{\tau}_{\mu i}\phi_{\mu}(\mathbf{r})
\label{e_016} \ .
\end{align}
Thus, for $N$ basis functions,\textbf{} there are 2$N$ terms in the sum in Eq.\,\eqref{e_016} and the Fock and density matrices are blocked 2$N\times$2$N$ matrices in the atomic orbtial (AO) basis. The Fock matrix in a uniform magnetic field can be written as 
\begin{align}
F^{\tau\kappa}_{\mu\nu} &= h_{\mu\nu}\delta_{\tau\kappa} + G^{\tau\kappa}_{\mu\nu}(\mathbf{P}) + {}^{\mathrm Z}F^{\tau\kappa}_{\mu\nu}
\label{cphf_Fmat_def} \ .
\end{align}
where $h_{\mu\nu}$ is the one-electron matrix element in the presence of a magnetic field, $G^{\tau\kappa}_{\mu\nu}(\mathbf{P})$ is the two-electron matrix element given by 
\begin{align}
G^{\tau\kappa}_{\mu\nu}(\mathbf{P}) &= \sum\limits_{\lambda\sigma\eta\xi}\mathrm{P}^{\eta\xi}_{\lambda\sigma}[(\mu\nu|\sigma\lambda)\delta_{\tau\kappa}\delta_{\eta\xi} - (\mu\lambda|\sigma\nu)\delta_{\tau\eta}\delta_{\kappa\xi}]
\label{cphf_G_def} \ , \\
P^{\eta\xi}_{\mu\nu} &= \sum\limits_{i}c^{\eta}_{\mu i}c^{\xi *}_{\nu i}
\label{cphf_P_def} \ ,
\end{align}
and the blocked AO basis representation of the spin-Zeeman contribution to Eq.\,\eqref{cphf_Fmat_def} is given by\cite{Sen2018}
\begin{align}
{}^{\mathrm Z}\mathbf{F} = \dfrac{1}{2}\theFZone
\label{cphf_FZone} \ ,
\end{align}
where the overlap matrix elements are given by
\begin{align}
S_{\mu\nu}=\int\! \mathrm d\mathbf{r} \phi^{*}_{\mu}(\mathbf{r})\phi_{\nu}(\mathbf{r})
\label{cphf_overlap} \ .
\end{align}

Taking the derivative of a spinor coefficient with respect to some component of the nuclear position vector $R_x$ we have
\begin{align}
\dfrac{\partial c_{\mu p}^{\tau}}{\partial R_x}
= c_{\mu p}^{\tau, x} = \sum \limits_{n} c_{\mu n}^{\tau} U_{np}^{x} 
\label{cphf_001} \ ,
\end{align}
where the sum is over all spinors $n$ and the $\mathbf{U}^{x}$ matrix is the solution to the CPHF equations in the MO basis. In the case of noncanonical MOs, the overlap matrix is diagonal and the Fock matrix block diagonal;
\begin{align}
S_{pq} &= (\mathbf{C}^{\dagger}\mathbf{S}\mathbf{C})_{pq}= \delta_{pq}
\label{cphf_S1}, \\
F_{ai} &= (\mathbf{C}^{\dagger}\mathbf{F}\mathbf{C})_{ai} = 0.
\label{cphf_F1}
\end{align}
Since these equations must hold at all geometries, their differentiation gives the equations that determine the $\mathbf{U}^{x}$ matrix.

Differentiating the overlap matrix in Eq.\,\eqref{cphf_S1} in the spinor representation and setting the resulting expression equal to zero gives the conditions
\begin{align}
S_{pq}^{x}&=U^{x *}_{qp} + \mathscr{S}^{x}_{pq} + U^{x}_{pq} = 0
\label{cphf_002} \ , \\
\mathscr{S}_{pq}^{x}&=\sum\limits_{\mu\nu\tau\kappa}c_{\mu p}^{\tau *}S_{\mu\nu}^{x}c_{\nu q}^{\kappa}\delta_{\tau\kappa}
\label{cphf_002a} \ ,
\end{align}
and we have made use of Eq.\,\eqref{cphf_001} to arrive at Eq.\,\eqref{cphf_002}. For the off-diagonal elements of the $\mathbf{U}^{x}$ matrix further conditions will be obtained by differentiating the virtual--occupied elements of the Fock matrix, while the diagonal elements $U^{x}_{pp}$ may be chosen freely to satisfy the conditions
\begin{align}
U^{x *}_{pp} + U^{x}_{pp} + \mathscr{S}^{x}_{pp} = 0
\label{cphf_003} \ ,
\end{align}
implying that they are determined only up to an imaginary constant (corresponding to a phase factor).

Differentiating the virtual--occupied elements of the Fock matrix in Eq.\,\eqref{cphf_F1} and using Eq.\,\eqref{cphf_001}, we obtain
\begin{align}
F_{ai}^{x}& =\sum\limits_{j}U^{x *}_{ja}F_{ji} + \mathscr{F}^{x}_{ai} + \sum\limits_{b}U^{x}_{bi}F_{ab} = 0,
\label{cphf_005}
\end{align}
which may be rearranged to give
\begin{equation}
\sum\limits_{j}U^{x}_{aj}F_{ji} - \sum\limits_{b}U^{x}_{bi}F_{ab} = \mathscr{F}^{x}_{ai} -\sum\limits_{j}\mathscr{S}^{x}_{aj}F_{ji}
\label{cphf_006} \ ,
\end{equation}
where the first term on the right-hand side contains the contributions from the derivative AO Fock matrix:
\begin{align}
\mathscr{F}_{ai}^{x}&=\sum\limits_{\mu\nu\tau\kappa}c_{\mu a}^{\tau *}F_{\mu\nu}^{\tau\kappa,x}c_{\nu i}^{\kappa} 
\label{cphf_Fpq_def} \ .
\end{align}
It may be decomposed in the manner
\begin{align}
\mathscr{F}^{x}_{ai} &= \mathscr{F}^{(x)}_{ai} + \mathscr{G}_{ai}(\mathbf{P}^{x})
\label{cphf_007} \ , \\
\mathscr{F}^{(x)}_{ai} &= h^{x}_{ai} + {}^{\mathrm Z}\mathscr{F}^{x}_{ai} + \mathscr{G}^{x}_{ai}(\mathbf{P})
\label{cphf_008} \ ,
\end{align}
where the terms entering $\mathscr{F}^{(x)}_{ai}$ depend only on derivative integrals, not on derivatives of coefficients:
\begin{align}
h^{x}_{ai} &= \sum\limits_{\mu\nu\tau\kappa} c^{\tau *}_{\mu a} (h^{x}_{\mu\nu}) c^{\kappa}_{\nu i}\delta_{\tau\kappa}
\label{cphf_hx} \ ,
\\
{}^{\mathrm Z}\mathscr{F}^{x}_{ai} &= \sum\limits_{\mu\nu\tau\kappa} c^{\tau *}_{\mu a} ({}^{\mathrm Z}F^{\tau\kappa,x}_{\mu\nu}) c^{\kappa}_{\nu i}
\label{cphf_FZ2} \ , \\
\mathscr{G}^{x}_{ai}(\mathbf{P}) &= \sum\limits_{\mu\nu\tau\kappa} c^{\tau *}_{\mu a} (G^{\tau\kappa,x}_{\mu\nu}(\mathbf{P})) c^{\kappa}_{\nu i}
\label{cphf_009} \, ,  \\
G^{\tau\kappa,x}_{\mu\nu}(\mathbf{P}) &= \sum\limits_{\lambda\sigma\eta\xi}\mathrm{P}^{\eta\xi}_{\lambda\sigma}[(\mu\nu|\sigma\lambda)^{x}\delta_{\tau\kappa}\delta_{\eta\xi} \nonumber \\ &\qquad \qquad \quad - (\mu\lambda|\sigma\nu)^{x}\delta_{\tau\eta}\delta_{\kappa\xi}] \, .
\label{cphf_Gx_def} 
\end{align}
By contrast, the second contribution to $\mathscr{F}^{x}_{ai}$ in Eq.\,\eqref{cphf_007},
\begin{align}
\mathscr{G}_{ai}(\mathbf{P}^{x}) &= \sum\limits_{\mu\nu\tau\kappa} c^{\tau *}_{\mu a} (G^{\tau\kappa}_{\mu\nu}(\mathbf{P}^{x})) c^{\kappa}_{\nu i} \, ,
\label{cphf_010} 
\end{align}
depends on $\mathbf U^x$ through its dependence on the derivative density matrix $\mathbf P^x$, whose elements are given by
\begin{align}
P^{\eta\xi,x}_{\mu\nu} &= \sum\limits_{i}c^{\eta,x}_{\mu i}c^{\xi *}_{\nu i} + \sum\limits_{i}c^{\eta}_{\mu i}c^{\xi,x *}_{\nu i}.
\label{cphf_011}
\end{align}
To make the dependence on $\mathbf U^x$ explicit, we use Eq.\,\eqref{cphf_001} to expand the coefficient derivatives in Eq.\,\eqref{cphf_011} in terms of the response matrices. Separating out the sum over $n$ in Eq.\,\eqref{cphf_001} into occupied and virtual parts and using Eq.\,\eqref{cphf_002}, we obtain
\begin{align}
P^{\eta\xi,x}_{\mu\nu}
&= - \sum\limits_{ij}c^{\eta}_{\mu i}\mathscr{S}_{ij}^{x}c^{\xi *}_{\nu j} +  \sum\limits_{bi} c^{\eta}_{\mu b} U_{bi}^{x}c^{\xi *}_{\nu i} \nonumber \\
& \quad + \sum\limits_{bi} c^{\eta}_{\mu i} U_{bi}^{x*} c^{\xi *}_{\nu b}
\label{cphf_012} \ .
\end{align}
Substituting this result into Eq.\,\eqref{cphf_010} and evaluating 
$G^{\tau\kappa}_{\mu\nu}(\mathbf{P}^{x})$ according to Eq.\,\eqref{cphf_G_def},
we obtain
\begin{align}
\mathscr{G}_{ai}(\mathbf{P}^{x}) &= -\sum\limits_{kl}\mathscr{S}_{kl}^{x}(ai\Vert lk) \nonumber \\
& \quad + \sum\limits_{bj}\left[(ai\Vert jb)U_{bj}^{x} + (ai\Vert bj)U_{bj}^{x*}\right]
\label{cphf_013} \ ,
\end{align}
completing our discussion of Eq.\,\eqref{cphf_007}. 

We are now ready to set up the CPHF equations. Substituting the expression for
$\mathscr{F}^{x}_{ai}$ given in Eq.\,\eqref{cphf_007} into Eq.\,\eqref{cphf_006} and rearranging, we obtain
\begin{align}
&\sum\limits_{bj}\left[F_{ab}\delta_{ij} - F_{ji}\delta_{ab} + (ai\Vert jb)\right]U_{bj}^{x} \nonumber \\
& \quad + \sum\limits_{bj} (ai\Vert bj)U_{bj}^{x*} = - \mathscr{F}^{(x)}_{ai} + \sum\limits_{j}\mathscr{S}^{x}_{aj}F_{ji} \nonumber \\
& \qquad\qquad\qquad\qquad\qquad\quad +\sum\limits_{kl}\mathscr{S}_{kl}^{x}(ai\Vert lk)
\label{cphf_014} \ .
\end{align}
These equations are not sufficient to determine $U_{bj}^x$ and $U_{bj}^{x\ast}$ as independent linear parameters. To obtain a sufficient set of equations, we take the complex conjugate of both sides of Eq.\,\eqref{cphf_014} and arrive at the following system of linear equations for $U_{bj}^x$ and $U_{bj}^{x\ast}$:
\begin{align}
\mathbf{H}\mathbf{X}=\mathbf{b} \ ,
\label{cphf_015}
\end{align}
where $\mathbf{H}$ is the complex GHF Hessian matrix\cite{Pople_1977} and has the block structure
\begin{align}
\mathbf{H} = 
\begin{pmatrix}
\mathbf{A} & \mathbf{B} \\
\mathbf{B}^{*} & \mathbf{A}^{*}
\end{pmatrix}
\label{cphf_016}
\end{align}
with matrix elements
\begin{align}
A_{ai,bj} &= F_{ab}\delta_{ij} - F_{ji}\delta_{ab} + (ai\Vert jb)
\label{cphf_017} \ , \\
B_{ai,bj} &= (ai\Vert bj)
\label{cphf_018} \ .
\end{align}
The solution and right-hand side vectors are blocked as
\begin{align}
\mathbf{X} = 
\begin{pmatrix}
\mathbf{U}^{x} \\
\mathbf{U}^{x *}
\end{pmatrix} , \quad
\mathbf{b} = 
\begin{pmatrix}
\mathbf{b}_0 \\
\mathbf{b}_0^{*}
\end{pmatrix}
\label{cphf_020} \ ,
\end{align}
where the elements $\mathbf{b}_0$ are given by 
\begin{align}
(b_0)_{ai} = - \mathscr{F}^{(x)}_{ai} + \sum\limits_{j}\mathscr{S}^{x}_{aj}F_{ji} +\sum\limits_{kl}\mathscr{S}_{kl}^{x}(ai\Vert lk)
\label{cphf_021} \ .
\end{align}
We observe that there is little difference in basic structure between the standard CPHF equations and those presented here. Aside from the spin Zeeman term, the main difference is that any simplifications predicated upon assuming real-valued quantities cannot be undertaken in the context of magnetic fields.

\subsubsection{Berry Curvature and DBOC}

To calculate the Berry curvature and DBOC, we will be interested in wave function overlaps of the form $\braket{\nabla_{I\alpha} \psi}{\nabla_{J\beta} \psi}$ and $\braket{\psi}{\nabla_{I\alpha}  \psi}$. Expressing the electronic wave function in terms of a Slater determinant of spinors and evaluating the overlaps, we obtain
\begin{align}
\braket{\psi}{\nabla_{I\alpha} \psi}
&= \sum \limits_{i} \braket{\Phi_i}{\Phi_i^{I\alpha}} 
\label{wfd_002} \ , \\
\braket{\nabla_{I\alpha} \psi}{\nabla_{J\beta} \psi}
&= \sum \limits_{i} \braket{\Phi_i^{I\alpha}}{\Phi_i^{J\beta}} \nonumber \\
&\quad -\sum\limits_{i}\sum\limits_{j \neq i}\braket{\Phi_i^{I\alpha}}{\Phi_j}\braket{\Phi_j}{\Phi_i^{J\beta}} \nonumber \\
&\quad - \sum\limits_{i}\sum\limits_{j \neq i}\braket{\Phi_i}{\Phi_i^{I\alpha}}\braket{\Phi_j}{\Phi_j^{J\beta}}
\label{wfd_001} \ ,
\end{align}
where we have introduced the notation
\begin{align}
\dfrac{\partial \Phi_{i}}{\partial R_{I\alpha}} = \nabla_{I\alpha} \Phi_{i}
= \Phi_{i}^{I\alpha}
\label{wfd_003} \ .
\end{align}
Note that the sum in Eq.\,\eqref{wfd_002}, as well as each term in the sum, must either be zero (for the real-valued case) or pure imaginary. This follows from differentiation of the normalization condition for the electronic wave function and the spinors.

We wish to evaluate the Berry curvature and DBOC in terms of CPHF quantities. To begin, we note
\begin{align}
\Phi_{i}^{I\alpha} &= \sum \limits_{\mu\tau} c^{\tau}_{\mu i}\dfrac{\partial\phi_{\mu}}{\partial R_{I\alpha}}
+ \sum \limits_{\mu\tau} \dfrac{\partial c^{\tau}_{\mu i}}{\partial R_{I\alpha}}\phi_{\mu} \nonumber \\
&=\Phi_{i}^{(I\alpha)} + \sum \limits_{r} U_{ri}^{I\alpha}\Phi_{r} 
\label{wfd_004} \ ,
\end{align}
where the term $\Phi_{i}^{(I\alpha)}$ is the derivative of the spinor differentiating only the basis functions and leaving the coefficients unchanged (denoted by the superscript $(I\alpha)$) and $U_{ri}^{I\alpha}$ is the solution to the CPHF equations. Substituting Eq.\,\eqref{wfd_004} into Eq.\,\eqref{wfd_001} and carrying out the requisite algebraic manipulations, we find
\begin{align}
\braket{\psi}{\nabla_{I\alpha} \psi} 
&= \sum\limits_{i}\left(\braket{\Phi_{i}}{\Phi_{i}^{(I\alpha)}} + U_{ii}^{I\alpha}\right)
\label{wfd_006} \, ,
\end{align}
and
\begin{align}
&\braket{\nabla_{I\alpha} \psi}{\nabla_{J\beta} \psi} 
= \sum \limits_{i}\braket{\Phi_i^{(I\alpha)}}{\Phi_i^{(J\beta)}} \nonumber \\
& \quad + \sum\limits_{ir}\!\!\left(\braket{\Phi_{r}}{\Phi_{i}^{(I\alpha)}} + U_{ri}^{I\alpha}\right)^{\!*}\!\!\left(\braket{\Phi_{r}}{\Phi_{i}^{(J\beta)}} + U_{ri}^{J\beta}\right)  \nonumber \\
& \quad - \sum\limits_{ir}\braket{\Phi_{i}^{(I\alpha)}}{\Phi_{r}}\braket{\Phi_{r}}{\Phi_{i}^{(J\beta)}} \nonumber \\
& \quad - \sum \limits_{i\neq j}\!\!\left(\braket{\Phi_{j}}{\Phi_{i}^{(I\alpha)}} + U_{ji}^{I\alpha}\right)^{\!*}\!\!\left(\braket{\Phi_{j}}{\Phi_{i}^{(J\beta)}} + U_{ji}^{J\beta}\right) \nonumber \\
& \quad -\sum \limits_{i\neq j}\!\!\left(\braket{\Phi_{i}}{\Phi_{i}^{(I\alpha)}} + U_{ii}^{I\alpha}\right) \!\!\left(\braket{\Phi_{j}}{\Phi_{j}^{(J\beta)}} + U_{jj}^{J\beta}\right),
\end{align}
which may be rearranged to give
\begin{align}
&\braket{\nabla_{I\alpha} \psi}{\nabla_{J\beta} \psi} = \sum \limits_{i}\braket{\Phi_i^{(I\alpha)}}{\Phi_i^{(J\beta)}} \nonumber \\
& \quad + \sum\limits_{ia}\left(\braket{\Phi_{a}}{\Phi_{i}^{(I\alpha)}} + U_{ai}^{I\alpha}\right)^{\!*}\!\!\left(\braket{\Phi_{a}}{\Phi_{i}^{(J\beta)}} + U_{ai}^{J\beta}\right)  \nonumber \\
& \quad - \sum\limits_{ir}\braket{\Phi_{i}^{(I\alpha)}}{\Phi_{r}}\braket{\Phi_{r}}{\Phi_{i}^{(J\beta)}} \nonumber \\
& \quad -\sum \limits_{ij}\left(\braket{\Phi_{i}}{\Phi_{i}^{(I\alpha)}} + U_{ii}^{I\alpha}\right) \!\!\left(\braket{\Phi_{j}}{\Phi_{j}^{(J\beta)}} + U_{jj}^{J\beta}\right)
\label{wfd_005} \, .
\end{align}
Since the diagonal elements $U_{ii}^{I\alpha}$ and $U_{ii}^{J\beta}$ are determined only up to an imaginary constant (phase factor), the first- and second-order nonadiabatic matrix elements are determine up to imaginary and real constants, respectively.

Focusing on the Berry curvature, we see from the expression $\Omega_{I\alpha J\beta} = -2\hbar\text{Im}\braket{\nabla_{I\alpha} \psi}{\nabla_{J\beta} \psi}$ given in
Eq.\,\eqref{eom_t006} that any provably real-valued contributions to Eq.\,\eqref{wfd_005} may be discarded. As such, the Berry curvature can be written
\begin{align}
&\Omega_{I\alpha J\beta}  = -2\hbar\text{Im}\bigg[\sum \limits_{i}\braket{\Phi_i^{(I\alpha)}}{\Phi_i^{(J\beta)}} \nonumber \\
& \quad + \sum\limits_{ia}\left(\braket{\Phi_{a}}{\Phi_{i}^{(I\alpha)}} + U_{ai}^{I\alpha}\right)^{\!\ast}\!\!\left(\braket{\Phi_{a}}{\Phi_{i}^{(J\beta)}} + U_{ai}^{J\beta}\right)  \nonumber \\
& \quad - \sum\limits_{ir}\braket{\Phi_{i}^{(I\alpha)}}{\Phi_{r}}\braket{\Phi_{r}}{\Phi_{i}^{(J\beta)}}\bigg] \nonumber \\
&= -2\hbar\text{Im}\bigg[\sum \limits_{i}\braket{\Phi_i^{(I\alpha)}}{\Phi_i^{(J\beta)}} \nonumber \\
& \quad + \sum\limits_{ia} \braket{\Phi_{a}}{\Phi_{i}^{(J\beta)}} U_{ai}^{I\alpha*} + \sum\limits_{ia} \braket{\Phi_{i}^{(I\alpha)}}{\Phi_{a}} U_{ai}^{J\beta} \nonumber \\
& \quad + \sum\limits_{ia} U_{ai}^{I\alpha*} U_{ai}^{J\beta} - \sum\limits_{ij}\braket{\Phi_{i}^{(I\alpha)}}{\Phi_{j}}\braket{\Phi_{j}}{\Phi_{i}^{(J\beta)}}\bigg]
\label{wfd_007} \ ,
\end{align}
which is independent of the phase factors in~Eq.\,\eqref{wfd_005}.

We finally turn to the analytical calculation of the DBOC, which according to Eq.\,\eqref{dboc_bar} is given by
\begin{equation}
\overline\Delta_{I\alpha} = \braket{\nabla_{I\alpha} \psi}{\nabla_{I\alpha}\psi} + \braket{\psi}{\nabla_{I\alpha} \psi}^2
\end{equation}
From Eqs.\,\eqref{wfd_006} and~\eqref{wfd_005}, we see that the first-order nonadiabatic matrix elements remove the phase-factors from the second-order nonadiabatic matrix element, giving the following DBOC:
\begin{align}
\overline\Delta_{I\alpha} 
&= \sum \limits_{i}\braket{\Phi_i^{(I\alpha)}}{\Phi_i^{(I\alpha)}}-  \sum\limits_{ir}\left\vert \braket{\Phi_{i}^{(I\alpha)}}{\Phi_{r}}\right\vert^2  \nonumber \\
&\quad + \sum\limits_{ia}\left\vert \braket{\Phi_{a}}{\Phi_{i}^{(I\alpha)}} + U_{ai}^{I\alpha}\right\vert^{2}
\label{wfd_008} \ .
\end{align}
Like the Berry curvature given in Eq.\,\eqref{wfd_007}, the DBOC depends only on the occupied--virtual elements of $\mathbf{U}^{I\alpha}$.

\subsubsection{Closed-Shell CPHF and Stability Considerations}

Beginning from the CPHF equations in the spinor basis in Section II.B., it is straightforward to generate the corresponding UHF or RHF equations. For the purposes of this work, we are interested in examining singlet surfaces in a magnetic field. As such, we have only implemented the closed-shell CPHF equations. These equations are identical in structure to those already presented, but for the sake of completeness we write them here. Restricting the indices $i,j,k,l$ to doubly occupied spatial orbitals and $a,b,c,d$ to doubly occupied virtual orbitals, the linear system of equations to be solved in the RHF case is given by 
\begin{align}
\mathbf{H}^1\mathbf{X}^1=\mathbf{b}^1 \ ,
\label{cphf_cs1}
\end{align}
where $\mathbf{H}^1$ is the complex RHF Hessian or stability matrix matrix\cite{Pople_1977} and has the block structure
\begin{align}
\mathbf{H}^1 = 
\begin{pmatrix}
\mathbf{A}^1 & \mathbf{B}^1 \\
\mathbf{B}^{1*} & \mathbf{A}^{1*}
\end{pmatrix}
\label{cphf_cs2}
\end{align}
with 
\begin{align}
A_{ai,bj}^1 &= F_{ab}\delta_{ij} - F_{ji}\delta_{ab} + 2(ai|jb) - (ab|ji)
\label{cphf_cs3} \ , \\
B_{ai,bj}^1 &= 2(ai|bj) - (aj|bi) 
\label{cphf_cs4} \ .
\end{align}
The solution vector $\mathbf{X}^1$ and right-hand side $\mathbf b$ are also blocked, with
\begin{align}
\mathbf{X}^1 = 
\begin{pmatrix}
\mathbf{U}^{x} \\
\mathbf{U}^{x *}
\end{pmatrix}\ , 
\quad 
\mathbf{b}^1 = \begin{pmatrix}
\mathbf{b}_0^1 \\
\mathbf{b}_0^{1*}
\end{pmatrix}
\label{cphf_cs6} \ ,
\end{align}
where the elements $\mathbf{b}_0^1$ are given by 
\begin{align}
(b_0^1)_{ai} &= - \mathscr{F}^{(x)}_{ai} + \sum\limits_{j}\mathscr{S}^{x}_{aj}F_{ji} \nonumber \\
&\quad + \sum\limits_{kl}\mathscr{S}_{kl}^{x}\big[2(ai|lk) - (ak|li)\big]
\label{cphf_cs7} \ .
\end{align}
Note that the derivative terms appearing in Eq.\,\eqref{cphf_cs7} can be calculated from the analogous expressions appearing in Section II.B., accounting for the obvious restrictions in spin. Additionally, the final RHF Berry curvature and DBOC expressions in terms of doubly occupied spatial orbitals are \textit{exactly} the same as those in terms of spinors, with the exception that in the RHF case each term appearing in Eqs.\,\eqref{wfd_007} and~\eqref{wfd_008} needs to be multiplied by a factor of two. 

There is an important benefit to solving Eq.\,\eqref{cphf_cs1} for singlet surfaces, which is access to the complex RHF stability matrix given in Eq.\,\eqref{cphf_cs2}. Diagonalization of the complex RHF stability matrix and subsequent examination of the eigenvalues allows for classification of the SCF stationary point with respect to other complex RHF solutions.\cite{Pople_1977} Positive-definite stability matrices indicate minima, mixed positive/negative eignenvalues indicate saddle points, and zero eigenvalues are indeterminate, though often associated with symmetry breaking.\cite{Cui_2013} 

\subsection{Diagonal Nonadiabatic Matrix Elements in Second Quantization}

In the present section, we derive expressions for nonadiabatic matrix elements using the formalism of second quantization. As we shall see, the resulting second-quantization expression is equivalent but not identical to the first-quantization expression derived above, with slightly different linear equations to be solved. Although the second-quantization formulation is here given only for Hartree--Fock theory, it can easily be extended to many-body theories such as coupled-cluster theory, for which the second-quantization treatment is particularly well suited. Second quantization has previously been used to derive an expression for the atomic axial tensor in vibrational circular dichroism (VCD) for multi-configuration self-consistent field (MCSCF) wave functions\cite{Bak_1993}, but without the use of the natural connection, as done here. For a discussion of the natural connection, see Refs.\,\onlinecite{NatCon1_1995,NatCon2_1995}.

\subsubsection{Electronic Hamiltonian in the Natural Connection}

Since the second-quantization formalism is most transparent and easy to manipulate in an orthonormal basis, we begin by constructing a set of MOs that are manifestly orthonormal at all
geometries.\cite{Helgaker_1984,Helgaker_1986,Helgaker_1988} For this purpose, let
\begin{equation}
\Phi_p (\mathbf r; \mathbf R) = \sum_\mu c_{\mu p}^{(0)} \phi_\mu(\mathbf r;\mathbf R) \label{MOs}
\end{equation}
be a set of MOs that are constructed to be orthonormal at the reference geometry $\mathbf R = \mathbf R_0$: 
\begin{equation}
\mathbf S(\mathbf R_0) = \mathbf I,
\end{equation}
while $\mathbf S(\mathbf R) \neq \mathbf I$ for $\mathbf R \neq \mathbf R_0$.
At each $\mathbf R \neq \mathbf R_0$, a set of \emph{orthonormalized MOs (OMOs)} are obtained by orthormalization of the unperturbed reference MOs:
\begin{equation}
\tPhi_p (\mathbf r; \mathbf R) =
\sum_q T_{qp}(\mathbf R) \Phi_q(\mathbf r;\mathbf R) \label{OMOdef}
\end{equation}
where the $T_{qp}(\mathbf R)$ are the elements of the \emph{connection matrix} $\mathbf T(\mathbf R)$,\cite{Helgaker_1986,Helgaker_1988,NatCon1_1995} which
satisfies
\begin{equation}
\mathbf T^\dagger(\mathbf R) \mathbf S(\mathbf R) \mathbf T(\mathbf R) = \mathbf I \!\!\iff\!\! \mathbf T(\mathbf R) \mathbf T^\dagger(\mathbf R) = \mathbf S^{-1}(\mathbf R) \label{ortho}
\end{equation}
at an arbitrary geometry $\mathbf R$ with the special case
\begin{equation}
\mathbf T(\mathbf R_0) = \mathbf I . \label{T0S0}
\end{equation}
We may choose
$\mathbf T(\mathbf R) = \mathbf S^{-1/2}(\mathbf R)$ but this \emph{symmetric connection}\cite{Helgaker_1984} is not optimal, introducing unnecessarily large changes in
the orbitals as we distort the geometry. Instead, we use the \emph{natural connection},\cite{NatCon1_1995} with connection matrix:
\begin{equation}
\mathbf T(\mathbf R) = \mathbf W^{-1}(\mathbf R) \left[\mathbf W(\mathbf R) \mathbf S^{-1}(\mathbf R) \mathbf W^\dagger(\mathbf R)\right]^{1/2},
\end{equation}
where $\mathbf W$ contains overlaps between MOs at $\mathbf R_0$ and $\mathbf R$:
\begin{equation}
W_{pq}(\mathbf R) = \langle \Phi_p(\mathbf R_0) \vert \Phi_q(\mathbf R) \rangle. \label{Wmat}
\end{equation}
In terms of the OMOs, we may now  construct the second-quantization Hamiltonian in the usual manner:\cite{Helgaker_1984}
\begin{align}
\tilde H(\mathbf R) &= \sum_{pq} \tilde h_{pq}(\mathbf R) \a_p^\dagger(\mathbf R) \a_q(\mathbf R) \nonumber \\ &+ \frac{1}{2} \sum_{pqrs} \tilde g_{pqrs}(\mathbf R)
\a_p^\dagger(\mathbf R)
\a_r^\dagger(\mathbf R)
\a_s(\mathbf R)
\a_q(\mathbf R)
\end{align}
associating a creation operator $\a_p^\dagger(\mathbf R)$ with
each OMO. Together with the corresponding
annihilation operators $\a_p(\mathbf R)$, they satisfy
the usual anticommutation relations of second quantization at all values of $\mathbf R$. Here and in the following, we use tilde to denote quantities (operators and integrals) in the OMO basis, at a general geometry $\mathbf R$.

At $\mathbf R \neq \mathbf R_0$, the MOs $\Phi_p(\mathbf R)$
at $\mathbf R$ in Eq.\;\eqref{MOs} cannot be exactly
represented in the basis of the MOs $\Phi_p(\mathbf R_0)$
at $\mathbf R_0$.  Introducing the orthogonal complement of the space spanned by
the MOs at $\mathbf R_0$, we can write the resolution of identity in the manner
\begin{equation}
\sum_p \vert \Phi_p(\mathbf R_0) \rangle \langle \Phi_p(\mathbf R_0)   \vert +\!
\sum_u \vert \Phi_u(\mathbf R_0)  \rangle \langle \Phi_u(\mathbf R_0)   \vert  =\!1,
\end{equation}
where index $p$ is used for MOs in the primary basis of Eq.\,\eqref{MOs}
and index $u$ for the MOs in its complement. Differentiating the
condition in~Eq.\,\eqref{ortho}  and  $\mathbf W$ as defined in Eq.\,\eqref{Wmat}
with respect to some (unspecified) nuclear coordinate at $\mathbf R_0$, we obtain
$\mathbf T^\prime + \mathbf S^\prime + ({\mathbf T}^\prime)^\dagger = \mathbf 0$ and
$\mathbf W^\prime + ({\mathbf W}^\prime)^\dagger = \mathbf S^\prime$, respectively, which in combination give the following
expression for the first derivative of the connection matrix at the unperturbed geometry:
\begin{equation} 
\mathbf T^\prime(\mathbf R_0) = -  \mathbf W^\prime(\mathbf R_0), \label{eqTW}
\end{equation}
where 
\begin{equation}
W_{rq}^\prime(\mathbf R_0) = \langle \Phi_r(\mathbf R_0) \vert \Phi_q^\prime (\mathbf R_0)\rangle.
\label{Wrq}
\end{equation}
Next, differentiating the OMOs in Eq.\,\eqref{OMOdef} and using the expression for the differentiated connection matrix in in Eq.\,\eqref{eqTW}, we arrive at
the following expression for the first derivative of the OMOs:
\begin{align}
\tPhi_q^\prime(\mathbf r;\mathbf R_0) \label{eq80}
= \Phi^\prime_q(\mathbf r;\mathbf R_0)& - \sum_{r} W_{rq}^\prime(\mathbf R_0) \Phi_r(\mathbf r;\mathbf R_0)
\end{align}
We now establish an important consequence of the natural connection.
Multiplying Eq.\,\eqref{eq80} from the left by $\Phi_p^\ast(\mathbf r;\mathbf R_0)$ and by
$\Phi_u^\ast(\mathbf r;\mathbf R_0)$, integrating and invoking the orthonormality of
MOs, we obtain, respectively,
\begin{alignat}{2}
\langle \Phi_p(\mathbf R_0) \vert \tPhi_q^\prime(\mathbf R_0) \rangle &= 0
\, , \\
\langle \Phi_u(\mathbf R_0) \vert \tPhi_q^\prime(\mathbf R_0) \rangle &= \langle \Phi_u(\mathbf R_0) \vert \Phi_q^\prime(\mathbf R_0) \rangle \, \label{fuder}
\end{alignat}
In the natural connection, therefore, the \emph{first-derivative OMOs at $\mathbf R_0$ have no component in the original
MOs basis, belonging entirely to the orthogonal complement}.\cite{NatCon2_1995}
We may now write the first derivative of the OMOs
and of the corresponding creation operators as
\begin{align}
\tPhi_p^\prime(\mathbf r;\mathbf R_0) &= \sum_u \langle \Phi_u(\mathbf R_0) \vert \Phi_p^\prime(\mathbf R_0) \rangle \Phi_u(\mathbf r;\mathbf R_0) , \label{eq82}\\
(\tilde a^\dagger_p)^\prime(\mathbf R_0) &= \sum_u \langle \Phi_u(\mathbf R_0) \vert \Phi_p^\prime(\mathbf R_0) \rangle a^\dagger_u(\mathbf R_0) ,
\label{creder}
\end{align}
where the summations are only over MOs in the orthogonal complement.

\subsubsection{Hartee--Fock Diagonal Nonadiabatic Matrix Elements}

Suppressing the dependence of the creation and annihilation operators on $\mathbf R$, an $N$-electron
single-determinant wave function at $\mathbf R$ may be written as a unitarily transformed product of $N$ OMO creation operators,
\begin{equation}
\vert \mathbf R, \boldsymbol \kappa \rangle = \mathrm e^{\mathrm i \tilde \kappa} \vert \tilde 0 \rangle
= \mathrm e^{\mathrm i \tilde \kappa} \prod_{i=1}^N \tilde a_i^\dagger \vert \text{vac} \rangle. \label{HFstate}
\end{equation}
The unitary operator $ \mathrm e^{\mathrm i \tilde \kappa}$ is expressed in terms of
the Hermitian orbital-rotation operator  $\tilde \kappa = \tilde \kappa^\dagger$ given by
\begin{equation}
\tilde \kappa = \sum_{p > q} \kappa_{pq} \a_p^\dagger \a_q + \sum_{p > q} \kappa_{pq}^\ast \a_q^\dagger \a_p + \sum_p \kappa_{pp} \a_p^\dagger \a_p
\label{hatkappa}
\end{equation}
where the $\kappa_{pq}$ are the elements of a Hermitian matrix $\boldsymbol \kappa$, whose off-diagonal elements are related as $\kappa_{pq} = \kappa_{qp}^\ast$,
while the diagonal elements $\kappa_{pp}$ are real. In the following,
we will treat $\kappa_{pq}$ and $\kappa_{pq}^\ast$ with $p>q$ and $\kappa_{pp}$ as independent variational parameters.

Differentiating the expression for the HF state given in Eq.\;\eqref{HFstate} with respect to some
(unspecified geometric parameter) at $\mathbf R_0$, we obtain
\begin{equation}
\vert 0^\prime \rangle =  \mathrm i \kappa^\prime  \vert 0 \rangle + \sum\nolimits_i (\a_i^\dagger)^\prime a_i \vert 0 \rangle
\label{eq:perthf0}
\end{equation}
where 
\begin{equation}
\vert 0 \rangle
=  \prod_{i=1}^N  a_i^\dagger \vert \text{vac} \rangle \,
\end{equation}
is the variationally optimized HF wave function at $\mathbf R = \mathbf R_0$, which is parameterized such that
$\tilde \kappa = 0$. Since $\vert 0^\prime \rangle$ is calculated at $\mathbf R_0$, the OMOs reduce to the MOs. We have therefore removed the tilde form the operators except in
$(\a_i^\dagger)^\prime$, since the derivative is calculated for the OMOs.

In Eq.\,\eqref{eq:perthf0}, 
the first term represents a unitary transformation of MOs within the orbital basis and the second term the change in the orbital basis. Note that the second term in Eq.\,\eqref{eq:perthf0} represents a sum of states, where in each state one of the creation operators in the HF state has been replaced by the corresponding differentiated operator (in agreement with the rule for differentiation of a product).

Using Eq.\,\eqref{hatkappa} (noting that only the first term contributes) and Eq.\,\eqref{creder} and noting that
\begin{equation}
a_i^\dagger a_j \vert 0 \rangle = a_i^\dagger a_a \vert 0 \rangle =  a_a^\dagger a_b \vert 0 \rangle = 0
\end{equation}
where $i \neq j$,
we arrive at the following more explicit expression for the first-order wave function:
\begin{align}
\vert 0^\prime \rangle &= \left(  \sum_{ai} \mathrm i \kappa_{ai}^\prime  a_a^\dagger a_i+  
 \sum_{ui} \langle \Phi_u \vert \Phi_i^\prime \rangle a^\dagger_u a_i  \! \! \right) \!\vert 0 \rangle
\label{eq:perthf}
\end{align}
The overlap with the unperturbed wave function becomes
\begin{equation}
\langle 0 \vert 0^\Ia \rangle = \sum_i  \mathrm i \kappa_{ii}^{\Iaa} 
,
\end{equation}
while the overlap between two such states is
\begin{align}
\langle 0^\Ia\vert 0^\Jb \rangle
&= \sum_{aibj} \kappa^{\Iaa \ast}_{ai} \kappa^\Jbb_{bj} \, \langle 0 \vert a^\dagger_i a_a a^\dagger_b a_j  \vert 0 \rangle + \sum_{ij} \kappa^{I\alpha}_{ii} \kappa^{J\beta}_{jj}
\nonumber \\
&\quad +\sum_{uivj} \langle \Phi_i^\Ia \vert \Phi_u \rangle\langle \Phi_v \vert \Phi^\Jb_j \rangle \, \langle 0 \vert a^\dagger_i a_u a^\dagger_v a_j  \vert 0 \rangle \nonumber \\
&= \sum_{ai} \kappa^{\Iaa \ast}_{ai} \kappa^\Jbb_{ai} + \sum_{ij} \kappa^{I\alpha}_{ii} \kappa^{J\beta}_{jj}\nonumber \\
&\quad +\sum_{ui} \langle \Phi_i^\Ia \vert \Phi_u \rangle\langle \Phi_u \vert \Phi^\Jb_i \rangle\, .
\end{align}
Invoking the resolution of the identity, we obtain 
\begin{align}
\langle &0^\Ia\vert 0^\Jb \rangle =
\sum_{ai} \kappa^{\Iaa \ast}_{ai} \kappa^\Jbb_{ai} + \sum_{ij} \kappa^{I\alpha}_{ii} \kappa^{J\beta}_{jj}\nonumber \\& +
\sum_i
\langle \Phi^\Ia_i \vert \Phi^\Jb_i \rangle 
- \sum_{ip} \langle \Phi_i^\Ia \vert \Phi_p \rangle \langle \Phi_p \vert \Phi_i^\Jb \rangle, \label{SQnacme}
\end{align}
which is our final expression for the second-order nonadiabatic matrix element. It remains to evaluate the derivatives of the orbital-rotation parameters $\kappa^{\Iaa}_{ai}$ and $\kappa^\Jbb_{ai}$.

\subsubsection{Hartree--Fock Response Equations}

For each $\mathbf R$ and each $\boldsymbol \kappa$, the expectation value of the Hamiltonian in the state
$\vert\mathbf R, \boldsymbol \kappa \rangle$ of Eq.\,\eqref{HFstate} may be expanded as
\begin{align}
E(\mathbf R, \boldsymbol\kappa) &= \langle \mathbf R, \boldsymbol \kappa \vert \tilde H \vert \mathbf R, \boldsymbol \kappa \rangle \nonumber \\ &=
\langle \tilde 0 \vert \tilde H \vert \tilde 0 \rangle
- \mathrm i \langle \tilde  0 \vert [\tilde \kappa, \tilde H] \vert\tilde  0 \rangle \nonumber \\
&\qquad \qquad - \tfrac{1}{2} \langle\tilde  0 \vert [\tilde \kappa, [ \tilde \kappa, \tilde H]] \vert\tilde  0 \rangle + \cdots \label{Eexp}
\end{align}
As is easily verified, the only terms in Eq.\,\eqref{hatkappa} that contribute to the expansion of the HF energy are occupied-virtual orbital-rotation operators:
\begin{equation}
\tilde \kappa = \sum_{ai} \kappa_{ai} \a_a^\dagger \a_i + \sum_{ai} \kappa_{ai}^\ast \a_i^\dagger \a_a.
\label{hatkappaAI}
\end{equation}
At the reference geometry $\mathbf R_0$, the HF energy satisfies the following zero- and first-order stationary conditions with respect to these parameters:
\begin{align}
\left.\frac{\partial E(\mathbf R_0,\boldsymbol \kappa)}{\partial \kappa_{ai}}\right\vert_{\tilde \kappa = 0} = 0, \quad
\left.{\frac{\mathrm d}{\mathrm d \mathbf R}\frac{\partial E(\mathbf R_0,\boldsymbol \kappa)}{\partial \kappa_{ai}}}
\right\vert_{\stackrel{\mathbf R = \mathbf R_0}{\tilde \kappa = 0}} \!\!= \mathbf 0. \label{cond12}
\end{align}
From the expansion in Eq.\,\eqref{Eexp}, we find that the zero-order conditions (Hartree--Fock stationary conditions) at $\mathbf R_0$ are given by
\begin{align}
\mathrm i  \langle  0 \vert [  a^\dagger_a a_i, H ] \vert  0 \rangle &= 0, \label{gai} \\
\mathrm i  \langle  0 \vert [  a^\dagger_i a_a, H ] \vert  0 \rangle  &= 0. \label{gia}
\end{align}
where  $H = \tilde H(\mathbf R_0)$.
Evaluating the commutators, we find that
\begin{equation}
 \langle  0 \vert [  a^\dagger_i a_a, H ] \vert  0 \rangle = -\langle  0 \vert [  a^\dagger_a a_i, H ] \vert  0 \rangle^\ast
 = F_{ai}
\end{equation}
where $F_{ai}$ is a virtual--occupied element of the Fock matrix, which vanishes for the optimized HF state.

We next consider the first-order stationary conditions in Eq.\,\eqref{cond12} at the reference geometry $\mathbf R = \mathbf R_0$. From the expansion in Eq.\,\eqref{Eexp}, we obtain the HF response equations
\begin{align}
\langle 0 \vert [  a^\dagger_a a_i, [ H, \tilde \kappa^\prime ]] \vert 0 \rangle &= \mathrm i  \!\!\left.\langle\tilde  0 \vert [  \a^\dagger_a \a_i, \tilde H ] \vert \tilde 0 \rangle\right\vert^\prime_{\mathbf R=\mathbf R_0}, \\
\langle 0 \vert [  a^\dagger_i a_a, [ H, \tilde \kappa^\prime ]] \vert 0 \rangle &= \mathrm i  \!\!\left.\langle\tilde  0 \vert [  \a^\dagger_i \a_a, \tilde H ] \vert \tilde 0 \rangle\right\vert^\prime_{\mathbf R=\mathbf R_0} .
\end{align}
In setting up these equations, we have used the fact that
$\langle 0 \vert [  \tilde \kappa, [ a^\dagger_a a_i,  H ]] \vert 0 \rangle =
\langle 0 \vert [  a^\dagger_a a_i, [ \tilde \kappa, H ]] \vert 0 \rangle$, as is easily verified. By some
further straightforward but tedious algebra, we find that
\begin{align}
\langle 0 \vert [a^\dagger_i a_a, [H, a^\dagger_b a_j]] \vert 0 \rangle &= A_{aibj},
\label{dA}\\
\langle 0 \vert [a^\dagger_i a_a, [H, a^\dagger_j a_b]] \vert 0 \rangle&= - B_{aibj},
\label{dB}
\end{align}
where $A_{aibj}$ and $B_{aibj}$ are defined in Eqs.\,\eqref{cphf_017} and~\eqref{cphf_018}, respectively. To evaluate the right-hand side of the response equations, we note that
\begin{equation}
\langle\tilde  0 \vert [  \a^\dagger_i \a_a, \tilde H ] \vert \tilde 0 \rangle =\tilde F_{ai} =
\tilde h_{ai} + \sum_k (ai \Vert kk) 
\end{equation}
is a virtual--occupied element of the Fock matrix in the OMO basis at $\mathbf R$. Differentiation at $\mathbf R_0$ gives
\begin{align}
 \tilde F_{ai}^\prime &= F_{ai}^\prime
- \sum_j \langle \Phi_a^\prime \vert \Phi_j \rangle F_{ji} 
- \sum_b F_{ab} \langle \Phi_b \vert \Phi_i^\prime \rangle\nonumber \\ 
& \quad -  \sum_{jp} (a i \Vert p j) \langle \Phi_j^\prime \vert \Phi_p \rangle
-  \sum_{jp} (a i \Vert j p) \langle \Phi_p \vert \Phi_j^\prime \rangle
\label{eq:ifai}
\end{align}
where the first term is the derivative of the MO Fock matrix and the remaining terms arise from the differentiation of the connection matrix. We may now write the response equations in the form
\begin{align}
-\sum_{bj} B^\ast_{aibj} \kappa_{bj}^\prime + \sum_{bj} A^\ast_{aibj} \kappa^{\ast\prime}_{bj}&= (\mathrm i \tilde F_{ai}^\prime)^{\ast} \,, \label{der1}\\
\sum_{bj} A_{aibj} \kappa_{bj}^\prime - \sum_{bj} B_{aibj} \kappa^{\ast\prime}_{bj}&= \mathrm i \tilde F_{ai}^\prime \,. \label{der2}
\end{align}
Letting $\mathbf f^\prime$ be the vector containing the elements $\tilde F_{ai}^\prime$, we obtain the matrix equations
\begin{equation}
\begin{pmatrix} \mathbf A & \mathbf B \\ \mathbf B^\ast & \mathbf A^\ast \end{pmatrix}
\begin{pmatrix}\mathrm i \boldsymbol \kappa^\prime \\ 
(\mathrm i \boldsymbol \kappa)^{\!\ast\prime} \end{pmatrix}
=   - \begin{pmatrix} \mathbf f^\prime \\  \mathbf f^{\prime\ast} \end{pmatrix} \label{response_1}
\end{equation}
whose solution are needed to evaluated the nonadiabiatic matrix elements in Eq.\,\eqref{SQnacme}.

\subsubsection{Comparison with First-Quantization Formulation}

To compare with the closed-shell CPHF equations, let $\mathbf w^\prime$ be the vector containing the unoccupied-occupied elements $\mathbf W^\prime$ in
Eq.\,\eqref{Wrq} so that $w_{ai}^\prime = \langle \Phi_a \vert \Phi_i^\prime \rangle$. Then
\begin{equation}
\begin{pmatrix} \mathbf A & \mathbf B \\ \mathbf B^\ast & \mathbf A^\ast \end{pmatrix}
\begin{pmatrix}  \mathbf w^\prime\\   \mathbf w^{\ast\prime} \end{pmatrix}
=    \begin{pmatrix} \boldsymbol  \Delta^\prime \\ {\boldsymbol  \Delta}^{\!\ast\prime} 
\label{response_2}
\end{pmatrix}
\end{equation}
where
\begin{align}                                                                                                       \Delta_{ai}^\prime &=
\sum_b F_{ab} \langle \Phi_b \vert \Phi_i^\prime\rangle - \sum_j F_{ji} \langle \Phi_a \vert \Phi_j^\prime\rangle 
\nonumber \\ &+ \sum_{bj} (ai \Vert jb) \langle \Phi_b \vert \Phi_j^\prime\rangle + \sum_{bj} (ai \Vert bj) \langle \Phi_j^\prime \vert \Phi_b\rangle.
\label{Dai}
\end{align}
Adding  Eqs.\,\eqref{eq:ifai} and~\eqref{Dai} and rearranging, we find
\begin{align}
f_{ai}^ \prime+ \Delta_{ai}^\prime &=  F_{ai}^\prime
- \sum_j S_{a     j}^\prime F_{ji}  - \sum_{jk}  (a i \Vert j k) S_{kj}^\prime,
\end{align}
which we recognize as minus the elements of the right-hand side $\mathbf b_0$ of the CPHF equations; see Eq.\,\eqref{cphf_021}:
\begin{equation}
f_{ai}^\prime + \Delta_{ai}^\prime = - (b_0)_{ai}.
\end{equation}
It follows that
\begin{equation}
\begin{pmatrix} \mathbf A & \mathbf B \\ \mathbf B^\ast & \mathbf A^\ast \end{pmatrix}
\begin{pmatrix} \mathrm i \boldsymbol \kappa^{\prime} - \mathbf w^\prime\\ ( \mathrm i \boldsymbol \kappa^\prime)^\ast - \mathbf w^{\prime^\ast} \end{pmatrix}
=    \begin{pmatrix} \mathbf b_0^{\prime} \\ \mathbf b_0^{\prime\ast} \end{pmatrix}.
\end{equation}
Comparing with the CPHF equations in Eq.\,\eqref{cphf_015}, we conclude that
\begin{equation}
\mathrm i \kappa_{ai}^\prime = U_{ai}^{\prime} + \langle\Phi_a \vert \Phi_i^\prime \rangle .
\end{equation} 
Introducing this expression in Eq.\,\eqref{SQnacme} and choosing the phase factors
\begin{equation}
\mathrm i \kappa_{ii}^\prime = U_{ii}^{\prime} + \langle\Phi_i \vert \Phi_i^\prime \rangle ,
\end{equation} 
we obtain the CPHF expression in Eq.\,\eqref{wfd_005}. We note that $U_{ii}^\prime$ are determined by Eq.\,\eqref{cphf_003} only up to an imaginary constant, in agreement with the fact that the phase factors $\kappa_{ii}^\prime$ are undetermined in the second-quantization formalism.

\section{Results}

\subsection{Computational details}

Calculations were performed on a series of diatomic molecules H$_2$, LiH, BH, and CH$^+$.  All results presented here were performed with the decontracted  Lcc-pVTZ basis set, denoted Lu-cc-pVTZ, where ``L" indicates the use of London orbitals and ``u" indicates decontraction. Berry curvature, RHF energy, and DBOC surface calculations were performed for a uniform magnetic field of varying strength orientated along the $z$-axis, with the molecular orientations perpendicular to the magnetic field along the $x$-axis.
The analytic evaluation of the Berry curvature and DBOC was validated against finite difference results.\cite{Culpitt2021}.

Calculations were performed for the RHF singlet state starting from a bond distance of 0.05 ${\text{\AA}}$ for H$_2$, 0.6\,\AA\ for LiH, and 0.58\,\AA{} for BH and CH$^+$. Data was generated on a grid using a step size of 0.04 ${\text{\AA}}$, with cubic spline interpolation being used for plotting purposes, except in the case of the DBOC values of CH$^+$ at $1 \times 10^{-4} B_0$, for which a linear spline was used. The strength of the magnetic field ranged from 0.0$B_0$ to 1.0$B_0$ for calculations of SCF energies and DBOC and from $1 \times 10^{-4} B_0$ to 1.0$B_0$ for the Berry curvature. As a result of the findings stated in Section III.B., the upper limit for bond distance ranges at different field strengths is $(6.0 - 2.75|B|/B_0)$ ${\text{\AA}}$, with the exception of $1 \times 10^{-4} B_0$ for BH and CH$^+$, for which the plot of the Berry curvature truncated earlier due to divergent behavior, as  discussed below. 

The following conversion factors were used: 1 a.m.u. = 1.66053906660 $\times 10^{-27}$\,kg, E$_\mathrm h$ = 219474.6313632\,cm$^{-1}$, $m_\mathrm e$ = 5.48579909065 $\times 10^{-4}$\,a.m.u. (2018 CODATA recommended values). The nuclear masses used in calculating the DBOC were derived by subtracting electron mass from atomic masses published in the online NIST database of atomic masses,\cite{Coursey_2015} except in the case of the protonic mass, which was converted directly from the 2018 CODATA value of $m_\mathrm p$ = 1.67262192369 $\times 10^{-27}$\,kg. For the magnetic field strength, one atomic unit is $B_0 = 2.35 \times 10^5$\,T.

\subsection{Stability Analysis}

Ideally, when examining singlet states, one wishes to generate a surface where every point maintains spin symmetry (guaranteed by the RHF calculation itself) and also corresponds to a global minimum in the RHF space. When generating potential-energy surfaces (PESs), so long as the curve appears smooth, it could be taken for granted that the underlying self-consistent-field (SCF) calculations have converged to a minimum at every step, whether or not this is actually the case as revealed by diagonalization of the relevant stability matrices.

With this in mind, we have examined the lowest eigenvalues of the complex RHF stability matrix given in Eq.\,\eqref{cphf_cs2} for all SCF calculations. As shown in Fig.\,\ref{fig_hess}, it was observed that in general the lowest eigenvalue of the complex RHF stability matrix decreases in magnitude as a function of bond distance and field strength, with the effect becoming amplified with increasing field strength. Additionally, depending on the molecular species and bond distance, it was possible to find RHF saddle points that were nearly degenerate with minima (on the order of a microhartree), making it difficult in practice to guarantee SCF minima along the entire curve. This situation does not always visibly manifest itself when examining the energy alone; indeed, many curves may have smooth PES profiles while converging to stationary points of different character at various points along the curve.

Additionally, we observed that the DBOC and Berry curvature can exhibit discontinuous and/or erratic behavior when the eigenvalues of the RHF stability matrix approach zero. This phenomenon was recently observed by Thorpe and Stanton\cite{Thorpe_2020} when studying the DBOC for the NO and NO$_2$ radicals. In Ref.\,\onlinecite{Thorpe_2020}, it was observed that it is not necessarily the sign of the eigenvalue but rather its magnitude that is important when calculating solutions to the CPHF equations. Our results agree with this observation.

Because the eigenvalues of the RHF stability matrix generally decrease as a function of increasing bond distance and increasing field strength, we have presented results for each field strength only within a bond-distance range for which true minima could be reliably obtained with standard convergence techniques, and for which the eigenvalues of the RHF stability matrix do not become extremely small. These ranges increase as field strength decreases. All important and notable features of the SCF energy, DBOC, and Berry curvature are retained within the truncated bond distance ranges at higher field due to molecular compression with increasing field strength.

Finally, for CH$^+$ and BH, it was observed at zero field that there was an occurrence of broken time-reversal symmetry. Beyond a certain bond distance, the lowest energy SCF solution became associated with a complex-valued RHF wave function and a singular  RHF stability matrix. This point is further elaborated on in the relevant subsection for each species.

\subsection{Berry Curvature Charge Interpretation}

For diatomic molecules with a uniform magnetic field along the $z$-axis and a molecular orientation perpendicular to the field along the $x$-axis, the general form of the Berry curvature is given by
\begin{align}
    \boldsymbol \Omega &= \begin{pmatrix} \boldsymbol \Omega_\text{11} & \boldsymbol \Omega_\text{12} \\ \boldsymbol \Omega_\text{21} & \boldsymbol \Omega_\text{22} \end{pmatrix}
    \nonumber \\&=
    \left(\begin{array}{cccccc}
    0 & -\omega_\text{11} & 0 & 0 &  - \omega_\text{21} & 0 \\ \omega_\text{11} & 0 & 0 & \omega_\text{12} & 0 & 0 \\
    0&0&0&0&0&0\\
    0& - \omega_\text{12}& 0 & 0& -\omega_\text{22} & 0 \\
    \omega_\text{21} & 0 & 0 & \omega_\text{22} & 0 & 0 \\
      0&0&0&0&0&0
    \end{array}\right)
    \label{omega_gen}
\end{align}
where the elements of Eq.\,\eqref{omega_gen} are calculated according to Eq.\,\eqref{eom_t006} and Eq.\,\eqref{omega_gen} obeys the overall symmetry $\boldsymbol \Omega_\text{12} = - \boldsymbol \Omega_\text{21}^{\mathrm{T}}$. In the homonuclear diatomic case, we have that $\mathbf{\Omega}_{11} = \mathbf{\Omega}_{22}$ and $\mathbf{\Omega}_{12} =-\boldsymbol \Omega_{21}^\mathrm T= \mathbf{\Omega}_{21}$. Therefore, in this particular case, all blocks are anti-symmetric and the equations of motion become\cite{Culpitt2021,Peters2021}
\begin{align}
M_{I} \ddot{\mathbf{R}}_{I}
&= \mathbf{F}^{\text{BO}}_{I} - 
\sum_{J} \left(e \delta_{IJ} Z_{I} \mathbf{B} - \boldsymbol{\omega}^{\text{A}}_{IJ} \right)
\times \dot{\mathbf R}_{J}.
\label{res_010}
\end{align}
where
\begin{equation}
\boldsymbol \omega_{11}^\text A = \boldsymbol \omega_{22}^\text A = \begin{pmatrix} 0 \\ 0 \\ \omega_{11} \end{pmatrix}, \quad
\boldsymbol \omega_{12}^\text A = \boldsymbol \omega_{21}^\text A = \begin{pmatrix} 0 \\ 0 \\ \omega_{12} \end{pmatrix}.
\end{equation}
Introducing screening charges $Q_{IJ}$ defined as
\begin{align}
Q_{II} = -\dfrac{\omega_{11}}{e B_z}, \quad
Q_{IJ} = -\dfrac{\omega_{12}}{e B_z}, 
\label{res_012}
\end{align}
the equations of motion can be rewritten as
\begin{align}
M_{I} \ddot{\mathbf{R}}_{I} 
& = \mathbf{F}^{\text{BO}}_{I} - 
\sum_{J} e \left(\delta_{IJ} \thezi + Q_{IJ}\right) \mathbf{B}
\times \dot{\mathbf{R}}_{J},
\label{res_010x}
\end{align}
where the screening charges add up to the partial charge on each atom
\begin{align}
q_{I} = \sum_{J} Q_{IJ} = \sum_J Q_{JI}.
\label{res_013}
\end{align}
In general, however, the symmetric part of the Berry curvature does not vanish and we cannot express the force in terms of screening charges.

In the heteronuclear diatomic case, unlike the homonuclear diatomic case, we have $\boldsymbol \Omega_\text{12} \neq - \boldsymbol \Omega_\text{12}^{\mathrm{T}}$ and $\boldsymbol \Omega_\text{21} \neq - \boldsymbol \Omega_\text{21}^{\mathrm{T}}$. The Berry force of a heteronuclear diatomic therfore cannot be expressed in cross-product form. For interpretation purposes, we nevertheless calculate the electronic partial charges of atoms 1 and 2 from an average of $\omega_\mathrm{12}/\omega_\mathrm{21}$:
\begin{align}
q_\text{1}&= Q_\mathrm{22} + Q_\mathrm{12} = 
-\dfrac{\omega_\mathrm{11}}{e B_z} -
\dfrac{\omega_\mathrm{12} + \omega_\mathrm{21}}{2e  B_z}\,,\label{q1}\\
q_\text{2}&= Q_\mathrm{22} + Q_\mathrm{21} = 
- \dfrac{\omega_\mathrm{22} }{e B_z} -
\dfrac{\omega_\mathrm{21} + \omega_\mathrm{12}}{2e  B_z}\,.\label{q2} 
\end{align}
Note that Eqs.\,\eqref{q1} and~\eqref{q2} reduce to Eq.\,\eqref{res_012} in the homonuclear diatomic case. Strictly speaking, our electronic partial charges $q_1$ and $q_2$ need only obey total charge conservation, or $q_1 + q_2 = -N_\text e$.

\subsection{H$_2$ Molecule}

\begin{figure*}[h]
\centering
\begin{tabular}{ll}
(a) & (b) \\
\includegraphics[width=0.48\textwidth]{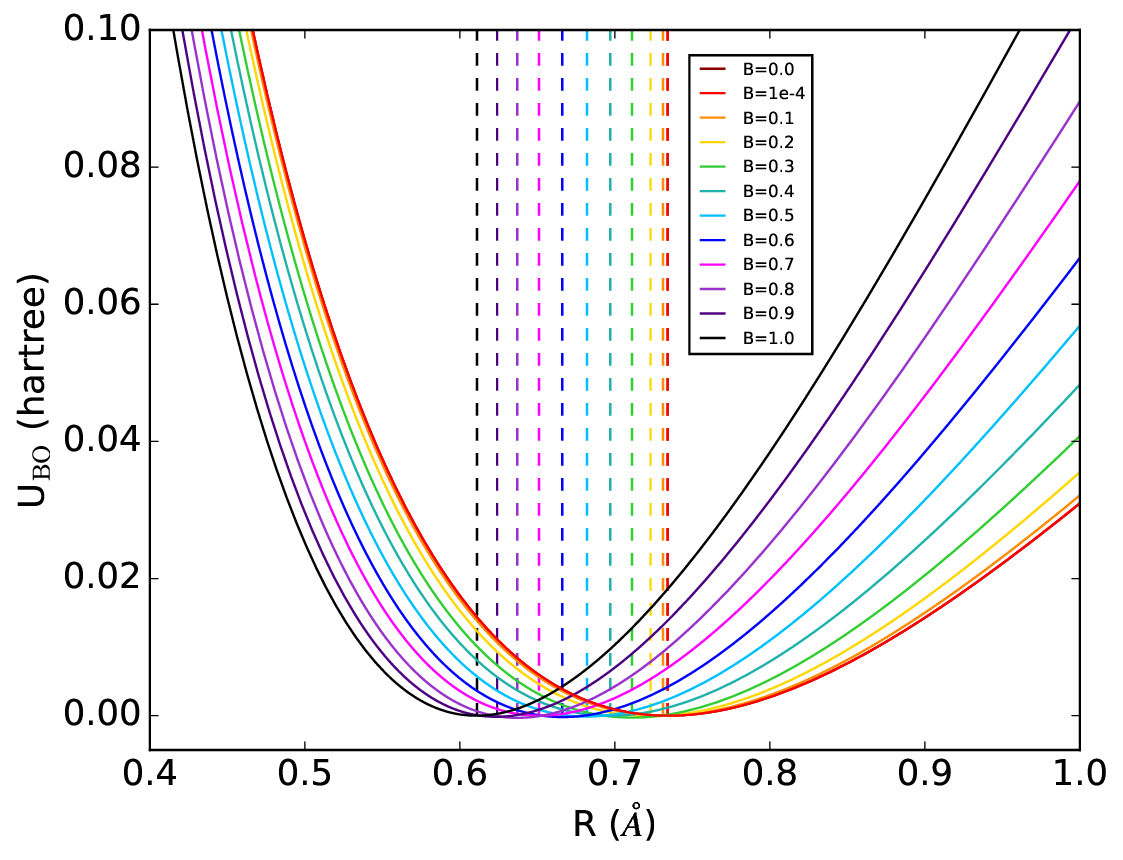} & \includegraphics[width=0.48\textwidth]{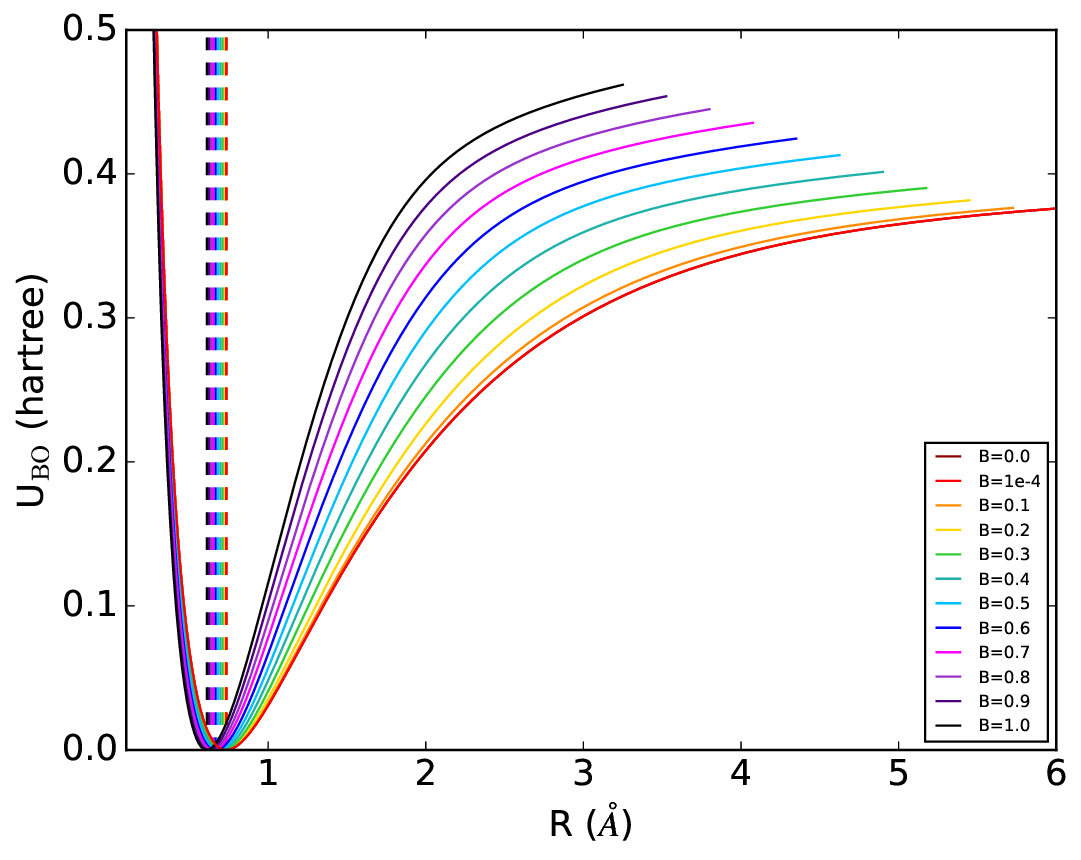}  \\
(c) & (d) \\
\includegraphics[width=0.48\textwidth]{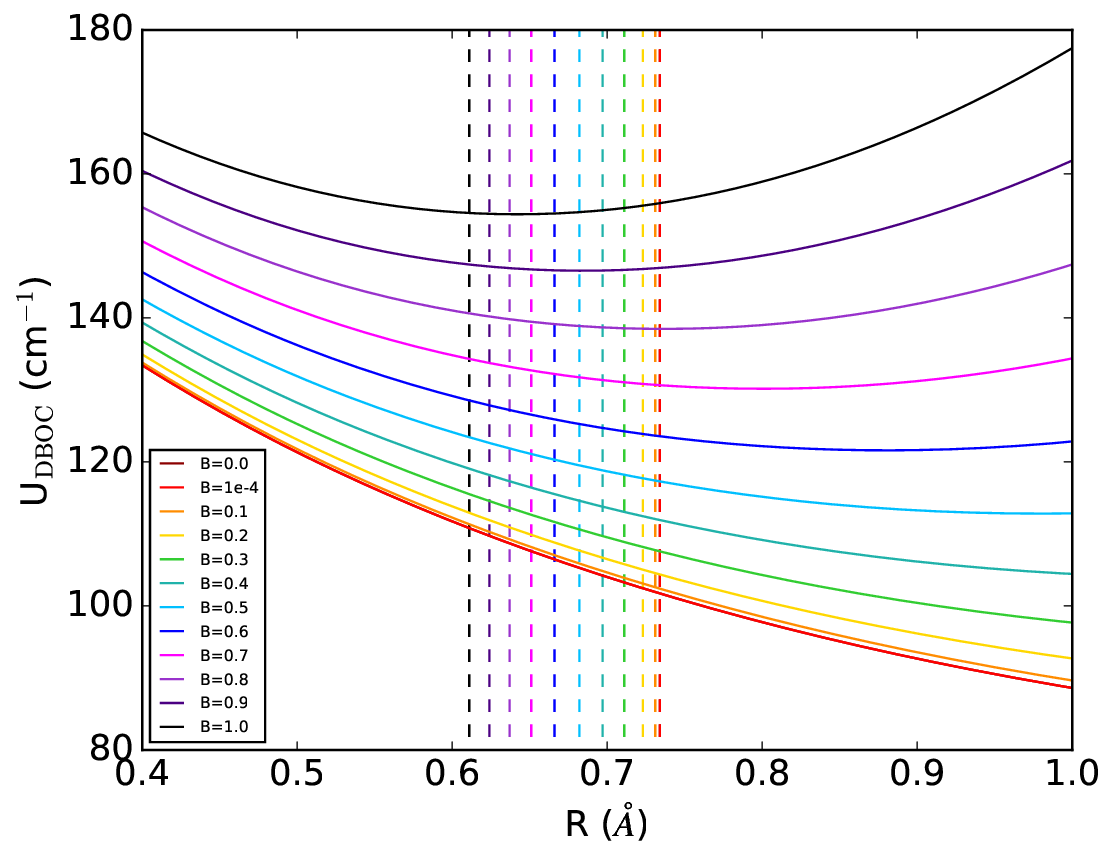} & \includegraphics[width=0.48\textwidth]{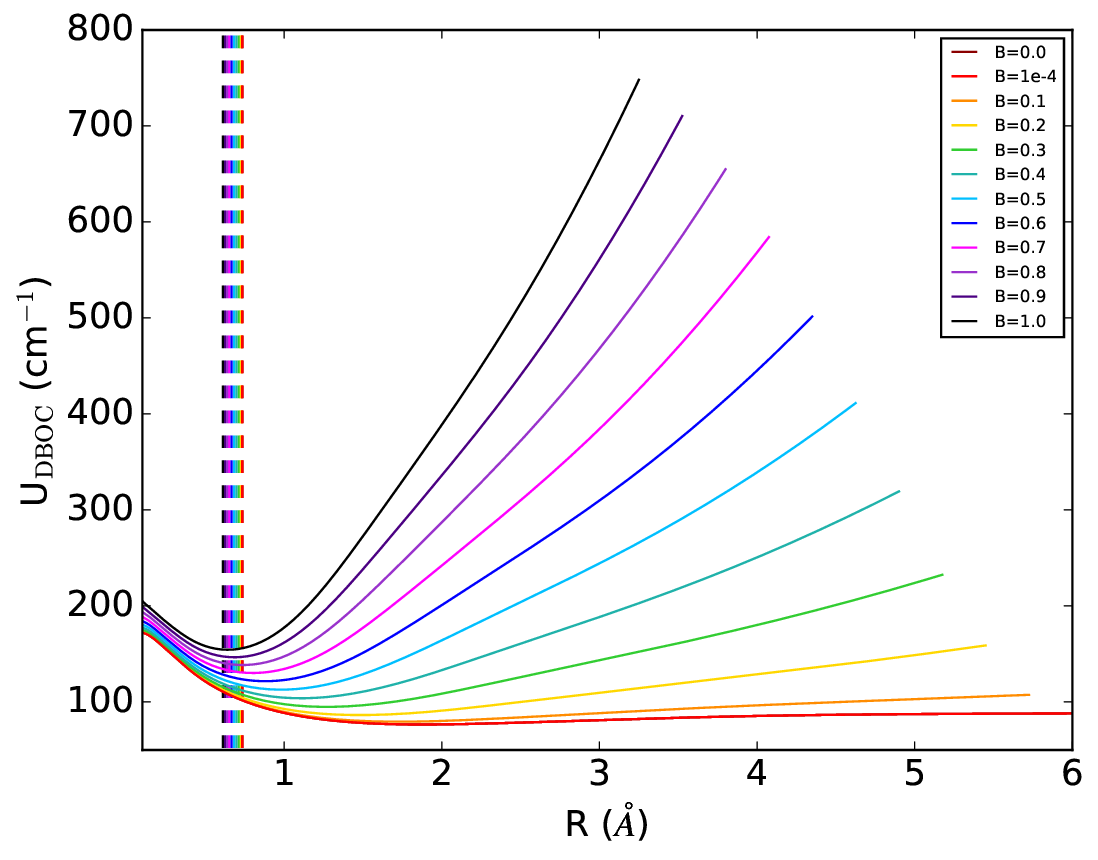} \\
\end{tabular}
\caption{SCF (a $\&$ b) and DBOC (c $\&$ d) energy of H$_2$ for the RHF singlet state with the Lu-cc-pVTZ basis set for a series of magnetic field strengths. The magnetic field is oriented along the z-axis with the molecule oriented along the x-axis. The Lu-cc-pVTZ basis set is comprised of the decontracted Lcc-pVTZ basis. Equilibrium bond distances given by the vertical dashed lines in each plot are 0.734 $\text{\AA}$ (B=0.0), 0.734 $\text{\AA}$ (B=$1\times 10^{-4}$), 0.731 $\text{\AA}$ (B=0.1), 0.723 $\text{\AA}$ (B=0.2), 0.711 $\text{\AA}$ (B=0.3), 0.697 $\text{\AA}$ (B=0.4), 0.682 $\text{\AA}$ (B=0.5), 0.666 $\text{\AA}$ (B=0.6), 0.651 $\text{\AA}$ (B=0.7), 0.637 $\text{\AA}$ (B=0.8), 0.624 $\text{\AA}$ (B=0.9) and 0.611 $\text{\AA}$ (B=1.0). Raw data was generated on a grid using a step size of 0.04 ${\text{\AA}}$ and a cubic spline interpolation was used for plotting purposes. The plot legends display magnetic field strengths given in units of $B_0$. The minima of the curves in panels (a $\&$ b) have been shifted to zero.}
\label{fig_dboc_h2_tz}
\end{figure*}

As shown in Figs.\,\ref{fig_dboc_h2_tz}.a and~\ref{fig_dboc_h2_tz}.b, the equilibrium bond distance systematically decreases as the field strength is increased, while the bond simultaneously stiffens. Both the shape and magnitude of the DBOC shown in Figs.\,\ref{fig_dboc_h2_tz}.c and~\ref{fig_dboc_h2_tz}.d  changes as a function of field strength. At zero field, the DBOC monotonically decreases in the vicinity of equilibrium. As the field strength is increased, the magnitude of the DBOC increases and the curvature of the DBOC changes. In a finite magnetic field, the DBOC values tend to increase as a function of bond distance, with the effect becoming amplified at higher field strengths. This behavior is correlated with the trend in Hessian eigenvalues observed in Fig.\,\ref{fig_hess}.a. The eigenvalues approach zero more quickly as a function of bond distance at higher field strengths.  

\begin{figure*}[h]
\centering
\begin{tabular}{ll}
\includegraphics[width=0.48\textwidth]{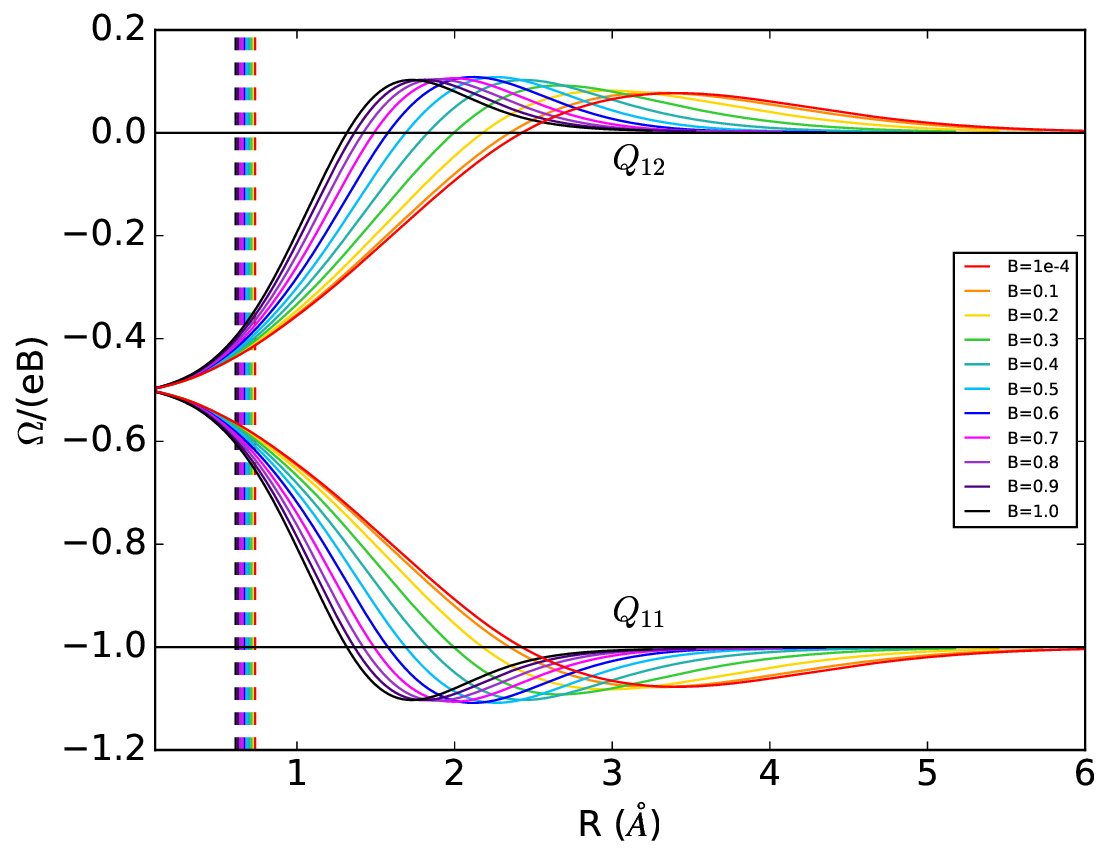} \\
\end{tabular}
\caption{Screening charges for H$_2$ as defined in Eqs.\,\eqref{q1} and~\eqref{q2} in the text. The partial charge on hydrogen 1 is given by q$_\mathrm{1}$ = Q$_\mathrm{11}$ + Q$_\mathrm{12}$  while the partial charge on hydrogen 2 is given by q$_\mathrm{2}$ = Q$_\mathrm{22}$ + Q$_\mathrm{21}$. From symmetry, Q$_\mathrm{11}$ $=$ Q$_\mathrm{22}$ and Q$_\mathrm{12}$ $=$ Q$_\mathrm{21}$. All calculations were performed on the RHF singlet state with the Lu-cc-pVTZ basis set for a series of magnetic field strengths. The magnetic field is oriented along the z-axis with the molecule oriented along the x-axis. The Lu-cc-pVTZ basis set is comprised of the decontracted Lcc-pVTZ basis. Equilibrium bond distances given by the vertical dashed lines in each plot are 0.734 $\text{\AA}$ (B=$1\times 10^{-4}$), 0.731 $\text{\AA}$ (B=0.1), 0.723 $\text{\AA}$ (B=0.2), 0.711 $\text{\AA}$ (B=0.3), 0.697 $\text{\AA}$ (B=0.4), 0.682 $\text{\AA}$ (B=0.5), 0.666 $\text{\AA}$ (B=0.6), 0.651 $\text{\AA}$ (B=0.7), 0.637 $\text{\AA}$ (B=0.8), 0.624 $\text{\AA}$ (B=0.9) and 0.611 $\text{\AA}$ (B=1.0). Raw data was generated on a grid using a step size of 0.04 ${\text{\AA}}$ and a cubic spline interpolation was used for plotting purposes. The plot legends display magnetic field strengths given in units of $B_0$.}
\label{fig_Q11_Q12_tz}
\end{figure*}

The partial charges corresponding to Berry curvature elements are shown in Fig.\,\ref{fig_Q11_Q12_tz}. The curves tend toward a value of -0.5 $\Omega/(eB)$ in the small bond-distance limit, corresponding to half the value of any one partial charge $q_1$ or $q_2$. In the dissociation limit, $Q_{11}$ tends to $-1\Omega/(eB)$ and the coupling term $Q_{12}$ tends to zero. All curves exhibit antiscreening/superscreening behavior, where at certain bond distances the values of $\Omega_{12}$ become positive (antiscreening) and the values of $\Omega_{11}$ become less than $-1 eB$ (superscreening). The domain over which either antiscreening or superscreening is exhibited becomes shorter as the field is increased, with the extrema also being shifted toward shorter distance as field is increased, indicating a correlation with the compression of the molecule. This behavior was observed previously for field strengths of 0.1$B_0$ and 1.0$B_0$ with the finite-difference procedure,\cite{Culpitt2021} and verified and expanded upon here with the analytic Berry curvature. 

To investigate the role the DBOC plays in dynamics, rovibrational spectra were generated for magnetic field strengths of 0.1$B_0$ and 1.0$B_0$ using analytic Berry curvature, both with and without DBOC contributions included in the PES. The simulations were conducted (as recommended for H$_2$ in Ref.\,\onlinecite{Peters2021}) using the auxiliary coordinates and momenta method with the RK4 integrator, with a step size was 0.5\,fs and a total simulation time of 200\,ps. As described in Ref.\,\onlinecite{Peters2021}, we precalculated all necessary ab initio data on a two-dimensional grid of the internal coordinates using the Lu-cc-pVTZ basis set. During the simulations the energy, forces, Berry curvature, and DBOC were obtained from spline fits generated from the precalculated data. The DBOC gradient was calculated directly from the DBOC energy spline fit as its derivative.

As see in Fig.\,\ref{fig_dboc_h2_tz}, the character of the DBOC depends on the magnetic field strength. Changes in the spectra as a result of including the DBOC can therefore manifest themselves differently from in the zero-field case. However, because the DBOC  of H$_2$ is still a relatively small correction near equilibrium (even at the upper end of the magnetic field strengths reported here), there is little impact on the spectra.

\begin{figure*}[h]
\centering
\begin{tabular}{ll}
(a) \\
\includegraphics[width=0.48\textwidth]{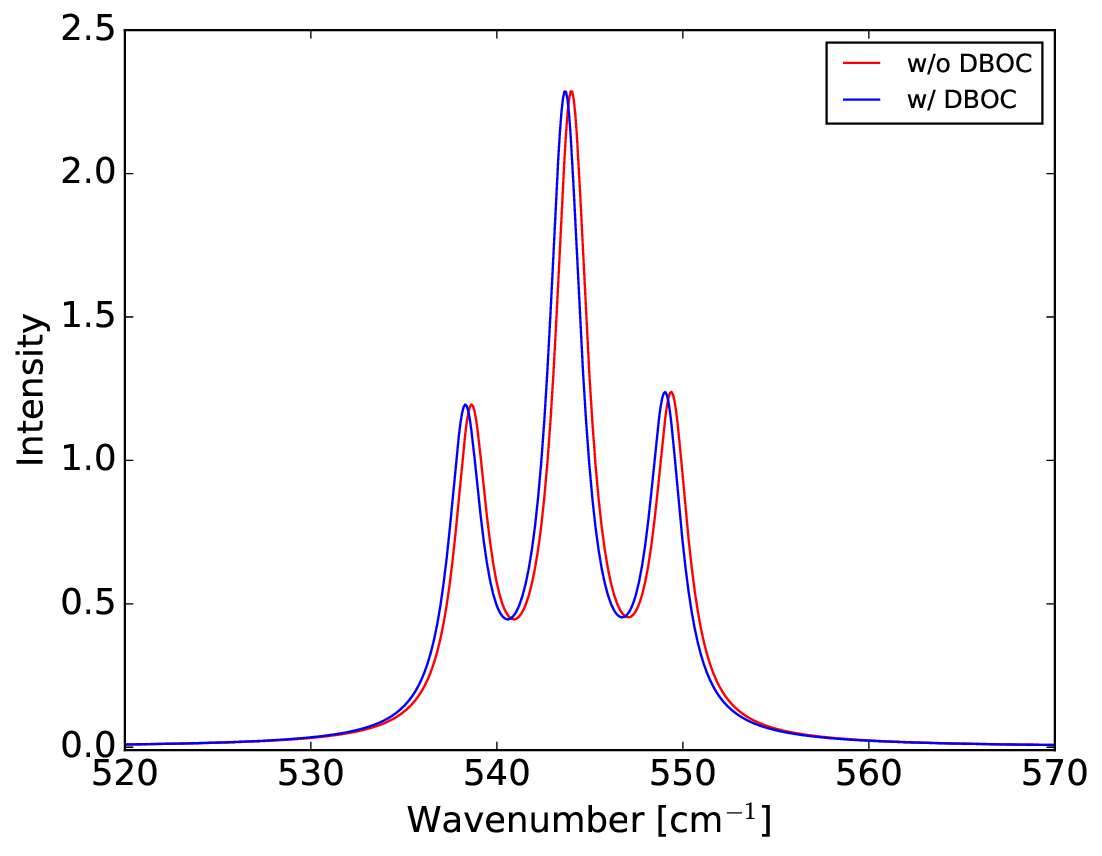} \\
(b) \\
\includegraphics[width=0.48\textwidth]{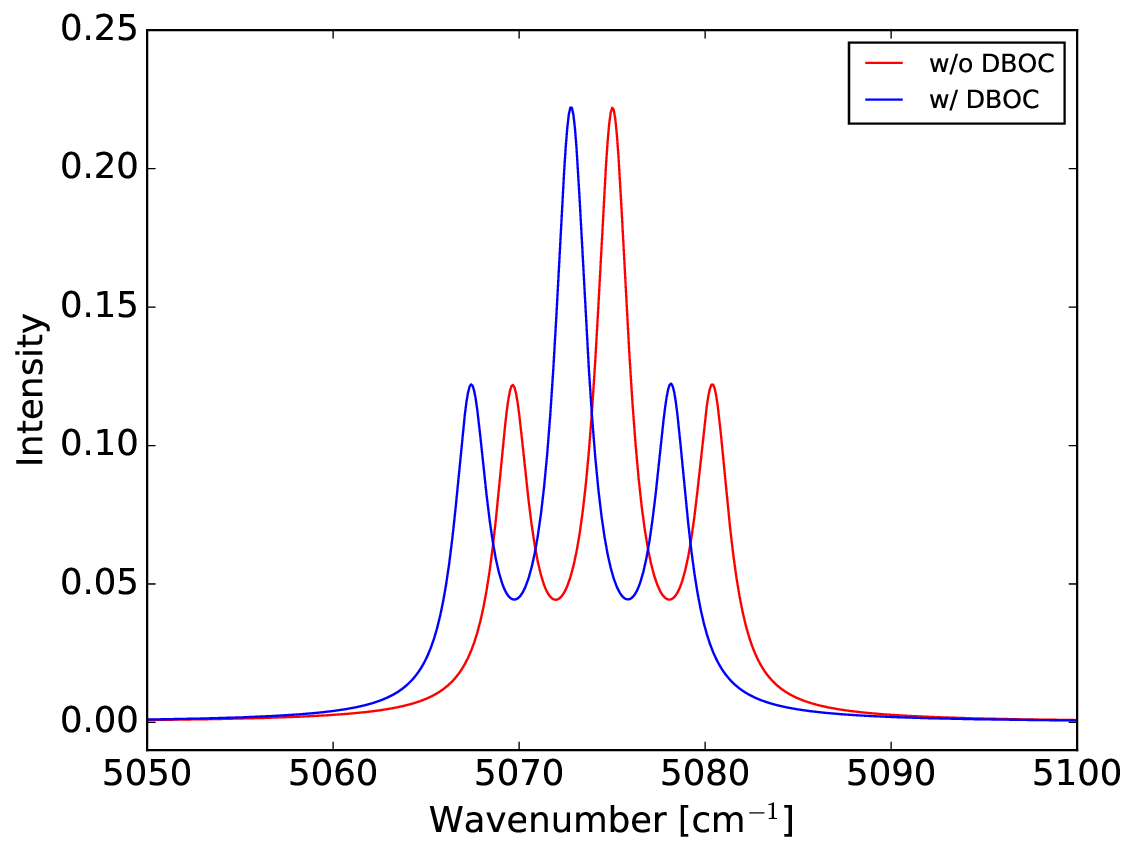} \\
\end{tabular}
\caption{Low energy (a) and high energy (b) vibrational fine sturcture for H$_2$ with (blue) and without (red) the inclusion of the DBOC in the PES. Calculations were performed with the analytic Berry curvature and the Lu-cc-pVTZ basis set for a magnetic field strength of 0.1B$_0$.}
\label{fig_spec_01}
\end{figure*}

\begin{figure*}[h]
\centering
\begin{tabular}{ll}
(a) \\
\includegraphics[width=0.48\textwidth]{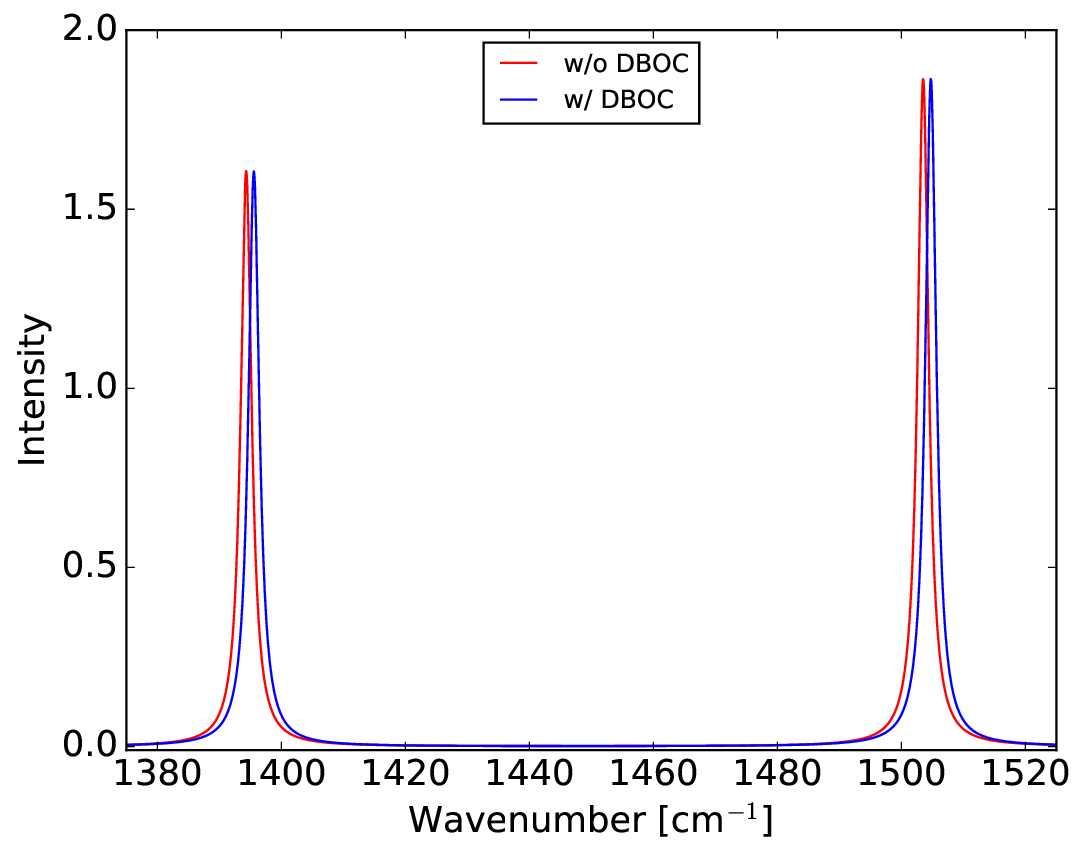} \\
(b) \\
\includegraphics[width=0.48\textwidth]{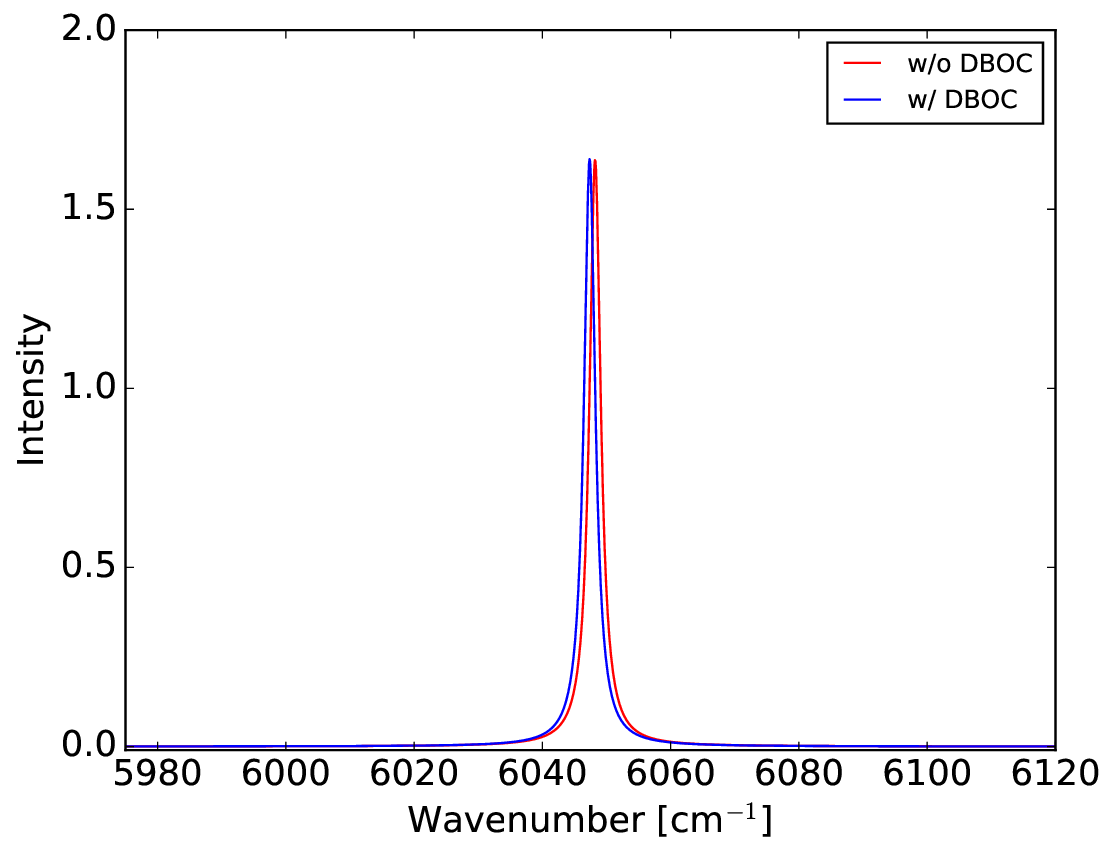} \\
\end{tabular}
\caption{Low energy (a) and high energy (b) vibrational fine sturcture for H$_2$ with (blue) and without (red) the inclusion of the DBOC in the PES. Calculations were performed with the analytic Berry curvature and the Lu-cc-pVTZ basis set for a magnetic field strength of 1.0B$_0$.}
\label{fig_spec_10}
\end{figure*}

For a magnetic field strength of 0.1$B_0$, the DBOC shifts the lower-energy (rotational) peaks in Fig.\,\ref{fig_spec_01} to a lower energy by roughly 1\,cm$^{-1}$, while the (vibrational) peaks at higher energy  in Fig.\,\ref{fig_spec_01}.b are blue-shifted by about 3\,cm$^{-1}$. Small changes are also observed in the spectrum at the stronger field of 1.0$B_0$ shown in Fig.\,\ref{fig_spec_10}, where the low-energy libration is red-shifted by about 1\,cm$^{-1}$, while the higher-energy vibrational peaks are blue-shifted with the inclusion of the DBOC. These observations are in line with the impact of the DBOC on the PES. For a more complete and detailed analysis of spectra, see Ref.\,\onlinecite{Peters2021}. 

\subsection{LiH Molecule}

\begin{figure*}[h]
\centering
\begin{tabular}{ll}
(a) & (b) \\
\includegraphics[width=0.48\textwidth]{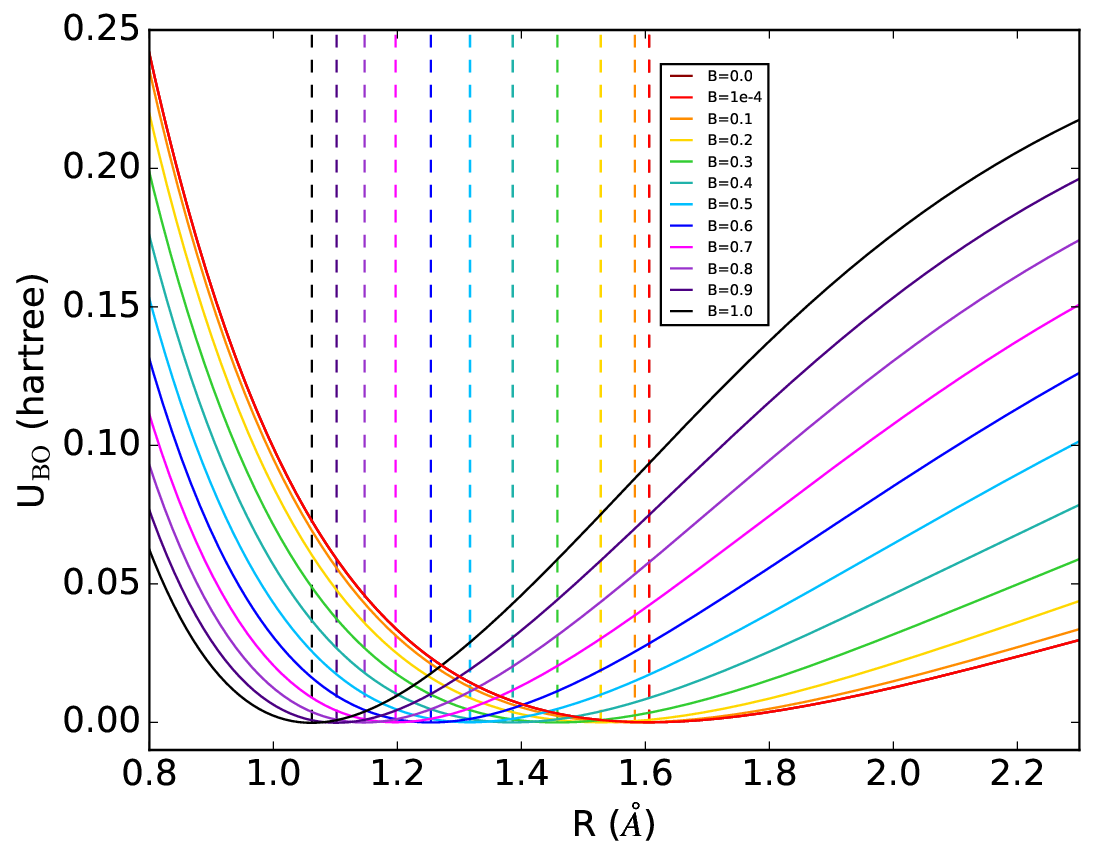} & \includegraphics[width=0.48\textwidth]{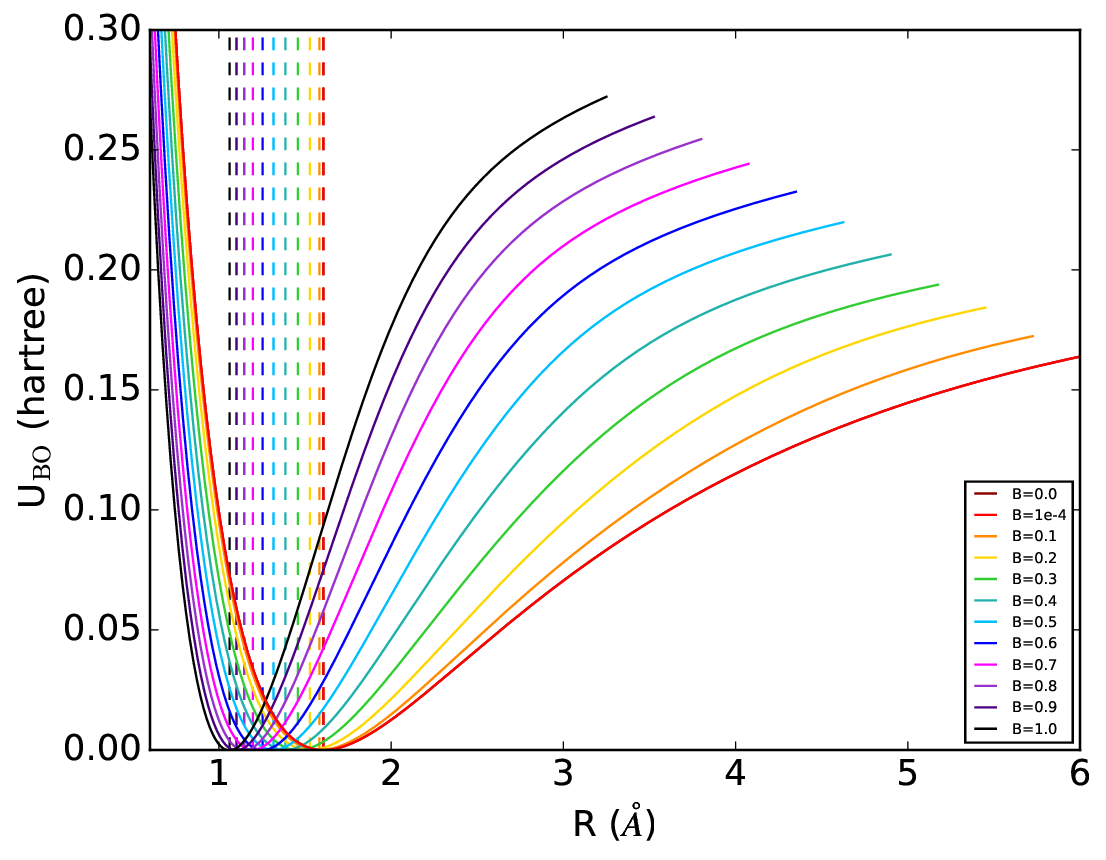}  \\
(c) & (d)\\
\includegraphics[width=0.48\textwidth]{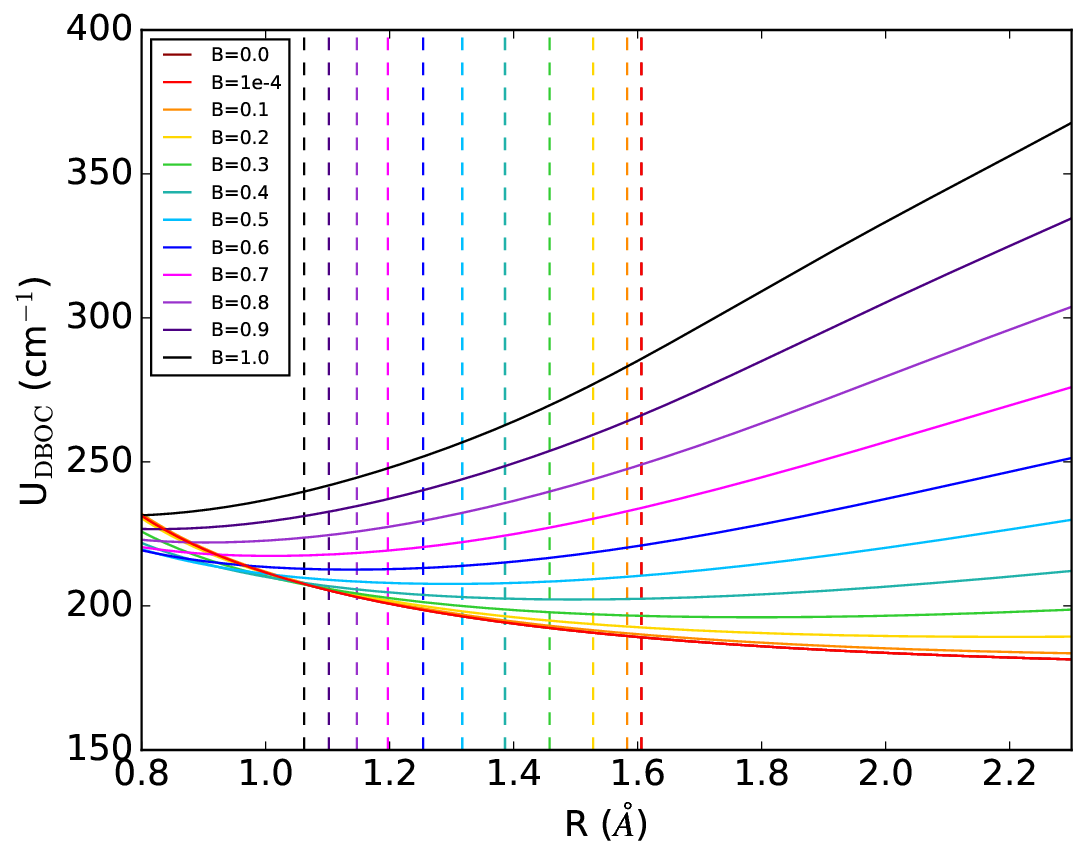} & \includegraphics[width=0.48\textwidth]{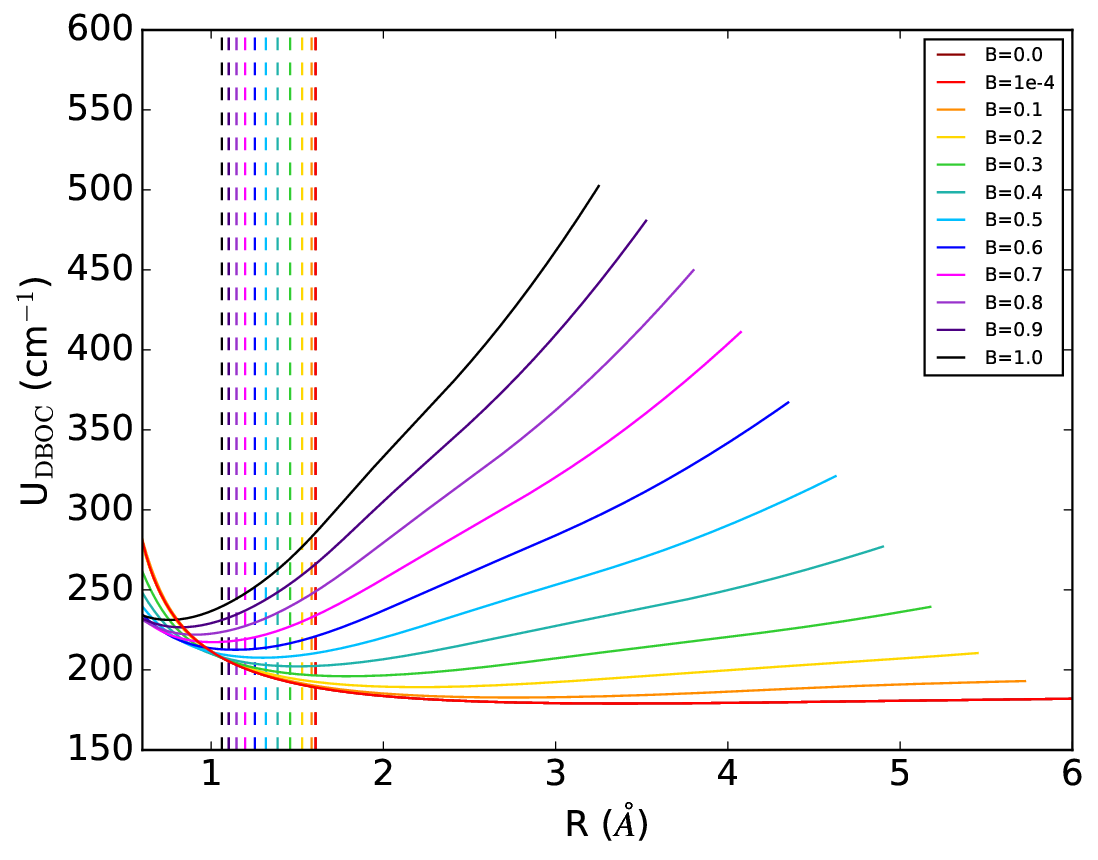} \\
\end{tabular}
\caption{SCF (a $\&$ b) and DBOC (c $\&$ d) energy of LiH for the RHF singlet state with the Lu-cc-pVTZ basis set for a series of magnetic field strengths. The magnetic field is oriented along the z-axis with the molecule oriented along the x-axis. The Lu-cc-pVTZ basis set is comprised of the decontracted Lcc-pVTZ basis. Equilibrium bond distances given by the vertical dashed lines in each plot are 1.606 $\text{\AA}$ (B=0.0), 1.606 $\text{\AA}$ (B=$1\times 10^{-4}$), 1.583 $\text{\AA}$ (B=0.1), 1.528 $\text{\AA}$ (B=0.2), 1.458 $\text{\AA}$ (B=0.3), 1.386 $\text{\AA}$ (B=0.4), 1.317 $\text{\AA}$ (B=0.5), 1.254 $\text{\AA}$ (B=0.6), 1.197 $\text{\AA}$ (B=0.7), 1.147 $\text{\AA}$ (B=0.8), 1.102 $\text{\AA}$ (B=0.9) and 1.062 $\text{\AA}$ (B=1.0). Raw data was generated on a grid using a step size of 0.04 ${\text{\AA}}$ and a cubic spline interpolation was used for plotting purposes. The plot legends display magnetic field strengths given in units of $B_0$. The minima of the curves in panels (a $\&$ b) have been shifted to zero.}
\label{fig_dboc_lih_tz}
\end{figure*}

As shown in Figs.\,\ref{fig_dboc_lih_tz}.a and~\ref{fig_dboc_lih_tz}.b, the equilibrium bond distance systematically decreases with increasing field strength, while the bond simultaneously stiffens, as also observed for H$_2$. The trends in the DBOC curves shown in Figs.\,\ref{fig_dboc_lih_tz}.c and~\ref{fig_dboc_lih_tz}.d also follow a similar pattern to that of H$_2$, where the zero-field DBOC monotonically decreases near equilibrium -- as the field strength is increased, the magnitude of the DBOC increases and the curvature of the DBOC changes, eventually becoming monotonically increasing in the vicinity of equilibrium. Additionally, the magnitude of the DBOC is larger for LiH than for H$_2$. In a  magnetic field, the DBOC values tend to increase as a function of bond distance, with the effect becoming amplified at higher field strengths. This behavior is correlated with the trend in Hessian eigenvalues observed in Fig.\,\ref{fig_hess}.b. The eigenvalues approach zero more quickly as a function of bond distance at higher field strengths.

\begin{figure*}[h]
\centering
\begin{tabular}{ll}
(a) \\
\includegraphics[width=0.48\textwidth]{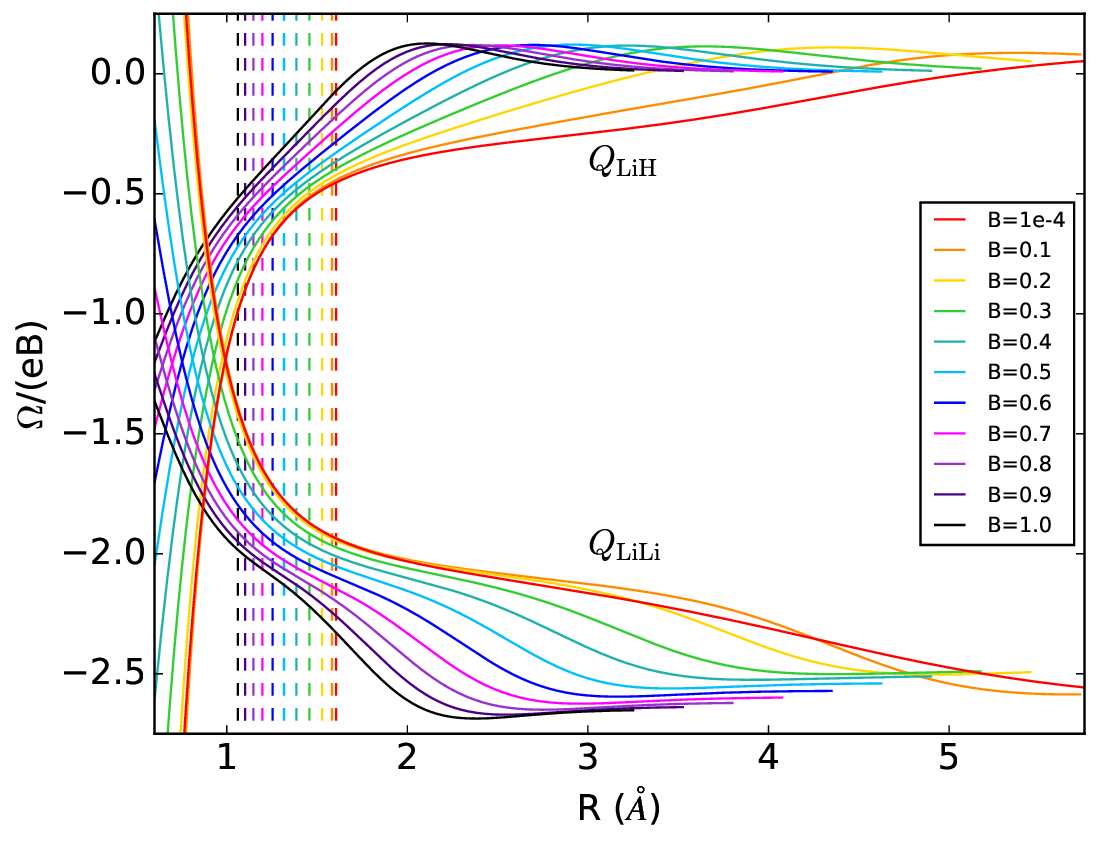} \\
(b) \\
\includegraphics[width=0.48\textwidth]{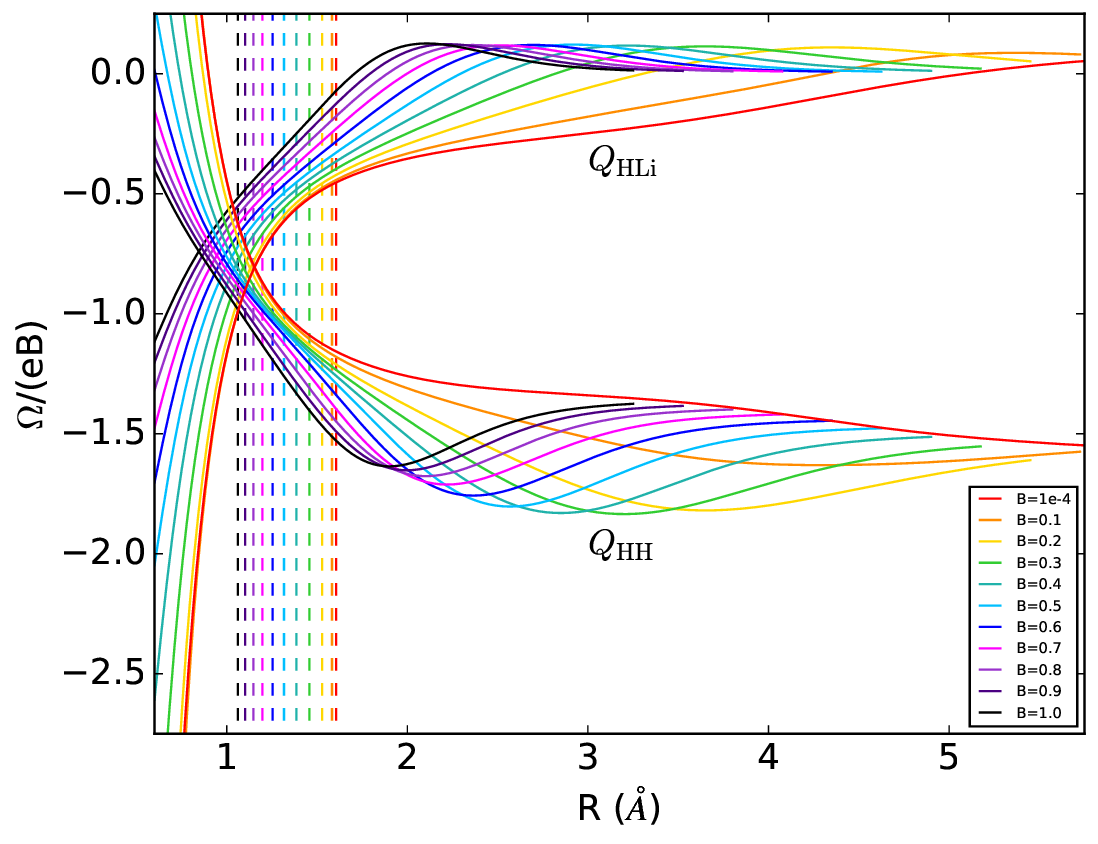} \\
\end{tabular}
\caption{Averages of Berry curvature tensor elements associated with partial charges of lithium (a) and hydrogen (b) as defined in Eqs.\,\eqref{q1} and~\eqref{q2} in the text. The partial charge on lithium is given by q$_\mathrm{Li}$ = Q$_\mathrm{LiH}$ + Q$_\mathrm{LiLi}$ while the partial charge on hydrogen is given by q$_\mathrm{H}$ = Q$_\mathrm{HLi}$ + Q$_\mathrm{HH}$. All calculations were performed on the RHF singlet state with the Lu-cc-pVTZ basis set for a series of magnetic field strengths. The magnetic field is oriented along the z-axis with the molecule oriented along the x-axis. The Lu-cc-pVTZ basis set is comprised of the decontracted Lcc-pVTZ basis. Equilibrium bond distances given by the vertical dashed lines in each plot are 1.606 $\text{\AA}$ (B=$1\times 10^{-4}$), 1.583 $\text{\AA}$ (B=0.1), 1.528 $\text{\AA}$ (B=0.2), 1.458 $\text{\AA}$ (B=0.3), 1.386 $\text{\AA}$ (B=0.4), 1.317 $\text{\AA}$ (B=0.5), 1.254 $\text{\AA}$ (B=0.6), 1.197 $\text{\AA}$ (B=0.7), 1.147 $\text{\AA}$ (B=0.8), 1.102 $\text{\AA}$ (B=0.9) and 1.062 $\text{\AA}$ (B=1.0). Raw data was generated on a grid using a step size of 0.04 ${\text{\AA}}$ and a cubic spline interpolation was used for plotting purposes. The plot legends display magnetic field strengths given in units of $B_0$.}
\label{fig_QhQLi_tz}
\end{figure*}

Examining the Berry curvature  in terms of screening charges in Fig.\,\ref{fig_QhQLi_tz}, we note again that the general shape of each curve is conserved as a function of field strength, but compressed towards the origin as field strength is increased, which may be expected given the trends in equilibrium bond distance. We have from Eqs.\,\eqref{q1} and~\eqref{q2} that $Q_{\mathrm{LiH}} = Q_{\mathrm{HLi}}$, which is observed in Fig.\,\ref{fig_QhQLi_tz}, and we also see that these values of the off-diagonal coupling tend toward zero as the bond is stretched. The off-diagonal elements $Q_{\mathrm{LiH}}$ become antiscreening beyond the equilibrium bond distance, while the diagonal elements $Q_{\mathrm{LiLi}}$ and $Q_{\mathrm{HH}}$ behave differently. For $Q_{\mathrm{LiLi}}$, supercreening is never achieved since the value of the data is never less than $-3$. By contrast, almost all $Q_{\mathrm{HH}}$ data exhibits superscreening for all investigated bond lengths, notably including the domain of equilibrium bond lengths. This is probably a result of the polarity of the Li-H bond.

\begin{figure*}[h]
\centering
\begin{tabular}{ll}
\includegraphics[width=0.48\textwidth]{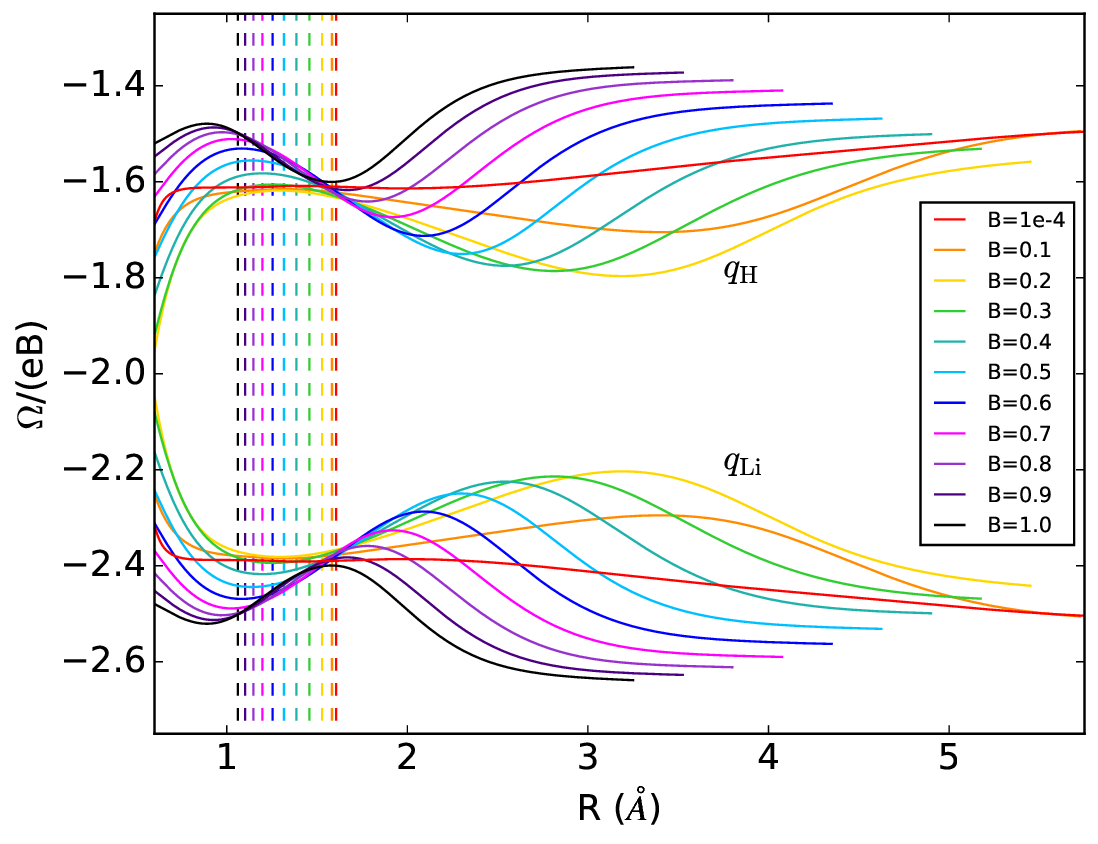} \\
\end{tabular}
\caption{Partial charges of lithium and hydrogen as defined in Eqs.\,\eqref{q1} and~\eqref{q2} in the text. The partial charge on lithium is given by q$_\mathrm{Li}$ = Q$_\mathrm{LiH}$ + Q$_\mathrm{LiLi}$  while the partial charge on hydrogen is given by q$_\mathrm{H}$ = Q$_\mathrm{HLi}$ + Q$_\mathrm{HH}$. All calculations were performed on the RHF singlet state with the Lu-cc-pVTZ basis set for a series of magnetic field strengths. The magnetic field is oriented along the z-axis with the molecule oriented along the x-axis. The Lu-cc-pVTZ basis set is comprised of the decontracted Lcc-pVTZ basis. Equilibrium bond distances given by the vertical dashed lines in each plot are 1.606 $\text{\AA}$ (B=$1\times 10^{-4}$), 1.583 $\text{\AA}$ (B=0.1), 1.528 $\text{\AA}$ (B=0.2), 1.458 $\text{\AA}$ (B=0.3), 1.386 $\text{\AA}$ (B=0.4), 1.317 $\text{\AA}$ (B=0.5), 1.254 $\text{\AA}$ (B=0.6), 1.197 $\text{\AA}$ (B=0.7), 1.147 $\text{\AA}$ (B=0.8), 1.102 $\text{\AA}$ (B=0.9) and 1.062 $\text{\AA}$ (B=1.0). Raw data was generated on a grid using a step size of 0.04 ${\text{\AA}}$ and a cubic spline interpolation was used for plotting purposes. The plot legends display magnetic field strengths given in units of $B_0$.}
\label{fig_qtot_lih_tz}
\end{figure*}

The partial charges $q_{\mathrm{H}}$ and $q_{\mathrm{Li}}$ are shown in Fig.\,\ref{fig_qtot_lih_tz}. Again, we note a compression of the curves as the field strength increases, as well as reflection symmetry across the axis of $-2\Omega/(eB) = - N_\mathrm e/2$, which is necessary because $q_{\mathrm{H}} + q_{\mathrm{Li}} = - N_\mathrm e$ must hold at any given bond distance. As the bond length changes, the partial charges fluctuate, unlike for homonuclear diatomics such as H$_2$, where the partial charges on each atom remain constant. As a general trend, the difference in magnitude between $q_{\mathrm{H}}$ and $q_{\mathrm{Li}}$ increases for increasing field, reflecting a less polar Li-H bond in this regime.

\subsection{BH Molecule}

\begin{figure*}[h]
\centering
\begin{tabular}{ll}
(a) & (b) \\
\includegraphics[width=0.48\textwidth]{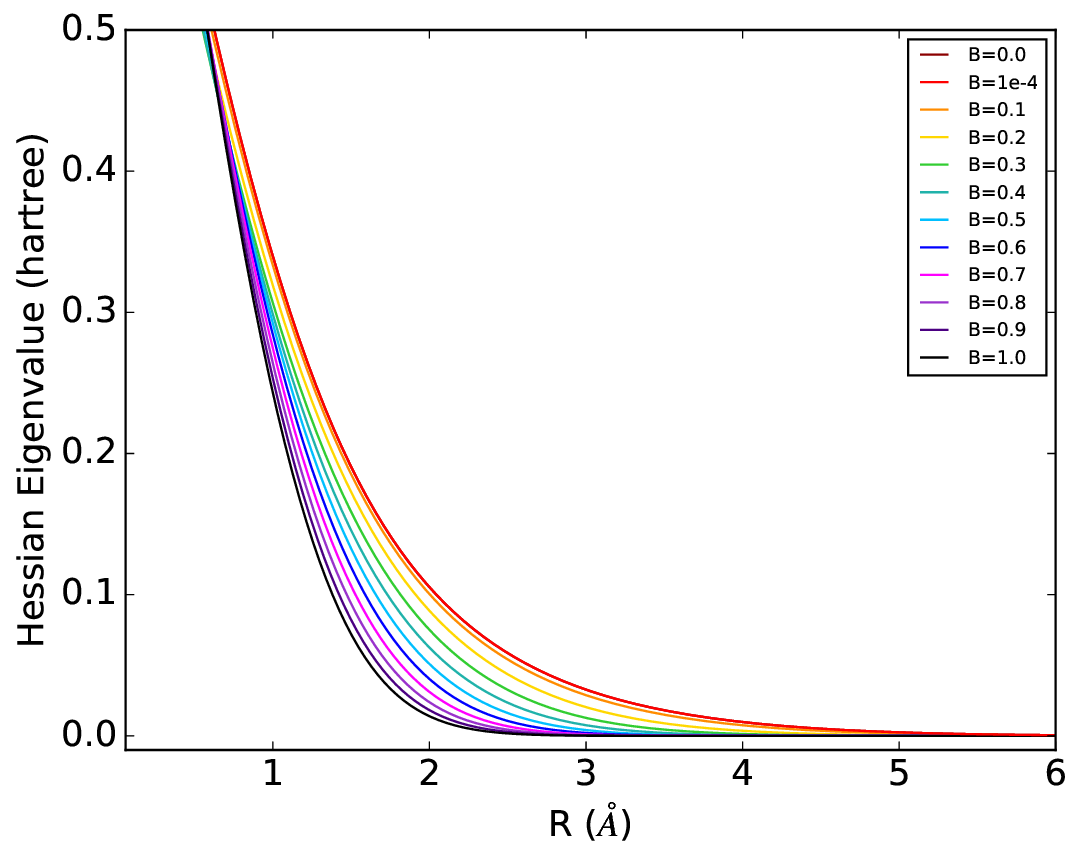} &
\includegraphics[width=0.48\textwidth]{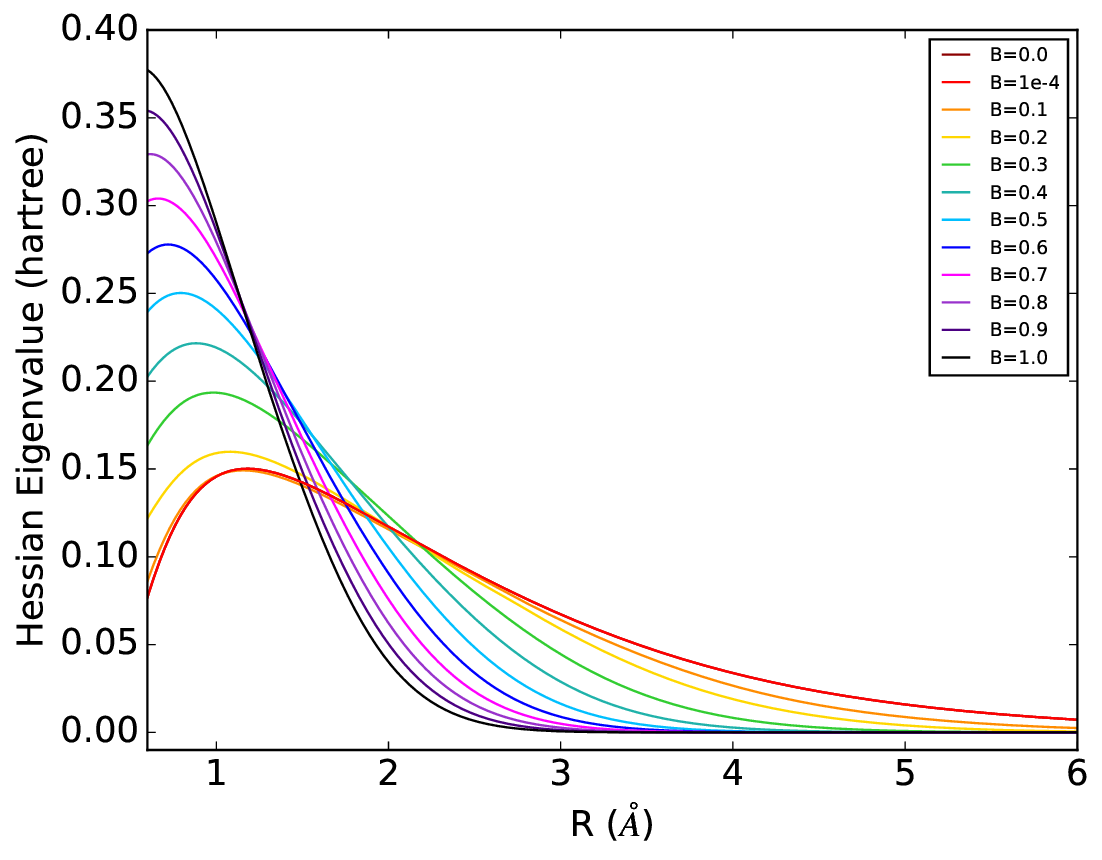} \\
(c) & (d) \\
\includegraphics[width=0.48\textwidth]{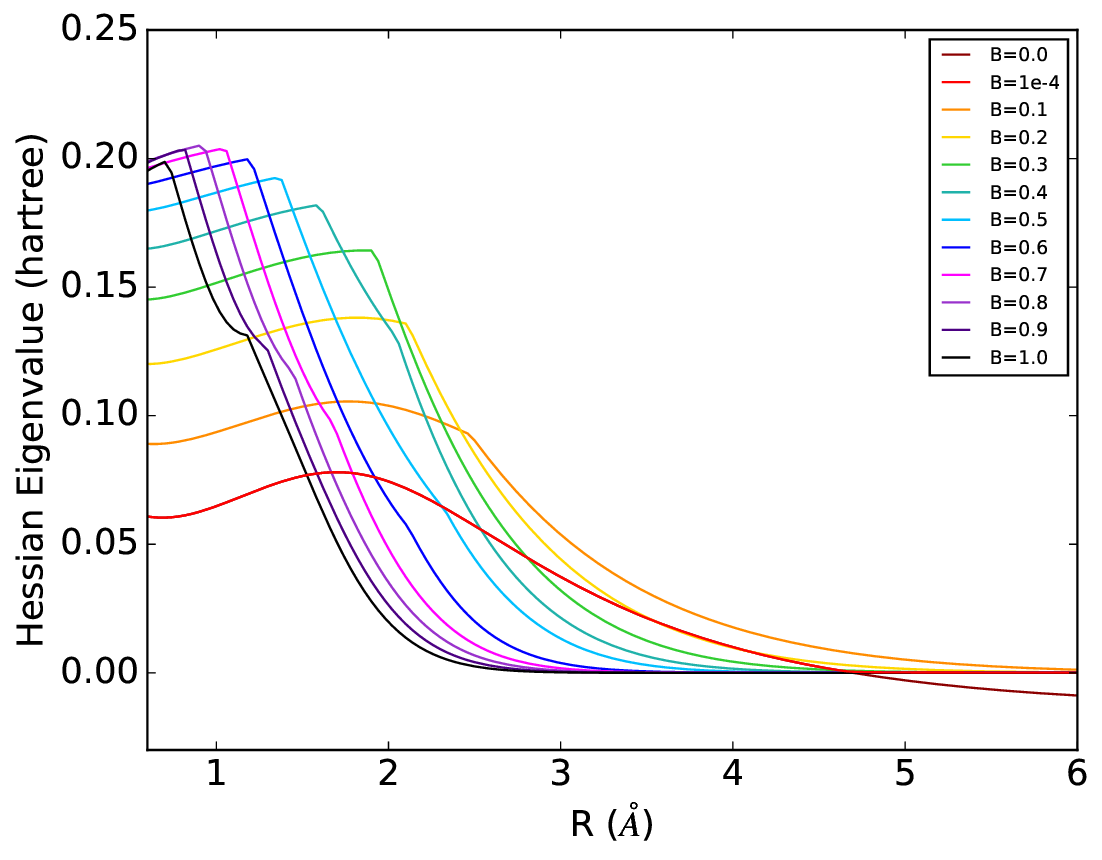} &
\includegraphics[width=0.48\textwidth]{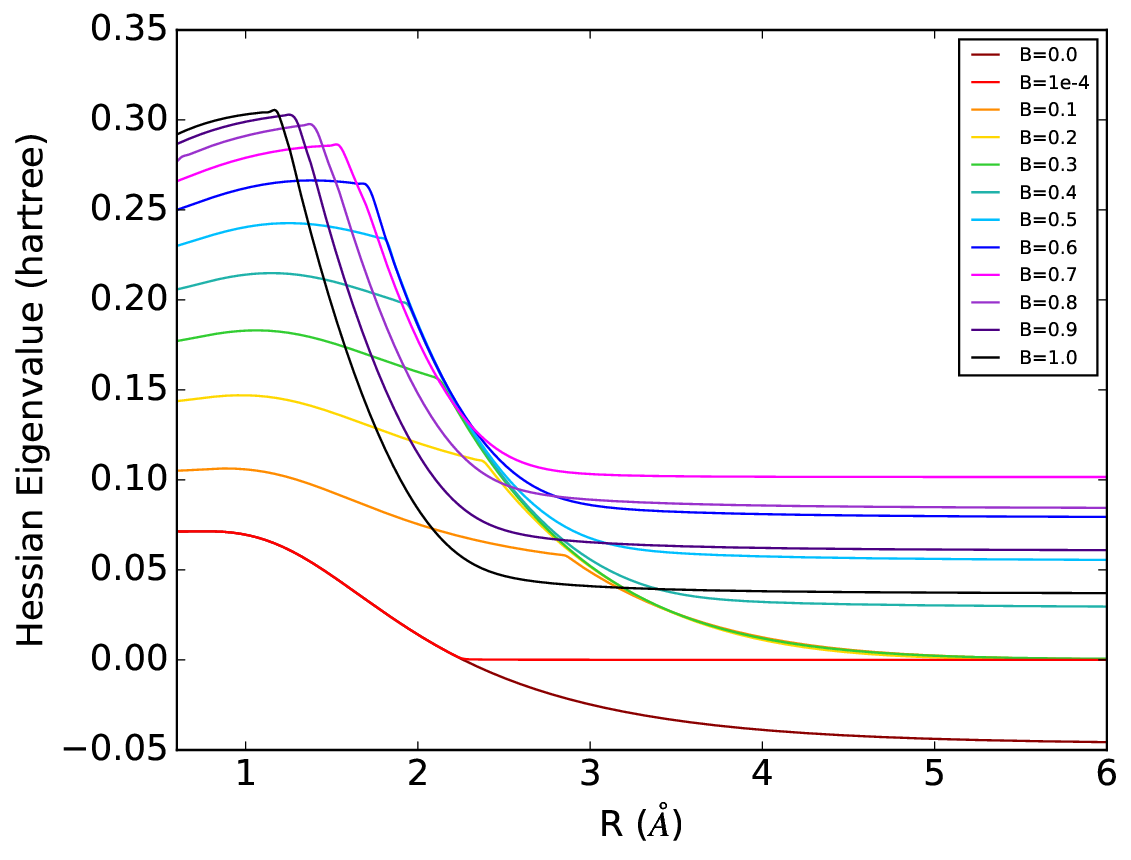} \\
\end{tabular}
\caption{Lowest eigenvalue of the RHF Hessian/stability matrix given in Eq.\,\eqref{cphf_cs2} for the H$_2$ (a), LiH (b), BH (c), and CH$^+$ (d) molecules calculated as a function of bond distance for various magnetic field strengths. Raw data was generated on a grid using a step size of 0.04 ${\text{\AA}}$ and a cubic spline interpolation was used for plotting purposes. The plot legends display magnetic field strengths given in units of $B_0$.}
\label{fig_hess}
\end{figure*}

\begin{figure*}[h]
\centering
\begin{tabular}{ll}
(a) & (b) \\
\includegraphics[width=0.48\textwidth]{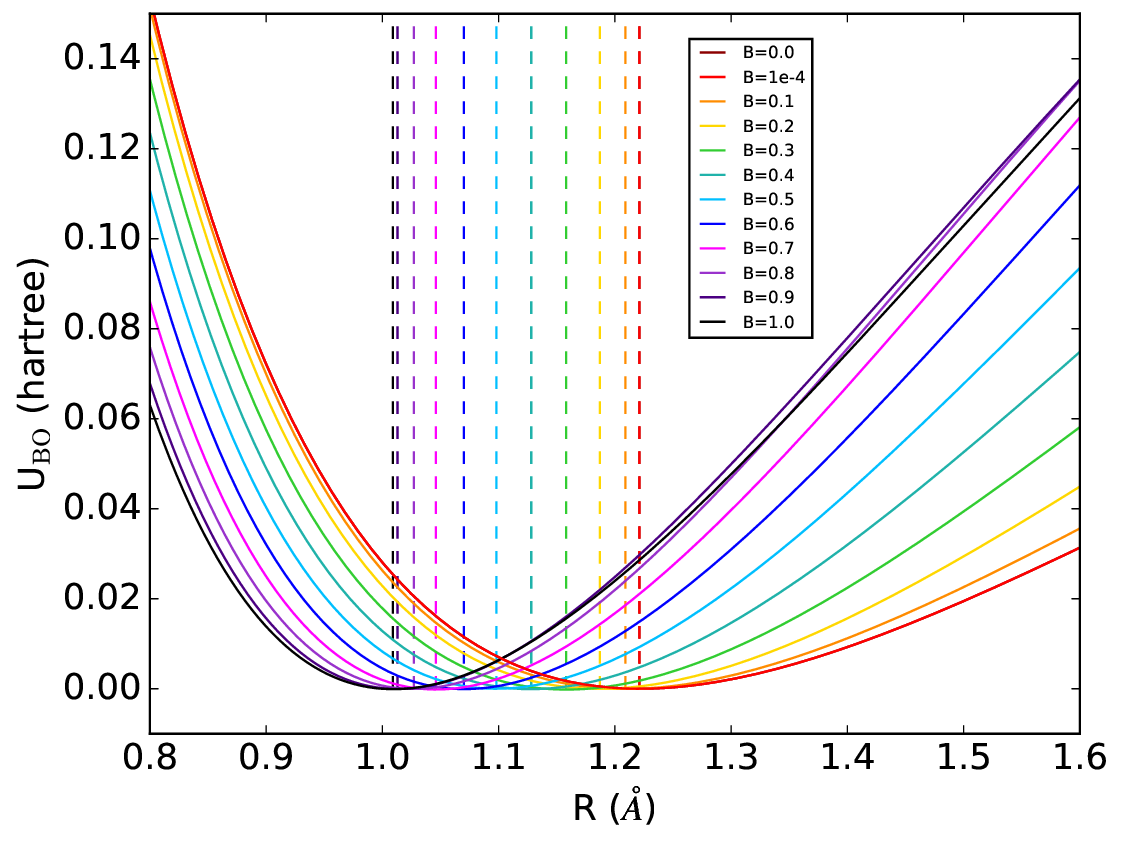} & \includegraphics[width=0.48\textwidth]{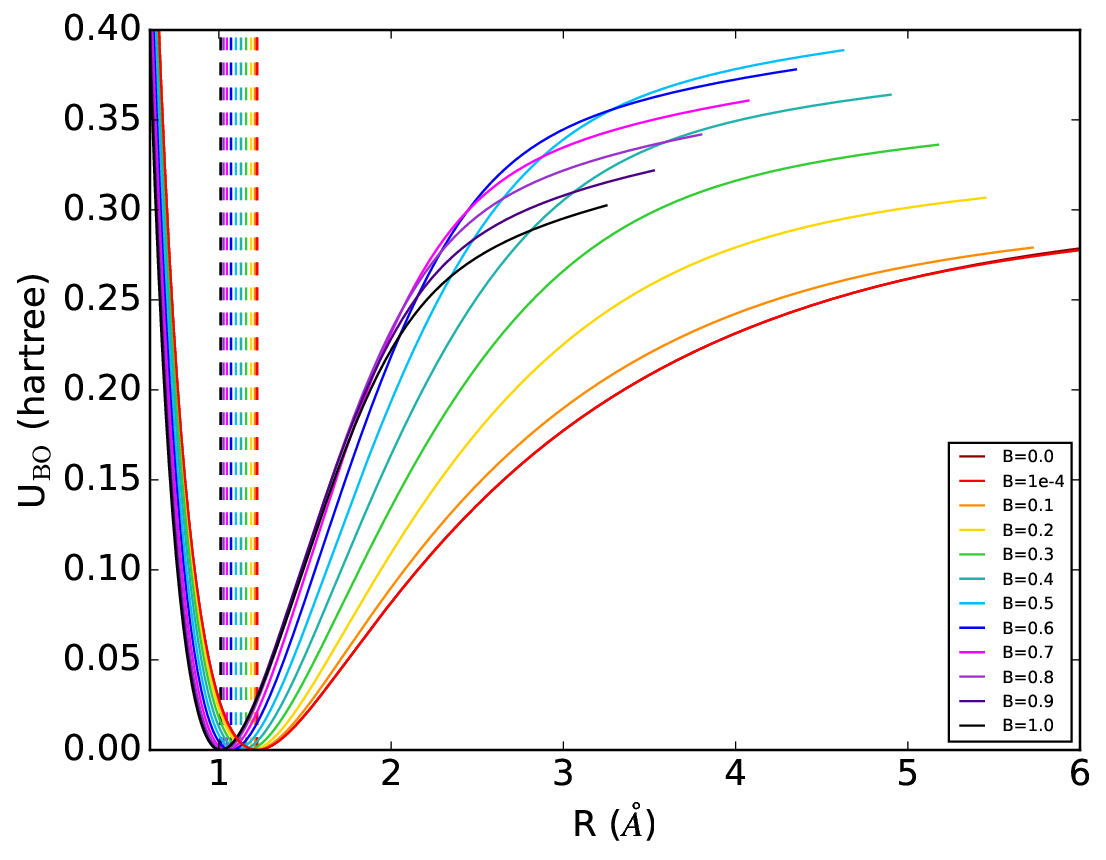} \\
(c) & (d) \\
\includegraphics[width=0.48\textwidth]{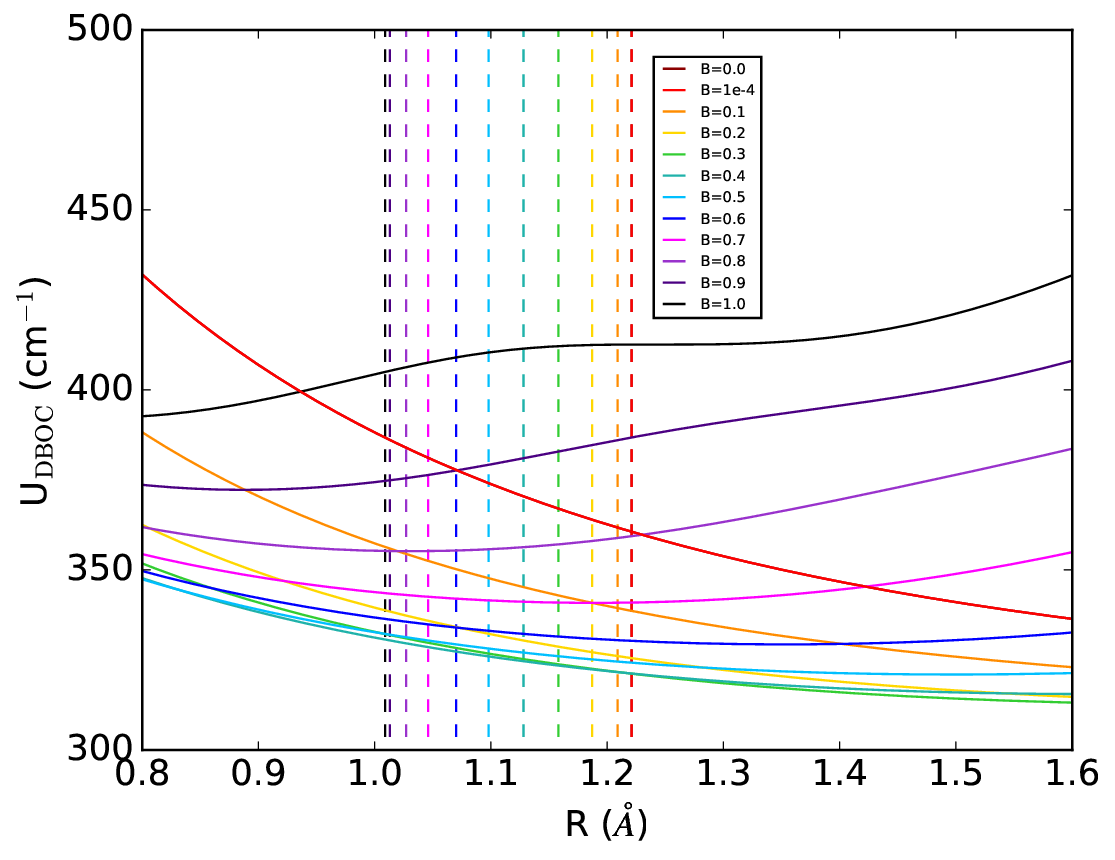} & \includegraphics[width=0.48\textwidth]{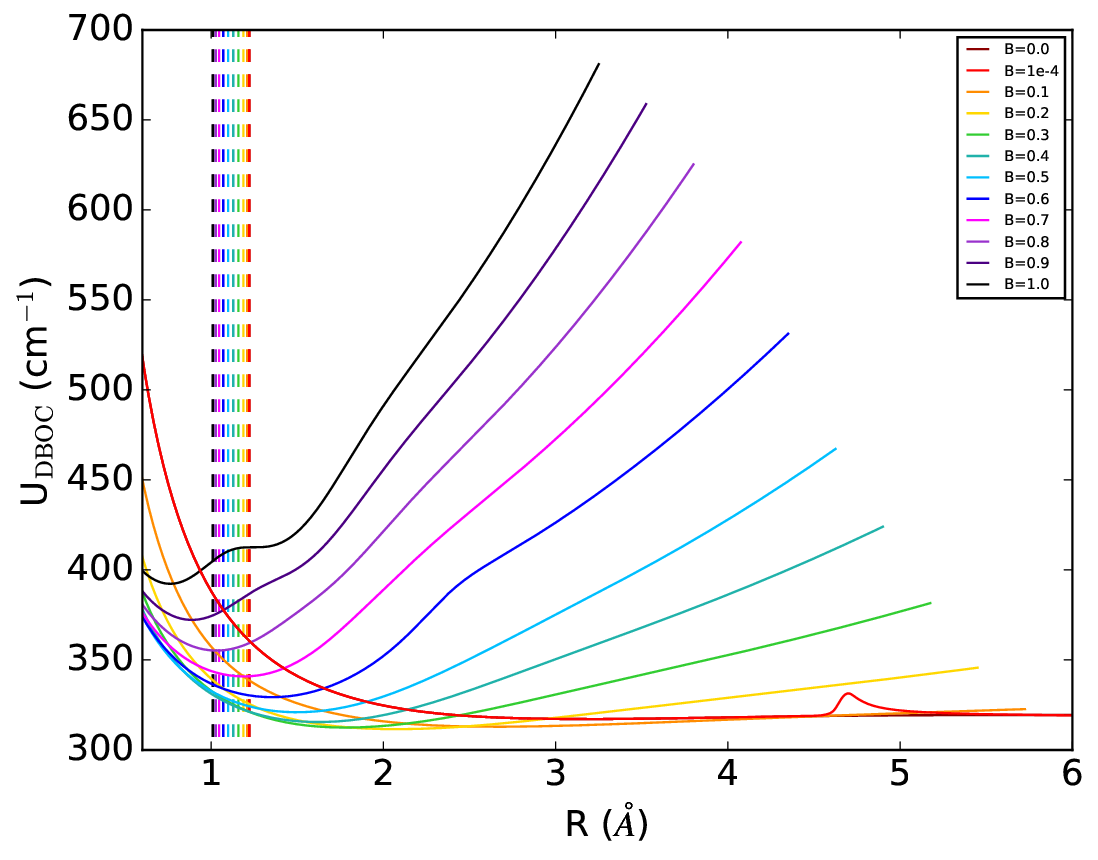} \\
\end{tabular}
\caption{SCF (a $\&$ b) and DBOC (c $\&$ d) energy of BH for the RHF singlet state with the Lu-cc-pVTZ basis set for a series of magnetic field strengths. The magnetic field is oriented along the z-axis with the molecule oriented along the x-axis. The Lu-cc-pVTZ basis set is comprised of the decontracted Lcc-pVTZ basis. Equilibrium bond distances given by the vertical dashed lines in each plot are 1.221 $\text{\AA}$ (B=0.0), 1.221 $\text{\AA}$ (B=$1\times 10^{-4}$), 1.209 $\text{\AA}$ (B=0.1), 1.187 $\text{\AA}$ (B=0.2), 1.158 $\text{\AA}$ (B=0.3), 1.128 $\text{\AA}$ (B=0.4), 1.098 $\text{\AA}$ (B=0.5), 1.070 $\text{\AA}$ (B=0.6), 1.046 $\text{\AA}$ (B=0.7), 1.027 $\text{\AA}$ (B=0.8), 1.013 $\text{\AA}$ (B=0.9) and 1.009 $\text{\AA}$ (B=1.0). Raw data was generated on a grid using a step size of 0.04 ${\text{\AA}}$ and a cubic spline interpolation was used for plotting purposes. The plot legends display magnetic field strengths given in units of $B_0$. The minima of the curves in panels (a $\&$ b) have been shifted to zero.}
\label{fig_dboc_bh_tz}
\end{figure*}

As shown in Figs.\,\ref{fig_dboc_bh_tz}.a and~\ref{fig_dboc_bh_tz}.b, the equilibrium bond distance of BH systematically decreases with increasing field strength. The trends in the DBOC curves of BH depicted in Figs.\,\ref{fig_dboc_bh_tz}.c and~\ref{fig_dboc_bh_tz}.d represent a significant departure  from those exhibited by H$_2$ and LiH. In Fig.~\ref{fig_dboc_bh_tz}.c, the curves become more closely spaced and even intersect, with the zero-field DBOC values being higher than those at field strengths 0.1$B_0$ to 0.6$B_0$. The magnitude of the DBOC decreases from 0.1$B_0$ to 0.3$B_0$, then remains fairly constant to about 0.6$B_0$, where it finally begins to increase with increasing field strength. As observed in Fig.~\ref{fig_dboc_bh_tz}.d, in a finite magnetic field the DBOC values tend to increase as a function of bond distance, with the effect becoming amplified at higher field strengths. This behavior is correlated with the trend in Hessian eigenvalues observed in Fig.\,\ref{fig_hess}.c. The eigenvalues approach zero more quickly as a function of bond distance at higher field strengths.

\begin{figure*}[h]
\centering
\begin{tabular}{ll}
(a) \\
\includegraphics[width=0.48\textwidth]{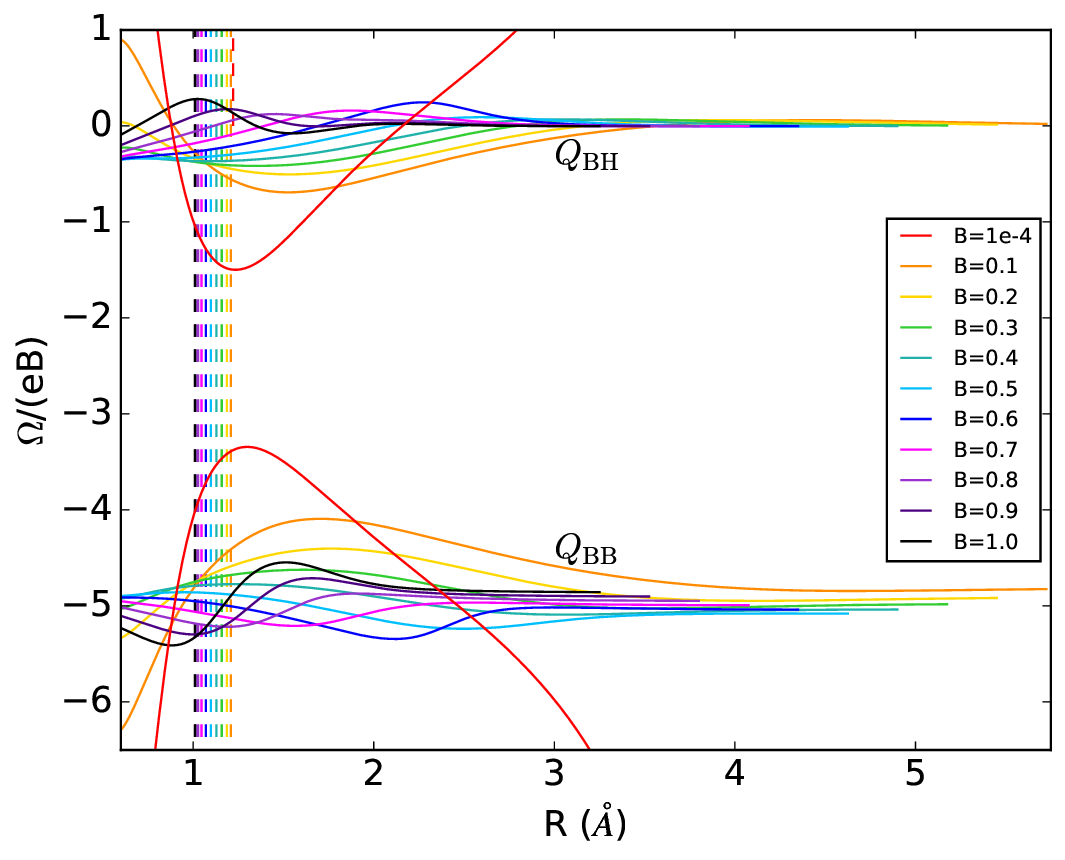} \\
(b) \\
\includegraphics[width=0.48\textwidth]{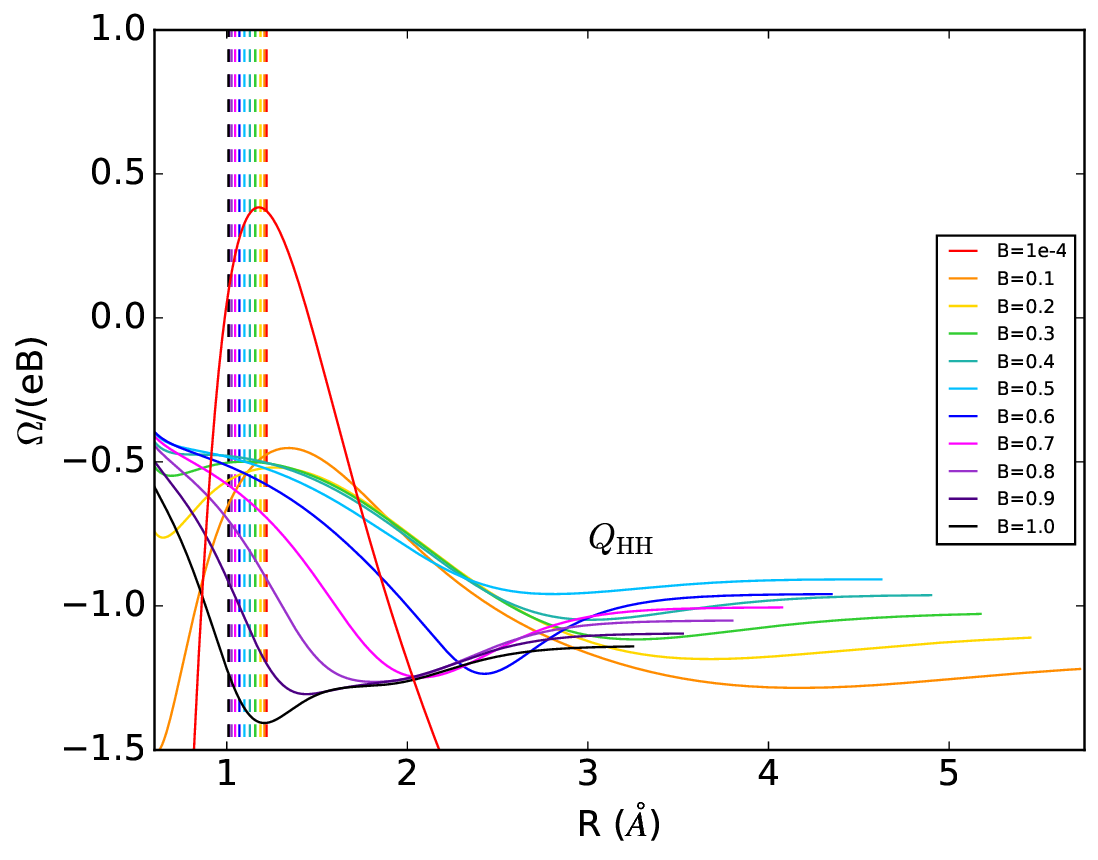} \\
\end{tabular}
\caption{Averages of Berry curvature tensor elements associated with partial charges of boron (a) and hydrogen (b) as defined in Eqs.\,\eqref{q1} and~\eqref{q2} in the text. The partial charge on boron is given by q$_\mathrm{B}$ = Q$_\mathrm{BH}$ + Q$_\mathrm{BB}$ while the partial charge on hydrogen is given by q$_\mathrm{H}$ = Q$_\mathrm{HB}$ + Q$_\mathrm{HH}$. All calculations were performed on the RHF singlet state with the Lu-cc-pVTZ basis set for a series of magnetic field strengths. The magnetic field is oriented along the z-axis with the molecule oriented along the x-axis. The Lu-cc-pVTZ basis set is comprised of the decontracted Lcc-pVTZ basis. Equilibrium bond distances given by the vertical dashed lines in each plot are 1.221 $\text{\AA}$ (B=$1\times 10^{-4}$), 1.209 $\text{\AA}$ (B=0.1), 1.187 $\text{\AA}$ (B=0.2), 1.158 $\text{\AA}$ (B=0.3), 1.128 $\text{\AA}$ (B=0.4), 1.098 $\text{\AA}$ (B=0.5), 1.070 $\text{\AA}$ (B=0.6), 1.046 $\text{\AA}$ (B=0.7), 1.027 $\text{\AA}$ (B=0.8), 1.013 $\text{\AA}$ (B=0.9) and 1.009 $\text{\AA}$ (B=1.0). Raw data was generated on a grid using a step size of 0.04 ${\text{\AA}}$ and a cubic spline interpolation was used for plotting purposes. The plot legends display magnetic field strengths given in units of $B_0$.}
\label{fig_QhQB_tz}
\end{figure*}

The Berry-curvature behavior in terms of screening charges is shown in Fig.\,\ref{fig_QhQB_tz}. The curves do not evolve as incrementally as those of H$_2$ and LiH. Instead, as for the DBOC, the evolution of the curves as a function of field strength shows differing trends for lower fields and higher fields, although certain features are always conserved. For example, we see that the values of the off-diagonal coupling $Q_{\mathrm{BH}}$ tend toward zero as the bond distance is increased (excepting $1 \times 10^{-4} B_0$). Additionally, we see from Eqs.\,\eqref{q1} and~\eqref{q2} that $Q_{\mathrm{BH}} = Q_{\mathrm{HB}}$, although $Q_{\mathrm{HB}}$ is not shown in Fig.\,\ref{fig_QhQB_tz}.b due to the large overlap of the data being difficult to visually parse. The off-diagonal elements $Q_{\mathrm{BH}}$ exhibit antiscreening behavior both near and far from equilibrium, depending on the field strength. This is in contrast to H$_2$ and LiH, which only exhibit antiscreening outside of equilibrium bond distance ranges. For $Q_{\mathrm{BB}}$, supercreening is achieved depending upon field strength and bond distance. In the vicinity of equilibrium, the weaker field strengths from 0.1$B_0$ to 0.5$B_0$ do not give values below $-5$, while the values at higher field are all in the superscreening range. The $Q_{\mathrm{HH}}$ curves also show superscreening depending on field strength and bond distance, although the domains are different than in the case of $Q_{\mathrm{BB}}$. 

\begin{figure*}[h]
\centering
\begin{tabular}{ll}
\includegraphics[width=0.48\textwidth]{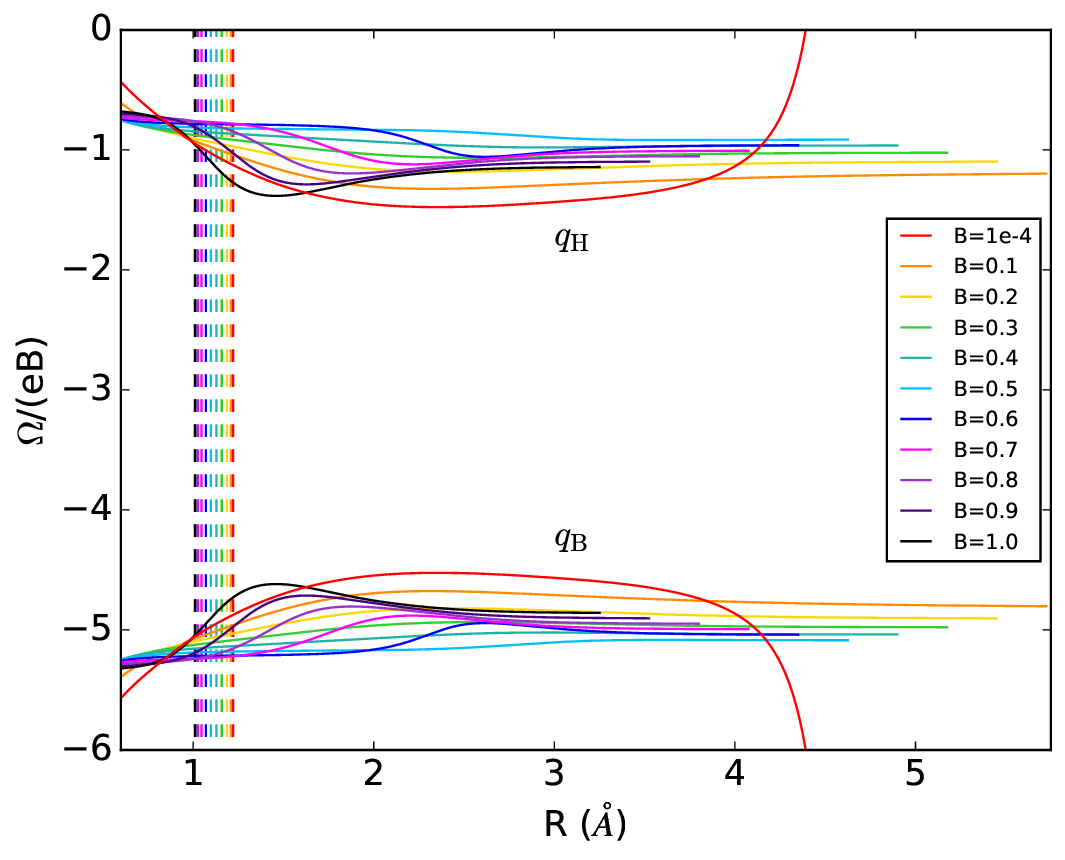} \\
\end{tabular}
\caption{Partial charges of boron and hydrogen as defined in Eqs.\,\eqref{q1} and~\eqref{q2} in the text. The partial charge on boron is given by q$_\mathrm{B}$ = Q$_\mathrm{BH}$ + Q$_\mathrm{BB}$ while the partial charge on hydrogen is given by q$_\mathrm{H}$ = Q$_\mathrm{HB}$ + Q$_\mathrm{HH}$. All calculations were performed on the RHF singlet state with the Lu-cc-pVTZ basis set for a series of magnetic field strengths. The magnetic field is oriented along the z-axis with the molecule oriented along the x-axis. The Lu-cc-pVTZ basis set is comprised of the decontracted Lcc-pVTZ basis. Equilibrium bond distances given by the vertical dashed lines in each plot are 1.221 $\text{\AA}$ (B=$1\times 10^{-4}$), 1.209 $\text{\AA}$ (B=0.1), 1.187 $\text{\AA}$ (B=0.2), 1.158 $\text{\AA}$ (B=0.3), 1.128 $\text{\AA}$ (B=0.4), 1.098 $\text{\AA}$ (B=0.5), 1.070 $\text{\AA}$ (B=0.6), 1.046 $\text{\AA}$ (B=0.7), 1.027 $\text{\AA}$ (B=0.8), 1.013 $\text{\AA}$ (B=0.9) and 1.009 $\text{\AA}$ (B=1.0). Raw data was generated on a grid using a step size of 0.04 ${\text{\AA}}$ and a cubic spline interpolation was used for plotting purposes. The plot legends display magnetic field strengths given in units of $B_0$.}
\label{fig_qtot_bh_tz}
\end{figure*}

The partial charges $q_{\mathrm{H}}$ and $q_{\mathrm{B}}$ are shown in Fig.\,\ref{fig_qtot_bh_tz}. We note the reflection symmetry across the axis of $-3$ $\Omega/(eB) = - N_\text e/2$, which is necessary because $q_{\mathrm{H}} + q_{\mathrm{B}} = -N_\text e$ must hold at any given bond distance. As the bond length changes, the partial charges fluctuate, and there is significant overlap of lower field strength $q_{\mathrm{H}}$ curves with higher field strength $q_{\mathrm{H}}$ curves, and of lower field strength $q_{\mathrm{B}}$ curves with higher field strength $q_{\mathrm{B}}$ curves. The difference in magnitude between $q_{\mathrm{H}}$ and $q_{\mathrm{B}}$ increases for increasing field strength until 0.6$B_0$, from which point this difference decreases with increasing field strength. Thus the polarity of the B-H bond seems to depend on the magnetic field strength. Note that the weak-field curves associated with $1 \times 10^{-4} B_0$ behave quite differently from those in stronger fields, as is evident for both the screening charges and partial charges. The partial charges begin to exhibit divergent behavior around 4 ${\text{\AA}}$. For this reason, data was not plotted past 4.5 ${\text{\AA}}$ for the screening charges and partial charges.

Finally, beyond 4.66 ${\text{\AA}}$, it was found that time-reversal symmetry of the system was broken in the case of zero magnetic field, with the wave function becoming complex valued and the RHF stability matrix becoming singular. Real-valued wave functions associated with higher energy beyond this point could be optimized by using the final density from geometry $n-1$ as a guess for geometry $n$. The RHF stability matrix has negative eigenvalues for real-valued wave functions in this regime (see Fig.\,\ref{fig_hess}.c). In Fig.\,\ref{fig_hess}.c, the weak-field curve at $1 \times 10^{-4} B_0$ deviates from the zero-field curve at 4.66 ${\text{\AA}}$, coinciding with a hump in the weak-field DBOC at roughly the same bond distance (Fig.\,\ref{fig_dboc_bh_tz}.d) as well as with discontinuities in the weak-field Berry curvature (Fig.\,\ref{fig_QhQB_tz} and Fig.\,\ref{fig_qtot_bh_tz}) as previously stated.

For the purposes of this work, broken time-reversal symmetry for zero magnetic field is of no practical importance as it occurs away from the equilibrium bond distance and is thus inconsequential for applications such as molecular dynamics. Nonetheless, it is an interesting observation, given that the Berry-curvature values in such a domain will become finite even though there is no magnetic field. Moreover, the onset of time-reversal symmetry breaking in the zero-field case is correlated with the onset of divergent behavior of the DBOC and partial charges in the weak-field case of $1 \times 10^{-4} B_0$. This phenomenon therefore has the potential for future study, but is beyond the scope of the present paper

\subsection{CH$^{+}$ Molecule}

\begin{figure*}[h]
\centering
\begin{tabular}{ll}
(a) & (b) \\
\includegraphics[width=0.48\textwidth]{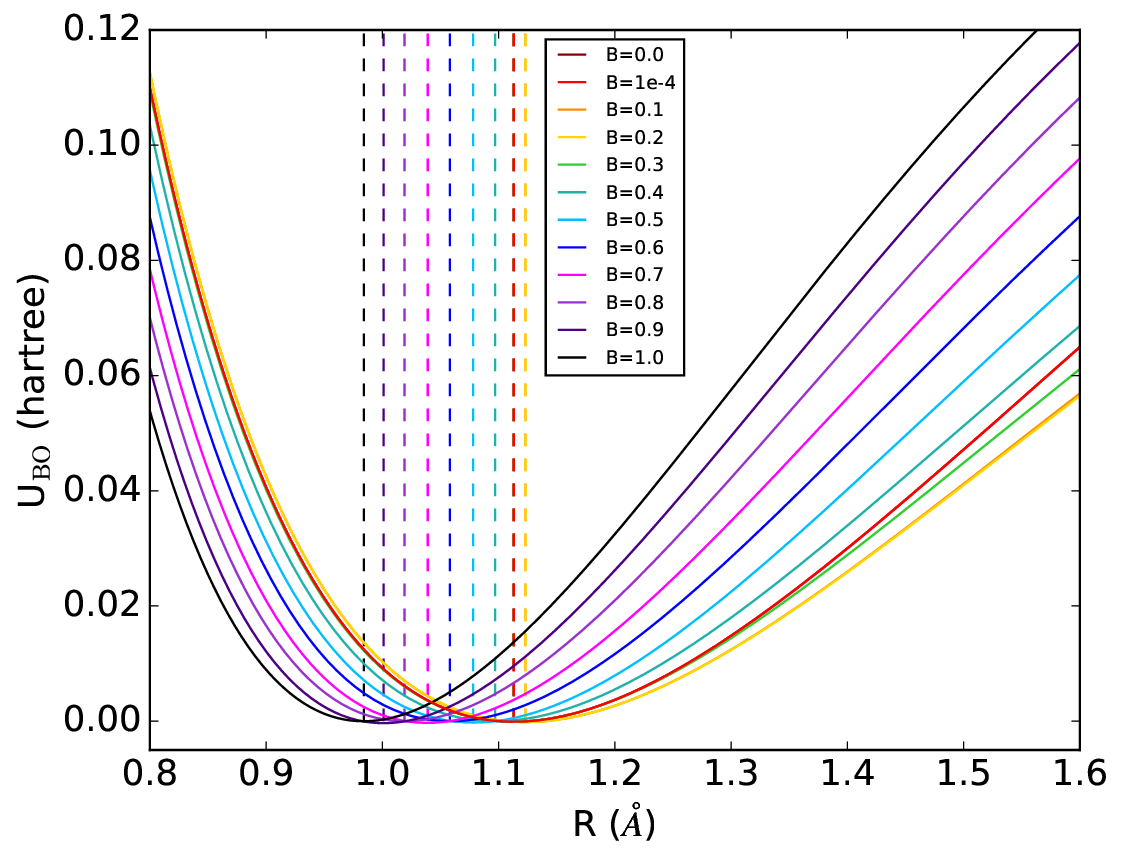} & \includegraphics[width=0.48\textwidth]{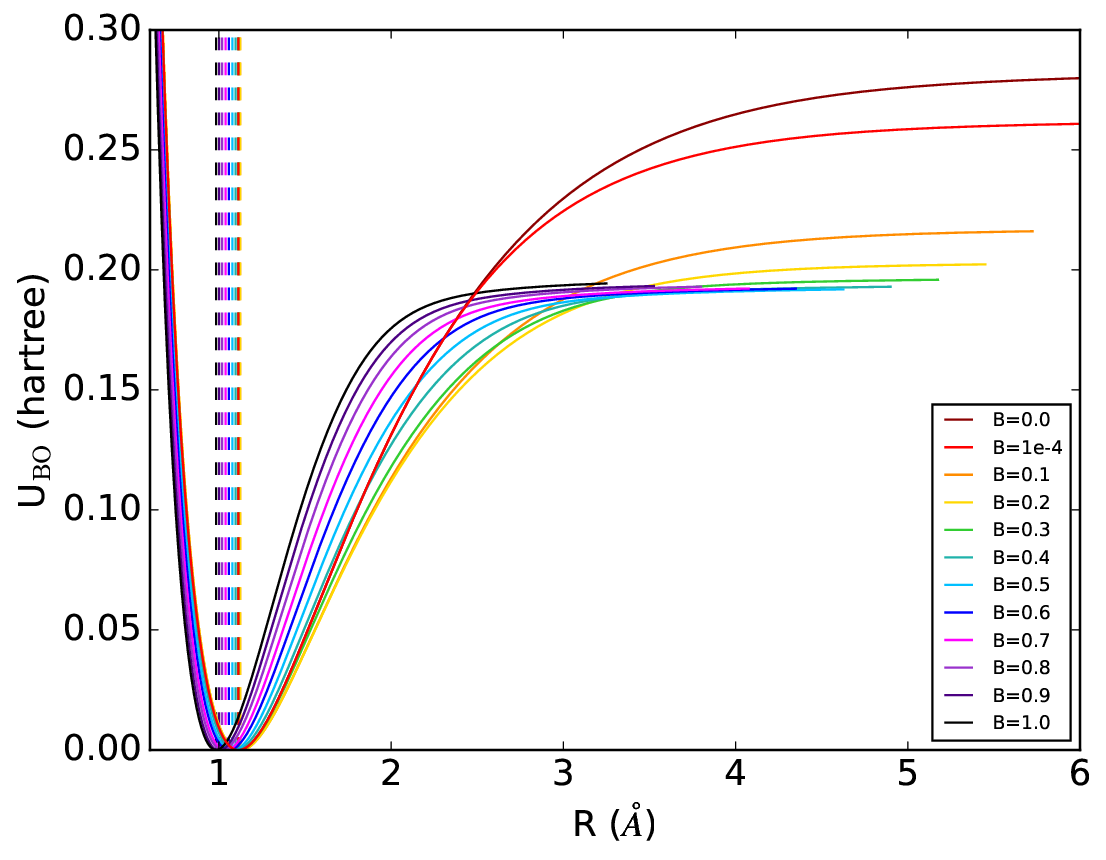}  \\
(c) & (d) \\
\includegraphics[width=0.48\textwidth]{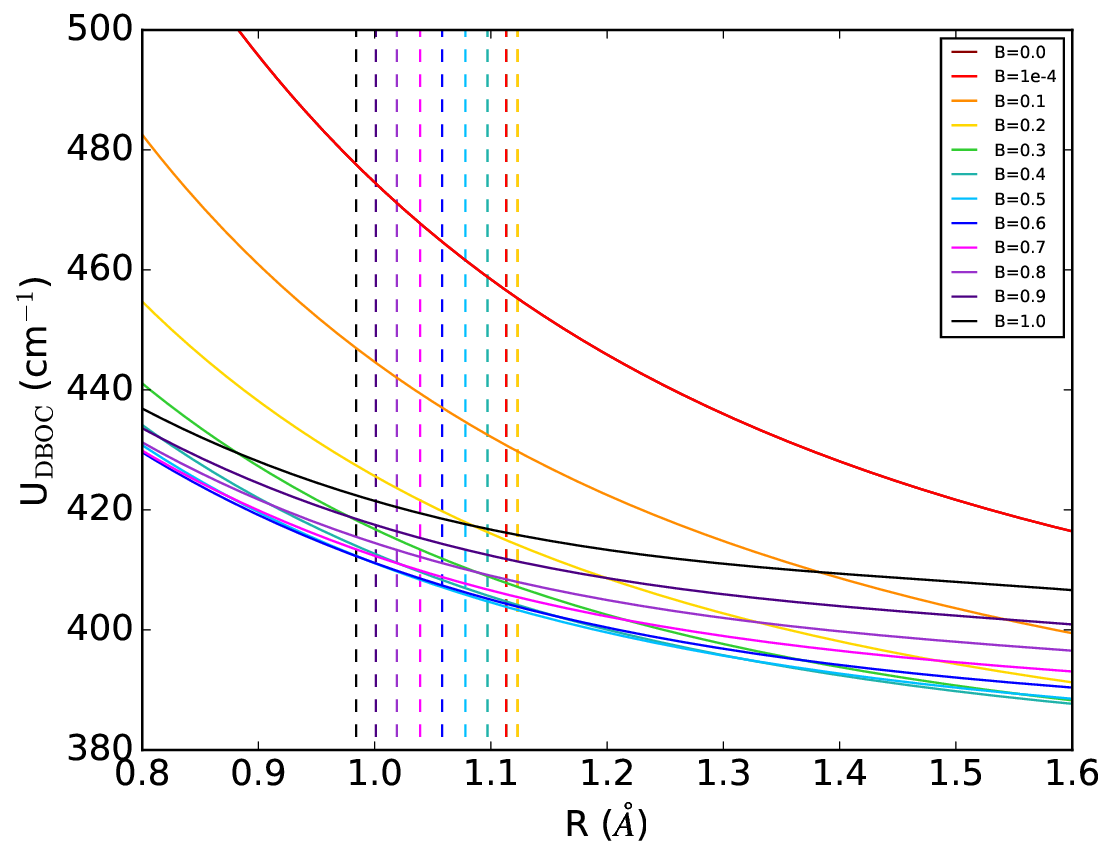} & \includegraphics[width=0.48\textwidth]{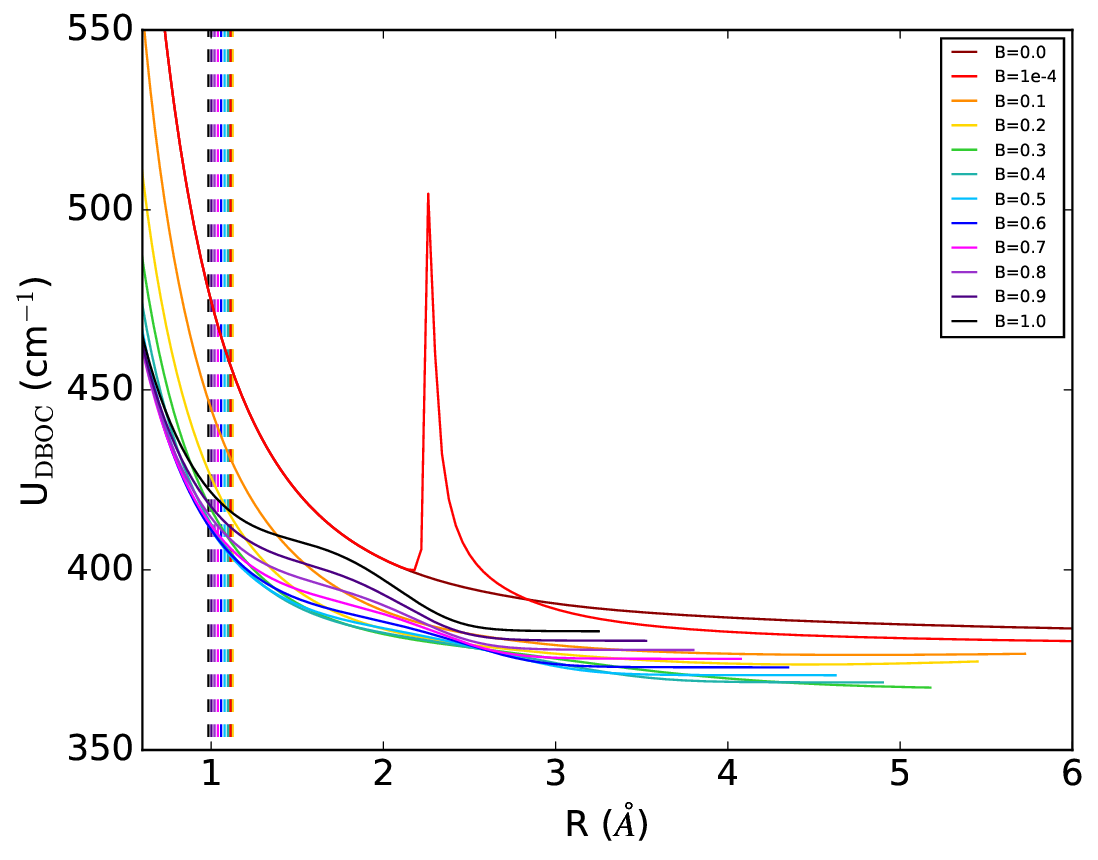}  \\
\end{tabular}
\caption{SCF (a $\&$ b) and DBOC (c $\&$ d) energy of CH$^{+}$ for the RHF singlet state with the Lu-cc-pVTZ basis set for a series of magnetic field strengths. The magnetic field is oriented along the z-axis with the molecule oriented along the x-axis. The Lu-cc-pVTZ basis set is comprised of the decontracted Lcc-pVTZ basis. Equilibrium bond distances given by the vertical dashed lines in each plot are 1.113 $\text{\AA}$ (B=0.0), 1.113 $\text{\AA}$ (B=$1\times 10^{-4}$), 1.123 $\text{\AA}$ (B=0.1), 1.123 $\text{\AA}$ (B=0.2), 1.113 $\text{\AA}$ (B=0.3), 1.097 $\text{\AA}$ (B=0.4), 1.078 $\text{\AA}$ (B=0.5), 1.058 $\text{\AA}$ (B=0.6), 1.039 $\text{\AA}$ (B=0.7), 1.019 $\text{\AA}$ (B=0.8), 1.001 $\text{\AA}$ (B=0.9) and 0.984 $\text{\AA}$ (B=1.0). Raw data was generated on a grid using a step size of 0.04 ${\text{\AA}}$ and a cubic spline interpolation was used for plotting purposes for all curves except the $1 \times 10^{-4} B_0$ curve in panel (d) which was not spline interpolated due to the divergent behavior. The plot legends display magnetic field strengths given in units of $B_0$. The minima of the curves in panels (a $\&$ b) have been shifted to zero.}
\label{fig_dboc_ch_tz}
\end{figure*}

Unlike for the other molecules, the bond distance of CH$^+$ first increases with increasing field strength until at about 0.3$B_0$ it begins to decrease as the bond stiffens --
see Fig.\,\ref{fig_dboc_ch_tz}.a and Fig.\,\ref{fig_dboc_ch_tz}.b.
The trends in the DBOC curves in Fig.\,\ref{fig_dboc_ch_tz}.c are similar to those exhibited by BH. 
The curves intersect, and the zero-field DBOC values are actually higher than all others, whereas for BH the stronger field DBOC values eventually became larger in magnitude than the zero field values over most of the domain. The magnitude of the DBOC decreases with increasing field strength from 0.1$B_0$ to about 0.5$B_0$. Beyond, 0.5$B_0$ the curves remain close in energy but do increase in magnitude with increasing field strength. 

Away from equilibrium, the DBOC trends for CH$^{+}$ in a magnetic field
differ  from the other molecules presented here; see Fig.\,\ref{fig_dboc_ch_tz}.d. Rather than increasing as a function of bond distance and  field strength, it  decreases. This is correlated with the behavior of the Hessian, and we observe in Fig.\,\ref{fig_hess}.d that the lowest Hessian eigenvalues do not systematically approach zero with increasing bond distance and field strength, instead approaching finite values at longer bond distances for most field strengths, or approaching zero relatively slowly.  

\begin{figure*}[h]
\centering
\begin{tabular}{ll}
(a) \\
\includegraphics[width=0.48\textwidth]{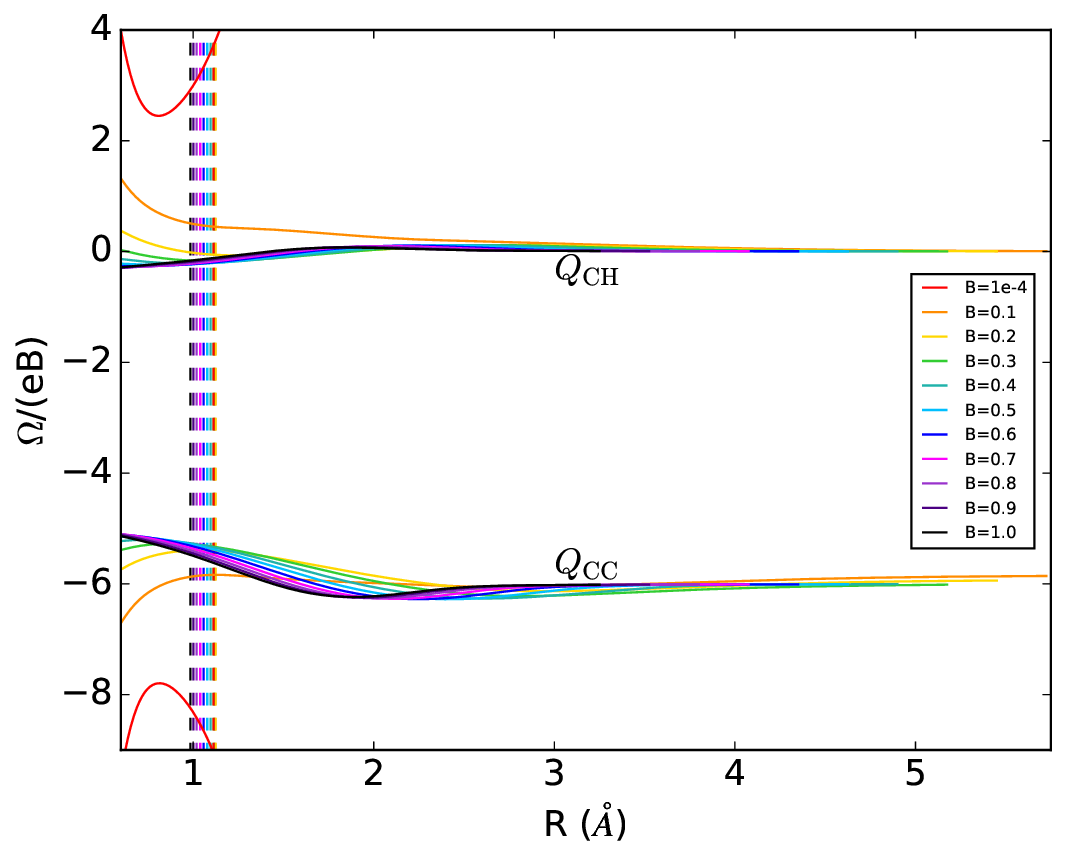} \\
(b) \\
\includegraphics[width=0.48\textwidth]{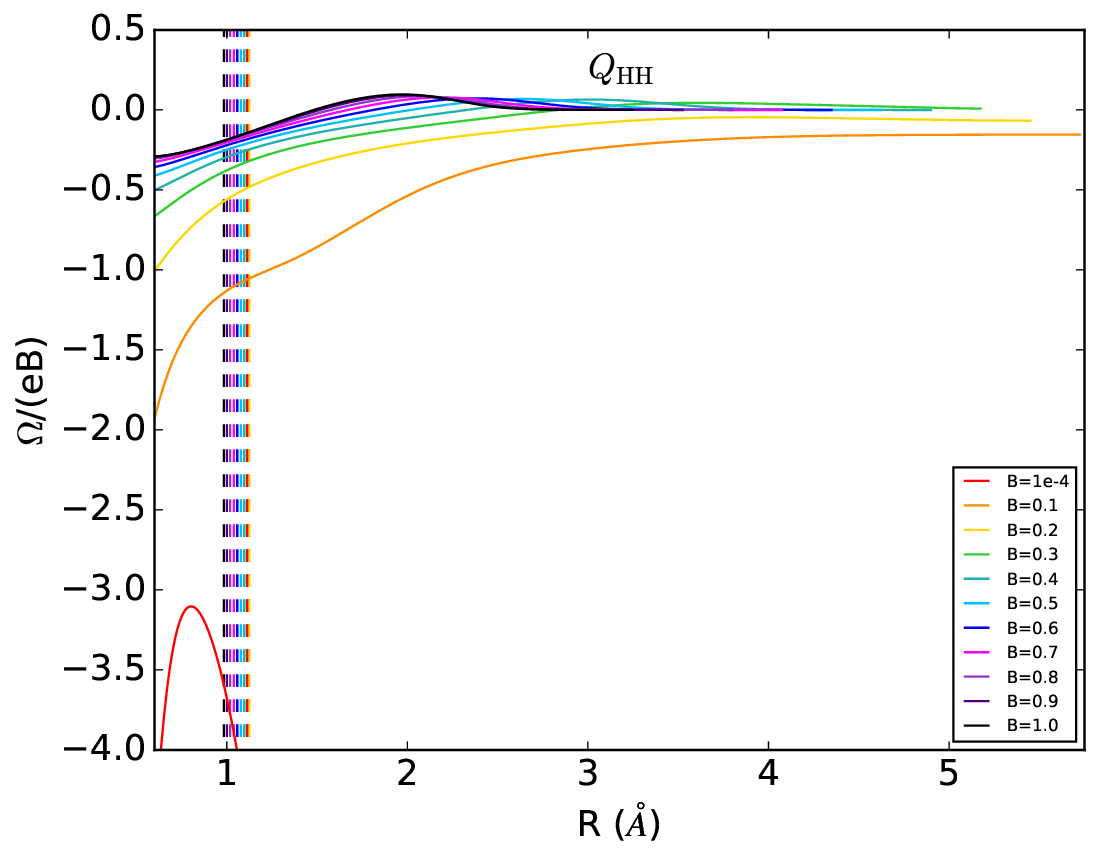} \\
\end{tabular}
\caption{Averages of Berry curvature tensor elements associated with partial charges of carbon (a) and hydrogen (b) as defined in Eqs.\,\eqref{q1} and~\eqref{q2} in the text. The partial charge on carbon is given by q$_\mathrm{C}$ = Q$_\mathrm{CH}$ + Q$_\mathrm{CC}$ while the partial charge on hydrogen is given by q$_\mathrm{H}$ = Q$_\mathrm{HC}$ + Q$_\mathrm{HH}$. All calculations were performed on the RHF singlet state with the Lu-cc-pVTZ basis set for a series of magnetic field strengths. The magnetic field is oriented along the z-axis with the molecule oriented along the x-axis. The Lu-cc-pVTZ basis set is comprised of the decontracted Lcc-pVTZ basis. Equilibrium bond distances given by the vertical dashed lines in each plot are 1.113 $\text{\AA}$ (B=$1\times 10^{-4}$), 1.123 $\text{\AA}$ (B=0.1), 1.123 $\text{\AA}$ (B=0.2), 1.113 $\text{\AA}$ (B=0.3), 1.097 $\text{\AA}$ (B=0.4), 1.078 $\text{\AA}$ (B=0.5), 1.058 $\text{\AA}$ (B=0.6), 1.039 $\text{\AA}$ (B=0.7), 1.019 $\text{\AA}$ (B=0.8), 1.001 $\text{\AA}$ (B=0.9) and 0.984 $\text{\AA}$ (B=1.0). Raw data was generated on a grid using a step size of 0.04 ${\text{\AA}}$ and a cubic spline interpolation was used for plotting purposes. The plot legends display magnetic field strengths given in units of $B_0$.}
\label{fig_QhQC_tz}
\end{figure*}

The Berry curvature behavior in terms of screening charges is shown in Fig.\,\ref{fig_QhQC_tz}. For Fig.\,\ref{fig_QhQC_tz}.a, the curves evolve as a function of field strength and bond distance in a more incremental fashion than the corresponding BH curves. The main differences, beyond the behavior in the short bond-distance region, are that the $Q_{\mathrm{CH}}$ and $Q_{\mathrm{CC}}$ curves are more tightly packed than their BH counterparts, with the $Q_{\mathrm{CH}}$ curves being especially close together. We again note that the values of the off-diagonal coupling $Q_{\mathrm{CH}}$ tend toward zero as the bond distance increases (excepting the case of $1 \times 10^{-4} B_0$). We have from Eqs.\,\eqref{q1} and~\eqref{q2} that $Q_{\mathrm{CH}} = Q_{\mathrm{HC}}$, although $Q_{\mathrm{HC}}$ is not shown in Fig.\,\ref{fig_QhQC_tz}.b due to the large overlap of the data being difficult to visually parse. The off-diagonal elements $Q_{\mathrm{CH}}$ exhibit antiscreening behavior away from equilibrium except at $1 \times 10^{-4} B_0$ and 0.1$B_0$, which are strongly antiscreening at equilibrium. For $Q_{\mathrm{CC}}$, supercreening values less than $-6$ are achieved depending upon field strength and bond distance, though always outside of equilibrium except for the case of $1 \times 10^{-4} B_0$. 

\begin{figure*}[h]
\centering
\begin{tabular}{ll}
\includegraphics[width=0.48\textwidth]{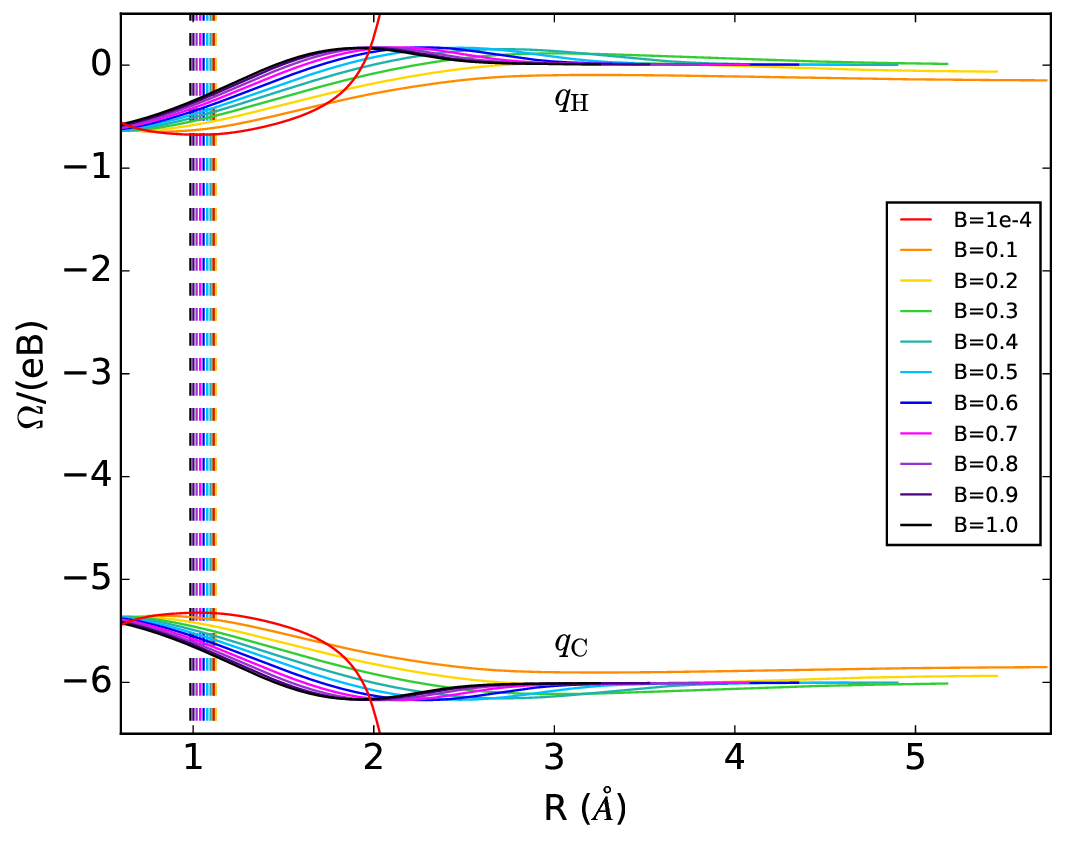} \\
\end{tabular}
\caption{Partial charges of carbon and hydrogen as defined in Eqs.\,\eqref{q1} and~\eqref{q2} in the text. The partial charge on carbon is given by q$_\mathrm{C}$ = Q$_\mathrm{CH}$ + Q$_\mathrm{CC}$ while the partial charge on hydrogen is given by q$_\mathrm{H}$ = Q$_\mathrm{HC}$ + Q$_\mathrm{HH}$. All calculations were performed on the RHF singlet state with the Lu-cc-pVTZ basis set for a series of magnetic field strengths. The magnetic field is oriented along the z-axis with the molecule oriented along the x-axis. The Lu-cc-pVTZ basis set is comprised of the decontracted Lcc-pVTZ basis. Equilibrium bond distances given by the vertical dashed lines in each plot are 1.113 $\text{\AA}$ (B=$1\times 10^{-4}$), 1.123 $\text{\AA}$ (B=0.1), 1.123 $\text{\AA}$ (B=0.2), 1.113 $\text{\AA}$ (B=0.3), 1.097 $\text{\AA}$ (B=0.4), 1.078 $\text{\AA}$ (B=0.5), 1.058 $\text{\AA}$ (B=0.6), 1.039 $\text{\AA}$ (B=0.7), 1.019 $\text{\AA}$ (B=0.8), 1.001 $\text{\AA}$ (B=0.9) and 0.984 $\text{\AA}$ (B=1.0). Raw data was generated on a grid using a step size of 0.04 ${\text{\AA}}$ and a cubic spline interpolation was used for plotting purposes. The plot legends display magnetic field strengths given in units of $B_0$.}
\label{fig_qtot_ch_tz}
\end{figure*}

The partial charges $q_{\mathrm{H}}$ and $q_{\mathrm{C}}$ are shown in Fig.\,\ref{fig_qtot_ch_tz}. We note the reflection symmetry across the axis of $-3$ $\Omega/(eB) = - N_\text e/2$, which is necessary because $q_{\mathrm{H}} + q_{\mathrm{C}} = - N_\text e$ must hold at any given bond distance. This axis of symmetry is the same for CH$^+$ and BH given that the two species are isoelectronic. As the bond length changes, the partial charges fluctuate, but there is far less overlap of low/high field strength $q_{\mathrm{H}}$ and $q_{\mathrm{C}}$ curves than in the case of BH. Note that the weak-field curves at $1 \times 10^{-4} B_0$ behave quite differently from those in stronger fields. This is evident for both the screening charges and partial charges. The partial charges begin to exhibit divergent behavior around 2\,${\text{\AA}}$. For this reason, data was not plotted past about 3\,${\text{\AA}}$.

Finally, beyond 2.26\,${\text{\AA}}$, time-reversal symmetry is broken at zero magnetic field, with the wave function becoming complex valued and the RHF stability matrix becoming singular. This is a similar trend as was observed for BH, although in the case of CH$^+$ the bond distance at which broken symmetry occurs is much closer to equilibrium. Real-valued wave functions associated with higher energy beyond this point could be achieved by using the final density from geometry $n-1$ as a guess for geometry $n$. The RHF stability matrix has negative eigenvalues for real valued wave functions in this regime (see Fig.\,\ref{fig_hess}.d). In Fig.\,\ref{fig_hess}.d, the weak field $1 \times 10^{-4} B_0$ curve deviates from the zero field curve at $\approx$ 2.26 ${\text{\AA}}$, and this coincides with a singularity in the weak field DBOC at roughly the same bond distance (Fig.\,\ref{fig_dboc_ch_tz}.d) as well as singularities in the weak field Berry curvature (Fig.\,\ref{fig_QhQC_tz} and Fig.\,\ref{fig_qtot_ch_tz}) as previously stated.

For the purposes of this work, broken time-reversal symmetry for zero magnetic field is of no practical importance as it occurs outside of the range of equilibrium, despite the fact that for CH$^+$ the onset is much closer to equilibrium than for BH. Additionally, we are presently interested in studying the Berry curvature only in the presence of a magnetic field. Nonetheless, as was the case for BH, it is an interesting observation, given that the Berry curvature values in such a domain will become finite even though there is no magnetic field. Moreover, the onset of time-reversal symmetry breaking in the zero field case is correlated with the onset of divergent behavior of the DBOC and partial charges in the weak field case of $1 \times 10^{-4} B_0$. This phenomenon therefore has the potential for future study, but is beyond the scope of the present paper.

\section{Conclusions}

In this work, we have derived the expressions for the analytical calculation of the DBOC and Berry curvature of a molecule in a uniform magnetic field using GHF and RHF wave functions. The RHF equations were implemented in the program package {\sc london},\cite{LondonProgram} which uses London atomic orbitals for gauge-origin invariant calculations of molecules in a magnetic field.

The RHF energy, DBOC, Berry curvature, and Hessian eigenvalues of H$_2$, LiH, BH, and CH$^{+}$ were studied as functions of bond distance and magnetic field strength up to 1.0$B_0$.  Additionally, for H$_2$, molecular dynamics simulations were performed at selected field strengths using the analytic Berry curvature, with and without inclusion of the DBOC. 

All investigated quantities are affected by varying the strength of the magnetic field, but the specific behavior is system dependent. The DBOC was found to depend on the external magnetic field but was in general small and behaved less systematically than the BO potential. Additionally, the magnitude of the DBOC was related to the magnitude of the eigenvalues of the corresponding Hessian matrices, with near-zero Hessian eigenvalues being correlated with erratic or singular DBOC values.

Molecular dynamics simulations were performed on H$_2$ using the analytic Berry curvature both with and without the inclusion of the analytic DBOC for field strengths of 0.1$B_0$ and 1.0$B_0$. It was found that inclusion of the DBOC had little impact on the spectra even at stronger fields where the DBOC was higher in magnitude. This is expected given that the absolute value of the energy correction represented by the DBOC is still quite small at higher field strengths, and the DBOC curves tend to flatten out around equilibrium at higher field strengths for this system. However, given that the DBOC was observed to change in magnitude as a function of field strength, it may be more important to include in the dynamics of other molecular systems in a magnetic field. Also, the DBOC is calculated from the same ingredients as the Berry curvature and therefore can be included in the dynamics at no additional cost.

The Berry curvature results were in general highly system dependent, with each molecular species exhibiting unique screening charges and partial charges as a function of bond distance and magnetic field strength. However, certain common features are present. These are (1) conservation of total electronic charge, where the magnitude of the partial charges $q_1$ and $q_2$ must sum to the total number of electrons for any combination of field strength and bond distance and (2) off-diagonal screening charges $Q_{12} = Q_{21}$ approaching zero in the limit of long bond distances (excepting the $1 \times 10^{-4} B_0$ case for BH and CH$^{+}$, where each exhibited divergent behavior past which no data is reported). The presence of superscreening and antiscreening was also observed for the screening charges across all species.

Lastly, broken time-reversal symmetry was observed for zero field wave functions for BH and CH$^+$. The onset of broken time-reversal symmetry coincides with the deviation of the weak field ($1 \times 10^{-4} B_0$) SCF energy values, DBOC values, and lowest Hessian eigenvalues from their corresponding zero field counterparts. Additionally, the onset of broken time-reversal symmetry also coincides with divergent DBOC and Berry curvature values in the weak field case. These trends are partially explicable in terms of the lowest Hessian eigenvalues for the weak field case being near zero, and the fact that the broken time-reversal symmetry solution in zero field corresponds to a truly singular Hessian, with the weak field case approximating the broken symmetry solution. However, the presence of broken time-reversal symmetry itself in zero magnetic field, as well as the implications for the Berry curvature remain as a topic for future investigation. 

Calculation of the Berry curvature and DBOC analytically allows for the accurate inclusion of the screening force due to the electrons in \textit{ab initio} molecular dynamics simulations as demonstrated by the findings presented here, as well as previous work.\cite{Culpitt2021,Peters2021} Additionally, analytic calculation of the DBOC and Berry curvature provides benefits over the same quantities calculated via finite difference. These benefits are potential savings in computational expense, increased accuracy and overall stability, as well as circumvention of the phase problem that accompanies finite difference calculations with generally complex orbitals.\cite{Culpitt2021} For these reasons, analytic calculation of the Berry curvature is desirable in the context of molecular dynamics in magnetic fields.

\section*{Acknowledgements}

This work was supported by the Research Council of Norway through ‘‘Magnetic Chemistry’’ Grant No.\,287950 and CoE Hylleraas Centre for Quantum Molecular Sciences Grant No.\,262695. This work has also received support from the Norwegian Supercomputing Program (NOTUR) through a grant of computer time (Grant No.\ NN4654K).

\section*{Data Availability}

The data that support the findings of this study are available within the article.

\section*{References}

%

\clearpage




\end{document}